\documentclass[longauth]{aa}  
\usepackage{graphicx}
\usepackage{txfonts}
\usepackage{natbib}
\bibpunct{(}{)}{;}{a}{}{,} 

\usepackage[colorlinks=true,allcolors=blue]{hyperref}
\usepackage{lscape}
\usepackage{xcolor}
\usepackage{multicol}
\usepackage{placeins}

\newcommand{\ms}{$M_{\odot}$}
\newcommand{\rs}{$R_{\odot}$}
\newcommand{\mzams}{$M_{\rm ZAMS}$}
\newcommand{\mej}{$M_{\rm ej}$}
\newcommand{\mhenv}{$M_{\rm H,env}$}
\newcommand{\mh}{$M_{\rm H}$}
\newcommand{\ra}{\emph{R}}
\newcommand{\e}{\emph{E}}
\newcommand{\Ni}{{$^{56}$Ni}}
\newcommand{\mni}{$M_{\rm Ni}$}
\newcommand{\mix}{$^{56}\rm Ni$ mixing}
\newcommand{\te}{$t_{\rm exp}$}
\newcommand{\sneii}{SNe~II}
\newcommand{\snii}{SN~II}
\newcommand{\mmin}{$M_{\rm low}$}
\newcommand{\mmax}{$M_{\rm high}$}

\defcitealias{martinez+20}{M20}
\defcitealias{martinez+21}{Paper~I}
\defcitealias{paper3_submitted}{Paper~III}

\begin{document}

        \title{Type II supernovae from the Carnegie Supernova Project-I}
        \subtitle{II. Physical parameter distributions from hydrodynamical modelling}
        
                \author{L. Martinez \inst{1,2,3}
                \and M.~C. Bersten \inst{1,2,4}
        \and J.~P. Anderson \inst{5}
        \and M. Hamuy \inst{6,7}
        \and S. González-Gaitán \inst{8}
        \and F. Förster \inst{9,10,11,12}
        \and \\ M. Orellana \inst{3,13}
        \and M. Stritzinger \inst{14}
        \and M.~M. Phillips \inst{15}
        \and C.~P. Guti\'errez \inst{16,17}
        \and C.~Burns \inst{18}
        \and C. Contreras \inst{15}
        \and T. de Jaeger \inst{19,20}
        \and K. Ertini \inst{1,2}
        \and G. Folatelli \inst{1,2,4}
        \and L. Galbany \inst{21}
        \and P. Hoeflich \inst{22}
        \and E.~Y. Hsiao \inst{22}
        \and N. Morrell \inst{15}
        \and P.~J. Pessi \inst{2,5}
        \and \\ N.~B. Suntzeff \inst{23}
        }

        \institute{Instituto de Astrofísica de La Plata (IALP), CCT-CONICET-UNLP. Paseo del Bosque S/N, B1900FWA, La Plata, Argentina \\
                        \email{laureano@carina.fcaglp.unlp.edu.ar}
                \and
                        Facultad de Ciencias Astronómicas y Geofísicas,
                        Universidad Nacional de La Plata, Paseo del Bosque S/N, B1900FWA, La Plata, Argentina
                \and
            Universidad Nacional de Río Negro. Sede Andina, Mitre 630 (8400) Bariloche, Argentina
        \and
            Kavli Institute for the Physics and Mathematics of the Universe (WPI), The University of Tokyo, 5-1-5 Kashiwanoha, Kashiwa, Chiba 277-8583, Japan
        \and
            European Southern Observatory, Alonso de Córdova 3107, Casilla 19, Santiago, Chile
        \and
            Vice President and Head of Mission of AURA-O in Chile, Avda. Presidente Riesco 5335 Suite 507, Santiago, Chile
        \and
            Hagler Institute for Advanced Studies, Texas A\&M University, College Station, TX 77843, USA
        \and
            CENTRA-Centro de Astrofísica e Gravitaçäo and Departamento de Física, Instituto Superio Técnico, Universidade de Lisboa, Avenida Rovisco Pais, 1049-001 Lisboa, Portugal
        \and
            Data and Artificial Intelligence Initiative, Faculty of Physical and Mathematical Sciences, University of Chile
        \and
            Centre for Mathematical Modelling, Faculty of Physical and Mathematical Sciences, University of Chile
        \and
            Millennium Institute of Astrophysics, Chile.
        \and
            Department of Astronomy, Faculty of Physical and Mathematical Sciences, University of Chile
        \and
            Consejo Nacional de Investigaciones Científicas y Técnicas (CONICET), Argentina.
        \and
            Department of Physics and Astronomy, Aarhus University, Ny Munkegade 120, DK-8000 Aarhus C, Denmark
        \and
            Carnegie Observatories, Las Campanas Observatory, Casilla 601, La Serena, Chile
        \and 
            Finnish Centre for Astronomy with ESO (FINCA), FI-20014 University of Turku, Finland
        \and
            Tuorla Observatory, Department of Physics and Astronomy, FI-20014 University of Turku, Finland
        \and 
            Observatories of the Carnegie Institution for Science, 813 Santa Barbara St., Pasadena, CA 91101, USA
        \and
            Institute for Astronomy, University of Hawaii, 2680 Woodlawn Drive, Honolulu, HI 96822, USA
        \and
            Department of Astronomy, University of California, 501 Campbell Hall, Berkeley, CA 94720-3411, USA.
        \and
            Institute of Space Sciences (ICE, CSIC), Campus UAB, Carrer de Can Magrans, s/n, E-08193 Barcelona, Spain.
        \and
            Department of Physics, Florida State University, 77 Chieftan Way, Tallahassee, FL 32306, USA
        \and
            George P. and Cynthia Woods Mitchell Institute for Fundamental Physics and Astronomy, Department of Physics and Astronomy, Texas A\&M University, College Station, TX 77843
            }

\titlerunning{Progenitor and explosion properties of SNe~II}

\date{Received XXX; accepted XXX}
 
\abstract 
{Linking supernovae to their progenitors is a powerful method for furthering our understanding of the physical origin of their observed differences while at the same time testing stellar evolution theory.
In this second study of a series of three papers where we characterise type II supernovae (\sneii) to understand their diversity, 
we derive progenitor properties (initial and ejecta masses and radius), explosion energy, and \Ni\ mass and its degree of mixing within the ejecta for a large sample of \sneii. 
This dataset was obtained by the Carnegie Supernova Project-I and is characterised by a high cadence of SNe II optical and near-infrared light curves and optical spectra that were homogeneously observed and processed.
A large grid of hydrodynamical models and a fitting procedure based on Markov chain Monte Carlo methods were used to fit the bolometric light curve and the evolution of the photospheric velocity of 53 \sneii.
We infer ejecta masses of between 7.9 and 14.8~\ms, explosion energies between 0.15 and 1.40~foe, and \Ni\ masses between 0.006 and 0.069~\ms. 
We define a subset of 24~SNe (the `gold sample') with well-sampled bolometric light curves and expansion velocities for which we consider the results more robust.
Most \sneii\ in the gold sample ($\sim$88\%) are found with ejecta masses in the range of $\sim$8$-$10~\ms, coming from low zero-age main-sequence masses (9$-$12~\ms). 
The modelling of the initial-mass distribution of the gold sample gives an upper mass limit of 21.3$^{+3.8}_{-0.4}$~\ms\ and a much steeper distribution than that for a Salpeter massive-star initial mass function (IMF). 
This IMF incompatibility is due to the large number of low-mass progenitors found -- when assuming standard stellar evolution. This may imply that high-mass progenitors lose more mass during their lives than predicted. However, a deeper analysis of all stellar evolution assumptions is required to test this hypothesis.}


\maketitle
\section{Introduction}
\label{sec:intro} 

Core-collapse supernovae (CCSNe) are produced by the explosion of massive stars (>8$-$10~\ms). The collapse is initiated when the iron core achieves the Chandrasekhar mass \citep{chandrasekhar1939}. When the collapsing core reaches nuclear densities, the implosion rebounds, generating a shock wave that stalls into accretion due to the flat density profile around the iron core.
However, neutrino heating from the proto-neutron star revives the shock and produces the supernova (SN) explosion. This is the so-called delayed neutrino-driven mechanism, and it is the most favoured explosion mechanism for CCSNe \citep[see e.g.][for a recent review]{burrows+21}.

Type II supernovae (\sneii) are the most common type of CCSN in nature \citep{li+11a,shivvers+17}. In the context of single stellar evolution, they are believed to arise from the least massive stars among the core-collapse mechanism.
Type II SNe are classified by the presence of hydrogen lines in their optical spectra \citep{minkowski41}.
Previous theoretical studies have shown that an extensive hydrogen-rich envelope, typical of red supergiant (RSG) stars, is required to reproduce \snii\ light-curve (LC) morphologies \citep[e.g.][]{grassberg+71,chevalier76}.
More recently, this picture was confirmed by the detection of a significant number of RSG stars at the position of \snii\ explosion sites on archival images prior to explosion, demonstrating that they are the progenitors of most \sneii\ \citep[e.g.][]{vandyk+03,smartt09}.
Type II SNe\ were historically grouped into SNe~IIP and SNe~IIL categories based on the shape of their LCs \citep{barbon+79}. However, in this paper we use `\sneii' to refer to both groups together since recent studies indicate that they come from a continuous population \citep[][although see \citealt{davis+19} for distinct populations in near-infrared spectral features]{anderson+14_lc,sanders+15,galbany+16,valenti+16,rubin+16,dejaeger+19}.
Other hydrogen-rich SNe (SNe~IIb, SNe~IIn, and SN~1987A-like events) are not analysed in this study, and they will not be discussed.

During the last few decades, several works have focused on the analysis of large samples of \sneii\ to examine their photometric and spectroscopic diversity \citep[e.g.][]{patat+94,hamuy03,bersten+09,arcavi+12,anderson+14_ha,anderson+14_lc,faran+14a,faran+14b,spiro+14,gonzalez+15,gutierrez+14,gutierrez+17II,gutierrez+17I,dejaeger+18,davis+19}.
The great diversity observed in \sneii\ is indicative of a large variety of progenitor and explosion properties. 
A key aim of SN research is to determine the full range of parameters that produce these events and constrain the predominant physical properties that yield the observed diversity. Obtaining such knowledge is critical to furthering our understanding of how massive stars evolve through their lives to produce hydrogen-rich events, along with the properties of the explosion mechanism.

There are different approaches to connecting the characteristics of SNe with their progenitors (e.g. pre-SN imaging, LC modelling, and nebular-phase spectral modelling), but here we concentrate on those methods that use LCs and spectral information to extract the physical properties of SNe II.
Since the pioneering works of \citet{litvinova+83,litvinova+85}, who presented a set of analytical relations connecting SN observables (the absolute $V$ magnitude and photospheric velocity at mid-plateau and the plateau duration) with progenitor and explosion properties (ejecta mass, progenitor radius, and explosion energy), significant efforts have been made to improve the estimation of confident physical parameters.
While the concept of using simple relations to derive progenitor properties is in principle powerful, atypical values have sometimes been obtained \citep[see e.g.][]{hamuy03}, motivating advances in this methodology.
The hydrodynamical models utilised to calibrate these analytical relations were based on simplified physical assumptions (e.g. simple polytropic models as initial configurations, old opacity tables, and the omission of \Ni\ heating in the calculations).
Similar relations were presented in \citet{popov93}, although they were built from a two-zone analytical model for hydrogen recombination in \snii\ ejecta \citep[see also][for an analytic analysis for \snii\ LCs]{arnett80}.
New analytical relations based on modern hydrodynamical models (including the effects of \Ni\ heating and contemporary opacity tables) were presented by \citet{kasen+09} and more recently by \citet{goldberg+19}. 
While these methods enable the user to extract physical parameters of \sneii\ in a relatively simple manner, it is not clear how accurate such constraints are because they are obtained using only three observables.

An alternative method for deriving the physical parameters of SNe II\  consists of detailed modelling of the complete LC evolution -- sometimes modelled together with spectral information. In such techniques, the explosion is often simulated by artificially adding internal energy near the centre of the star, known as a `thermal bomb'. 
Here, the structure of the progenitor prior to core collapse has to be assumed, together with the explosion energy, \Ni\ yields, and \Ni\ distribution within the ejecta \citep[e.g.][]{utrobin07,bersten+11,morozova+15}. These parameters are freely chosen to reproduce the observations. 
Most studies use hydrodynamical codes that solve the radiation transport in the diffusion approximation, thus producing bolometric LCs.
More sophisticated codes incorporate radiation-hydrodynamical modelling with LC and spectral information \citep{pumo+11} or multi-group radiative transfer that produces multi-band LC simulations \citep{blinnikov+98}.
Additionally, radiative transfer codes have been developed to calculate spectra in rapidly expanding SN atmospheres, assuming homologous expansion \citep[e.g.][]{dessart+05,kasen+06}.
Further sophistication can be attained by calculating more realistic explosions using CCSN simulations, parameterising the delayed neutrino-driven mechanism, and coupling this to explosive nucleosynthesis calculations. In this context, the explosion energy, \Ni\ mass, and distributions are no longer free parameters. 
These calculations simulate the collapse and explosion, but they need alternative codes to reproduce SN observables.
Recent efforts have been made in this direction \citep{sukhbold+16,curtis+21,barker+21}, where multi-dimensional explosions are sometimes mimicked in 1D models and mapped to different codes to further obtain SN observables.
While this methodology is more consistent (given that it only needs the progenitor structure as input), sometimes it cannot reproduce the observations and is restricted to a limited parameter space.

Progenitor models can also be constructed in different ways, either as an ad hoc configuration or with detailed stellar evolution from main sequence to core collapse.
In general, non-rotating single-star models are used, whereas binarity or stellar evolution with enhanced mass loss could have a significant effect on the structure of the progenitor.
Stellar evolution produces a final structure for each star, for which the pre-explosion mass and the progenitor size are not independent.
To treat the progenitor final mass and radius as independent parameters, non-evolutionary models -- polytropic models -- are used \citep[e.g.][]{utrobin07,bersten+11}.
These models can also mimic Rayleigh-Taylor (RT) mixing during shock propagation by smoothing the transitions between zones of different chemical abundances \citep{utrobin+08}.
It is important to note that different approaches assumed in the construction of pre-SN models and the subsequent explosion and transient modelling may produce different results because of their assumed physics and the assumed free parameters.

Many studies have been published that show hydrodynamical modelling of individual \sneii. Recently, the number of multiple studies of \sneii\ has increased \citep[e.g.][]{pumo+17,ricks+19,utrobin+19,eldridge+19b,martinez+19,martinez+20}. 
However, there are still only a few analyses of statistically significant samples \citep{gonzalez+15,morozova+18,forster+18}.
In \citet[][hereafter \citetalias{martinez+20}]{martinez+20}, we analysed a sample of eight \sneii\ with observed progenitors by fitting their bolometric LC and photospheric velocity evolution. For this purpose, we presented a large grid of explosion models and a fitting procedure based on Markov chain Monte Carlo (MCMC) methods. We found that our results are consistent with those from pre-SN imaging and late-time spectral modelling.
The analysis by \citetalias{martinez+20} served as the basis for the present study, where we derive physical parameters for a large sample of \sneii\ from a homogeneous dataset. 

The present work is the second in a series of three papers based on the study of a sample of 74~\sneii\ observed by the Carnegie Supernova Project-I \citep[CSP-I;][]{hamuy+06}. CSP-I was a SN follow-up programme based at the Las Campanas Observatory that obtained high-quality optical and near-infrared (NIR) LCs and optical spectra with high observational cadence.
In \citet[][hereafter Paper~I]{martinez+21}, we calculated the bolometric LCs for this sample of \sneii,\ discussing our methodology in detail and presenting an analysis of the observed bolometric parameters. An additional conclusion of \citetalias{martinez+21} is the importance of having NIR observations to derive reliable bolometric luminosities.
Here, in Paper~II, we assess the progenitor and explosion properties for this sample of events and characterise the physical parameter distributions.
Finally, in \citet[][hereafter Paper~III]{paper3_submitted}, we study correlations between physical parameters and different LC and spectroscopic measurements, and analyse \snii\ diversity, tying this to the physics of massive star explosions and their progenitors.

The current paper is organised as follows. In the following section we briefly outline the physical processes involved in \snii\ explosions.
Section~\ref{sec:sample} describes the data sample. In Sect.~\ref{sec:methods} we present the hydrodynamical simulations and the fitting procedure used to derive the physical parameters of the sample. Section~\ref{sec:results} presents the distributions of the physical parameters for our sample of \sneii. 
In Sect.~\ref{sec:analysis} we compare our findings with those presented in previous works. In Sect.~\ref{sec:discussion} we discuss possible explanations for the initial mass distribution found. We provide our concluding remarks in Sect.~\ref{sec:conclusions}.
Further analysis and figures not included in the main body of the manuscript are presented in the appendices.

\section{\snii\ physics}
\label{sec:physics}

The observed behaviour of \snii\ LCs and expansion velocities is related to physical properties of the progenitor star and the explosion; such as the ejecta mass ($M_{\rm ej}$), the progenitor radius ($R$), the explosion energy ($E$), the amount of \Ni\ (\mni), and the distribution of \Ni\ in the ejecta (\mix).
Below, different phases of \sneii\ evolution are discussed in the context of their physical processes.
Detailed reviews can be found in the literature \citep[e.g.][]{arnett96book,woosley+02,zampieri17book}.

During the core collapse, a powerful shock wave is generated near the centre of the star. The shock wave heats and accelerates the progenitor envelope until it arrives at the stellar surface, where photons begin to diffuse outwards (a phase known as the `shock breakout'). 
After the shock breakout, the outermost layers of the ejecta expand at great speed leading to a fast decrease in the temperature and, therefore, causing a rapid decline in the bolometric luminosity.
This phase is commonly known as the `cooling phase', and it is directly related to the progenitor size. 
However, this phase is also closely related to any material near the stellar surface. 

Type II SNe\ enter the recombination phase when the temperature drops to $\sim$6000~K. This phase is commonly referred to in the literature as the `plateau', although it is not necessarily a phase of constant luminosity.
During this phase, hydrogen recombination takes place at different layers of the ejecta as a recombination wave recedes (in mass coordinate) through the expanding ejecta \citep[e.g.][]{grassberg+71,bersten+11}.
Therefore, the hydrogen-rich envelope mass of the progenitor star at the time of explosion (\mhenv) is related to the duration of the plateau phase, although other physical parameters also play a role \citepalias[see][]{paper3_submitted}.
In addition, the explosion energy significantly drives SN~II observational properties \citep{kasen+09,dessart+13,bersten13phd}.
More energetic explosions produce more luminous \sneii\ expanding at higher velocities, thus cooling and recombining the ejecta more rapidly, and producing shorter plateau phases.
On the other hand, more extended progenitors produce more luminous and longer plateau phases.
The additional heating of the \snii\ ejecta at late times by the \Ni\ decay chain extends the duration of the plateau \citep{kasen+09} and increases the luminosity in the late-plateau phase \citep{bersten13phd}.
The mixing of \Ni\ determines when energy deposition from radioactive decay starts to influence the LC, affecting the duration of the plateau and its shape \citep{bersten+11,kozyreva+19}. An extensively mixed \Ni\ impacts the LC earlier \citep[see][their Fig.~12]{bersten+11}, that is, \Ni\ powers the LC sooner, producing a slowly declining plateau (typical SN~IIP). 
The cooling and plateau phases together are also known as the optically thick or photospheric phase.

When the hydrogen-rich ejecta is recombined, \sneii\ enter a transition phase, which is marked by a rapid decline in luminosity. After this, the luminosity is mainly powered by the $^{56}$Co decay. This phase is known as the radioactive tail and its luminosity primarily depends on the amount of \Ni\ in the ejecta.

\section{Supernova sample}
\label{sec:sample}

The sample of \sneii\ used in this study is the same as that analysed in \citetalias{martinez+21} where we present bolometric LCs for 74~\sneii\ observed by the CSP-I \citep[][PIs: Phillips and Hamuy; 2004$-$2009]{hamuy+06}.
The sample is characterised by a high cadence and quality of the observations. In addition, CSP-I LCs cover a wide wavelength range from optical ($uBgVri$) to NIR ($YJH$) bands. 
The data were homogeneously observed and processed \citep[see][for a detailed description of the data reduction and photometric calibration]{hamuy+06,contreras+10,stritzinger+11,folatelli+13,krisciunas+17}.
CSP-I \snii\ photometry is presented in Anderson et al. (in prep.), while the optical spectra were published by \citet{gutierrez+17II,gutierrez+17I}.

CSP-I was a SN follow-up programme that generally obtained observations for any SN that (a) was sufficiently bright to observe for a number of weeks; (b) had reasonable sky visibility to enable such observations; and (c) where the classification spectrum indicated a reasonably young explosion. This resulted in a magnitude-limited sample of SNe~II, which we use in the current study. Thus our sample will be biased towards the inclusion of intrinsically brighter SNe~II. We discuss how this may affect our results and conclusions in Sect.~\ref{sec:imf_incompatibility}.

In the present work, we use the bolometric LCs from \citetalias{martinez+21} and the \ion{Fe}{ii}~5169~\AA\ line velocities from \citet{gutierrez+17II} to infer the physical properties of the \sneii\ in the sample through hydrodynamic modelling (Sect.~\ref{sec:results}).
The explosion epochs are taken from \citet[][see also Table 1 from \citetalias{martinez+21} for details]{gutierrez+17I}.

\section{Methods}
\label{sec:methods}

\subsection{Hydrodynamical simulations}
\label{sec:hydro_models}

\begin{table}
\caption{Range of physical parameters used to compute the grid of explosion models.}
\label{tab:models}
\centering
\resizebox{0.5\textwidth}{!}{
\begin{tabular}{cccc}
\hline\hline\noalign{\smallskip}
Parameter & Range & In steps of & Additional values \\
\hline\noalign{\smallskip}
\mzams\ & 9$-$25~\ms\ & 1~\ms\ & --- \\
\e\ & 0.1$-$1.5~foe\tablefootmark{a} & 0.1~foe & --- \\
\mni\ & 0.01$-$0.08~\ms\ & 0.01~\ms\ & 10$^{-4}$, 5$\times$10$^{-3}$~\ms\ \\
\mix\ & 0.2$-$0.8\tablefootmark{b} & 0.3 & --- \\
\hline
\end{tabular}}
\tablefoot{
\tablefoottext{a}{Because of numerical difficulties, the models for the largest masses and lowest energies were not computed \citepalias[see][for details]{martinez+20}.}
\tablefoottext{b}{Given in fraction of the pre-SN mass.}
}
\end{table}

The determination of the physical properties of \sneii\ is based on comparing models constructed with different physical parameters with observations. 
In the current work, we used the grid of hydrodynamic models presented in \citetalias{martinez+20}. These models were calculated using a 1D Lagrangian code that simulates the explosion of the SN and produces bolometric LCs and expansion velocities at the photospheric layers \citep[see][for details]{bersten+11}. The grid of explosion models comprises a wide range of parameters \citepalias[Table~\ref{tab:models} and][for details]{martinez+20}.
Pre-SN models in hydrostatic equilibrium are necessary to initialise the explosion. The public stellar evolution code \texttt{MESA\footnote{\url{http://mesa.sourceforge.net/}}} \citep{paxton+11,paxton+13,paxton+15,paxton+18,paxton+19} was used to obtain non-rotating solar-metallicity pre-SN RSG models with initial masses (\mzams) in the 9$-$25~\ms\ range\footnote{The 9, 10, and 11~\ms\ progenitor models were calculated up to the end of core carbon burning since the evolution to core collapse for these stars is computationally expensive. However, we note that these massive stars develop an iron core that eventually collapses \citep[e.g.][]{sukhbold+16}.}.
Massive-star models are affected by the uncertainties in stellar modelling, such as the treatment of convection, rotation, and mass loss. In this context, we utilised the prescriptions adopted by the community to model these physical processes, and adopted standard values for the parameters involved to test standard single-star evolution against \snii\ LC and velocity modelling. 

Following the above, for convection we adopted the Ledoux criterion and a mixing-length parameter of $\alpha_{\rm mlt}$\,=\,2.0. We used exponential overshooting parameters of $f_{\rm ov}$\,=\,0.004 and $f_{\rm ov,D}$\,=\,0.001, a semiconvection efficiency of $\alpha_{\rm sc}$~=~0.01, and thermohaline mixing with the coefficient $\alpha_{\rm th}$~=~2 according to \citet{farmer+16}. We adopted the `Dutch' prescriptions for the wind mass loss defined in the \texttt{MESA} code with an efficiency $\eta$~=~1.
For each value of \mzams\ (9$-$25~\ms\ with 1~\ms\ increment), there is a corresponding value of \mej\ and \ra. In this grid of simulations, \mej\ cover a range of 7.9$-$15.7~\ms, while \ra\ is found in the range of 445$-$1085~\rs, similar to the values derived for RSGs \citep{levesque+05}.

\subsection{Fitting procedure}
\label{sec:mcmc}

A fitting procedure based on MCMC methods was employed to find the posterior probability of the model parameters given the observations. This technique was first used for SNe in \citet{forster+18}. A similar procedure to that used in the current work was presented in \citetalias{martinez+20}. The MCMC is implemented via the \texttt{emcee} package \citep{emcee}.
Since the models presented in \citetalias{martinez+20} may not be sufficient when fitting SN observations using statistical inference techniques, we used the interpolation method presented in \citet{forster+18}. This is a robust and quick method that allows for irregular grids of models in the space of parameters to be used.
For the MCMC runs, we used 400 parallel samplers (or walkers) and 10000 steps per sampler, with a burn-in period of 5000 steps. The walkers were randomly initialised covering the entire parameter space \citepalias[see Sect.~4 of][for details]{martinez+20}.

As in \citetalias{martinez+20}, there are six parameters in our model. Four of them are physical parameters of the explosions and their progenitors: \mzams, \e, \mni, and \mix. We included two additional parameters: the explosion epoch (\te) and a parameter named `scale'. 
The scale parameter multiplies the bolometric luminosity by a constant dimensionless factor to allow for errors in the bolometric LC because of the uncertainties in the distance and host-galaxy extinction. 
We used uniform priors for the following parameters: \te, \mzams, \e, \mni, and \mix. We allowed the sampler to run within the observational uncertainty of \te\ \citepalias[see Table~1 from][]{martinez+21} and within the ranges of the physical parameters in our grid of explosion models (Table~\ref{tab:models}).
We used a Gaussian prior for scale, where the Gaussian is centred at one. 
The uncertainty in the distance allows the SN to yield a more or less luminous bolometric LC. For this effect, we allowed variations to the bolometric LC produced by scale values within $\pm$1$\sigma_{d}$, where $\sigma_{d}$ is the relative difference in luminosity because of the uncertainty in the distance.
The effect of unaccounted host-galaxy extinction allows the SN to be only intrinsically more luminous, making the scale prior asymmetric.
To mimic this effect, we let the scale vary up to $+$1$\sigma_{A_{V}host}$ (i.e. only in the direction that would make the SN brighter), where $\sigma_{A_{V}host}$ is the relative difference in bolometric luminosity produced by taking a host-galaxy extinction equal to two standard deviations of the measurements of the host extinction from \citet{anderson+14_lc}\footnote{We note that the validity of different methods to estimate the host-galaxy extinction has been questioned \citep{poznanski+11,phillips+13}.}.
To summarise, the Gaussian prior for the scale is asymmetric. If the scale takes values below unity, the standard deviation is equal to $\sigma_{d}$, and $\sigma_{d}$~+~$\sigma_{A_{V}host}$ everywhere else.

Extinction also changes the shape of the bolometric LC\footnote{Extinction affects the bolometric luminosity differently as a function of time. When the SN is intrinsically blue, the effect on the bolometric luminosity is larger. On the contrary, when the SN in intrinsically red, the effect of extinction on the bolometric luminosity is smaller.}.
The scale prior does not consider different bolometric LC shapes.
However, we tested how much the shape of the bolometric LC changes when different host-galaxy extinctions are considered and found that the shape is only significantly affected at early times ($\lesssim$\,30~days after explosion) and for large extinction values. Given that we removed the first 30~days of evolution from the fitting (see Sect.~\ref{sec:modelling_background}), this effect is not considered in our prior.

\subsection{Modelling background}
\label{sec:modelling_background}

We used the models and the fitting procedure described in the previous subsections to derive the physical properties of the \sneii\ in our sample. 
Expansion velocities are important as they provide restrictions to the ejecta expansion rate.
In \citetalias{martinez+20}, we showed that in some cases, fits to the expansion velocities are crucial in breaking the LC degeneracies. In addition, the physical parameters derived from fits to the LC alone are significantly different to those that also include velocity data.

The bolometric LCs from the CSP-I sample were already presented in \citetalias{martinez+21}. The photospheric velocity is typically estimated by measuring the velocity at maximum absorption of optically thin lines \citep{leonard+02}. Consequently, we used the \ion{Fe}{ii}~5169~\AA\ line -- measured by \citet{gutierrez+17I} -- as a photospheric velocity indicator.
This assumption is extensively used in the literature given that \citet{dessart+05} showed that \ion{Fe}{ii} line delivers high accuracy in reproducing the photospheric velocity. However, these results are restricted to a minimum velocity of $\sim$4000~km~s$^{-1}$. 
Recently, \citet{paxton+18} proposed an alternative approximation to model the ejecta velocities. They calculated the \ion{Fe}{ii} line velocity in the Sobolev approximation and compared with the observed \ion{Fe}{ii} velocities. 
\citetalias{martinez+20} analysed the variation on the physical parameters if different model velocities are used to compare to observations. They found a tendency to larger \mej\ and \e\ values of, on average, $\sim$0.2~\ms\ and $\sim$0.2~foe, respectively.
While differences exist, they do not significantly alter the results. The reader is referred to Sect.~5.2 of \citetalias{martinez+20} for details.

Recently, it has been proposed that some \sneii\ show signatures of interaction with a dense circumstellar material (CSM) shell surrounding the star \citep[e.g.][]{gonzalez+15,khazov+16,yaron+17,forster+18,bruch+21}.
The interaction of the ejecta with a low-mass CSM is thought to only significantly affect the early evolution, with negligible effect at later epochs where the evolution is dominated by the hydrogen recombination and radioactive decay \citep{morozova+18,hillier+19}.
Given that we focus on inferring the intrinsic properties of \snii\ progenitors instead of characterising the CSM properties, our explosion models were calculated without including CSM. Therefore, our fitting is not valid during the early evolution of \sneii. 
For this reason, we did not consider the first 30~days of evolution of the observed LC in our fitting procedure. This value is similar to the median of the distribution of the cooling phase duration determined in \citetalias{martinez+21}.
Differences between our models and observations are therefore to be expected during the cooling phase, which may be strongly affected if CSM interaction exists. 
The results of LC and ejecta velocity modelling are presented in Sect.~\ref{sec:results}, but here we note that around 60\% of the \sneii\ in the CSP-I sample with photometric observations earlier than 30~days after explosion show differences in the early bolometric LC. Those \sneii\ that show differences between our models and observations during the first 30~days are good candidates to study the general properties of the CSM.
In spite of this, we did not remove the early data from the velocity evolution for two reasons. On the one hand, the effect of moderate CSM on the velocity evolution, as expected for normal \sneii, is less important than for the LC \citep{englert+20}.
On the other hand, its effect is to reduce the velocities only at early times \citep{moriya+18}.
When modelling \sneii\ with interaction signatures using models that do not account for CSM interaction, the models should display higher or similar early velocities than observed, given that these models may serve as the basis for future models considering interaction with a CSM. Then, these new models would show lower early velocities than our models, possibly consistent with observations.
If early velocities are not taken into account, our best-fitting model could display higher or lower velocities than early-time observations. 
From the above discussion, the former case is not an issue. 
However, the latter case would be incompatible within the above scenario, since any future models including CSM interaction would have early velocities even lower than observed. Therefore, to avoid this incompatibility, we decided not to remove the early velocities from the fitting.

\section{Results}
\label{sec:results}

\begin{figure}
\centering
\includegraphics[width=0.44\textwidth]{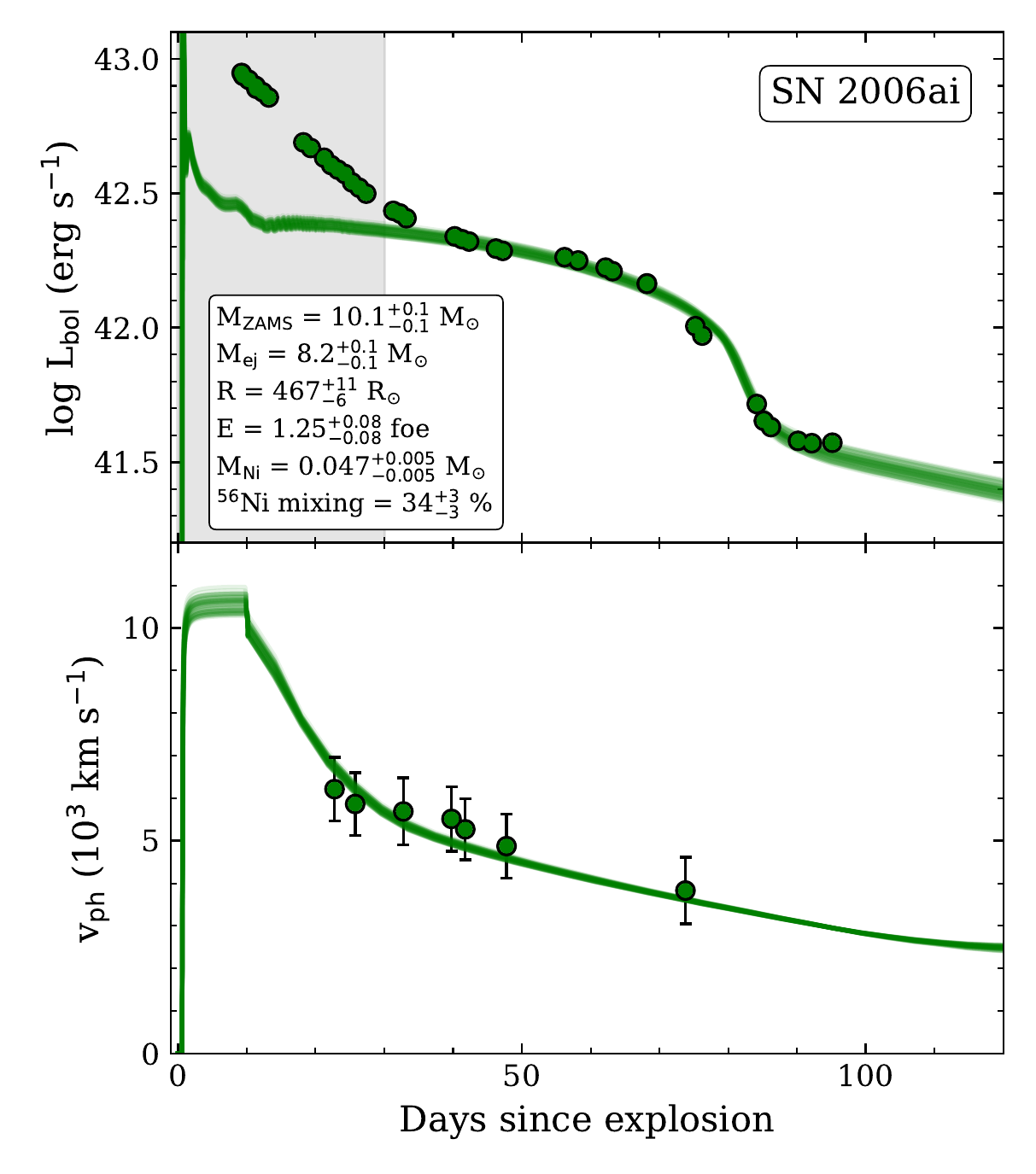}
\caption{Observed bolometric LC (top panel) and \ion{Fe}{ii} $\lambda$5169 line velocities (bottom panel) of SN~2006ai (filled markers) with models randomly sampled from the posterior distribution of the parameters (solid lines). The median value and 68\% confidence range for every physical parameter are also shown. The grey shaded region shows the early data removed from the fitting (first 30~days after explosion).}
\label{fig:2006ai}
\end{figure}

\begin{figure}
\centering
\includegraphics[width=0.45\textwidth]{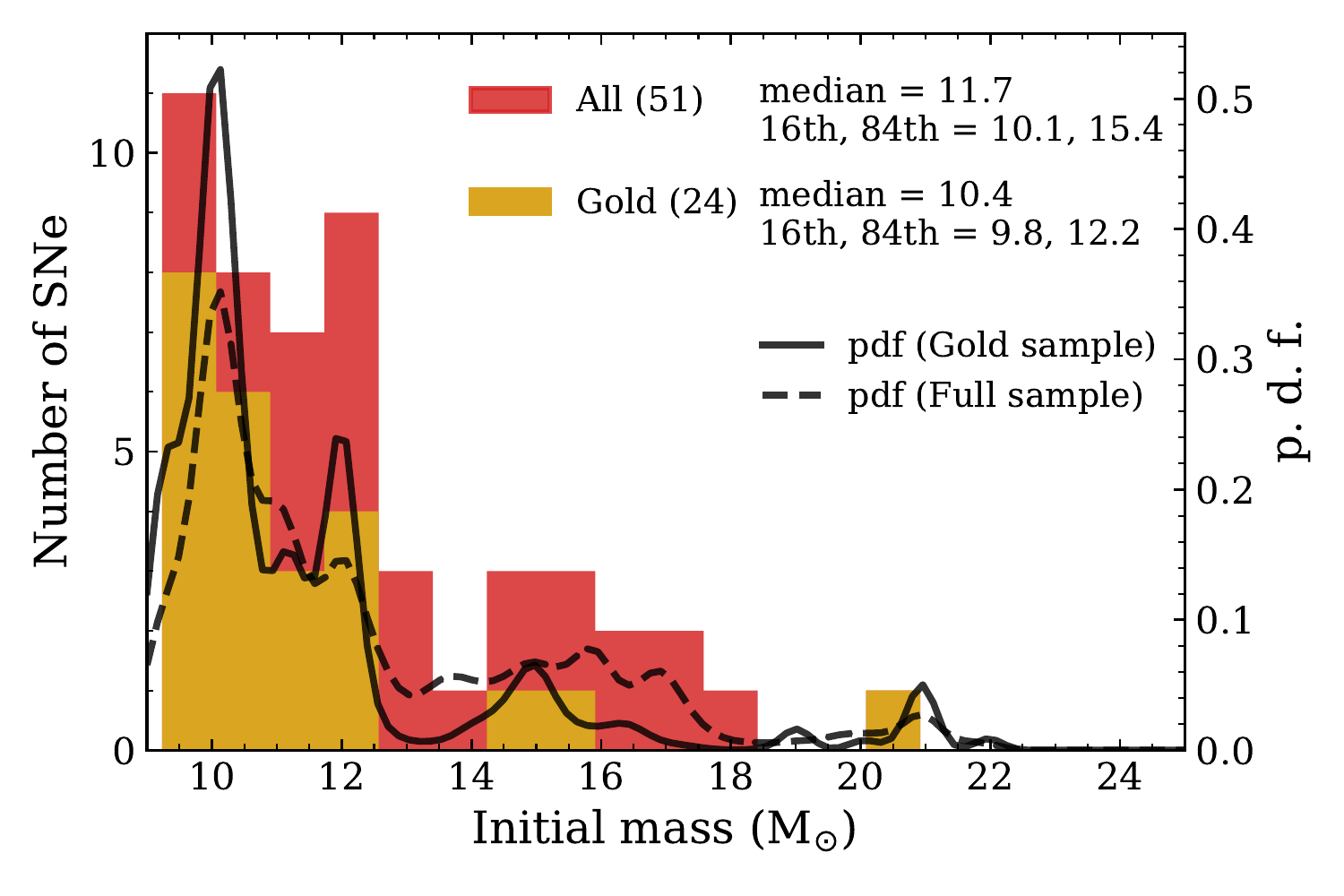} 

\includegraphics[width=0.45\textwidth]{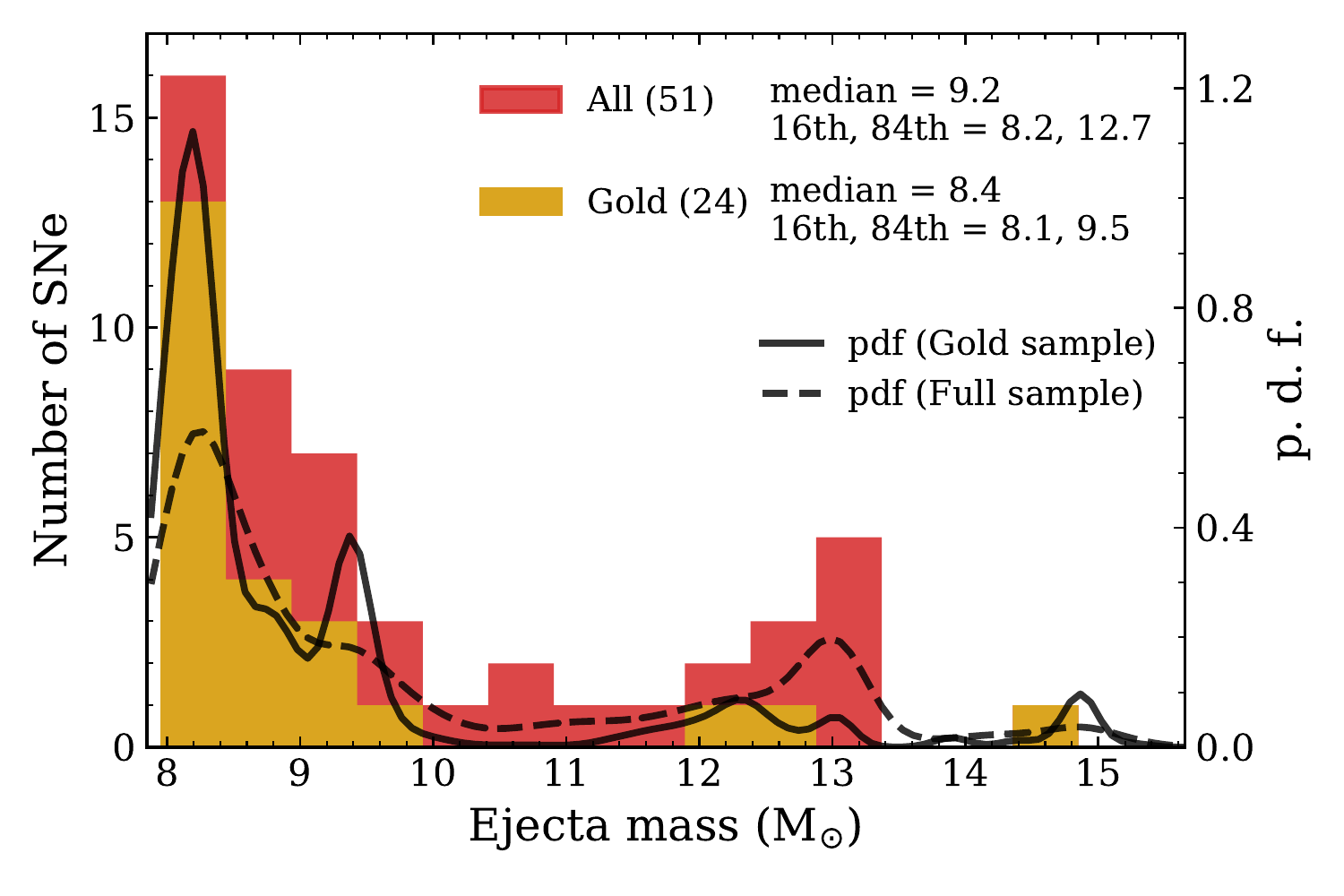} 

\includegraphics[width=0.45\textwidth]{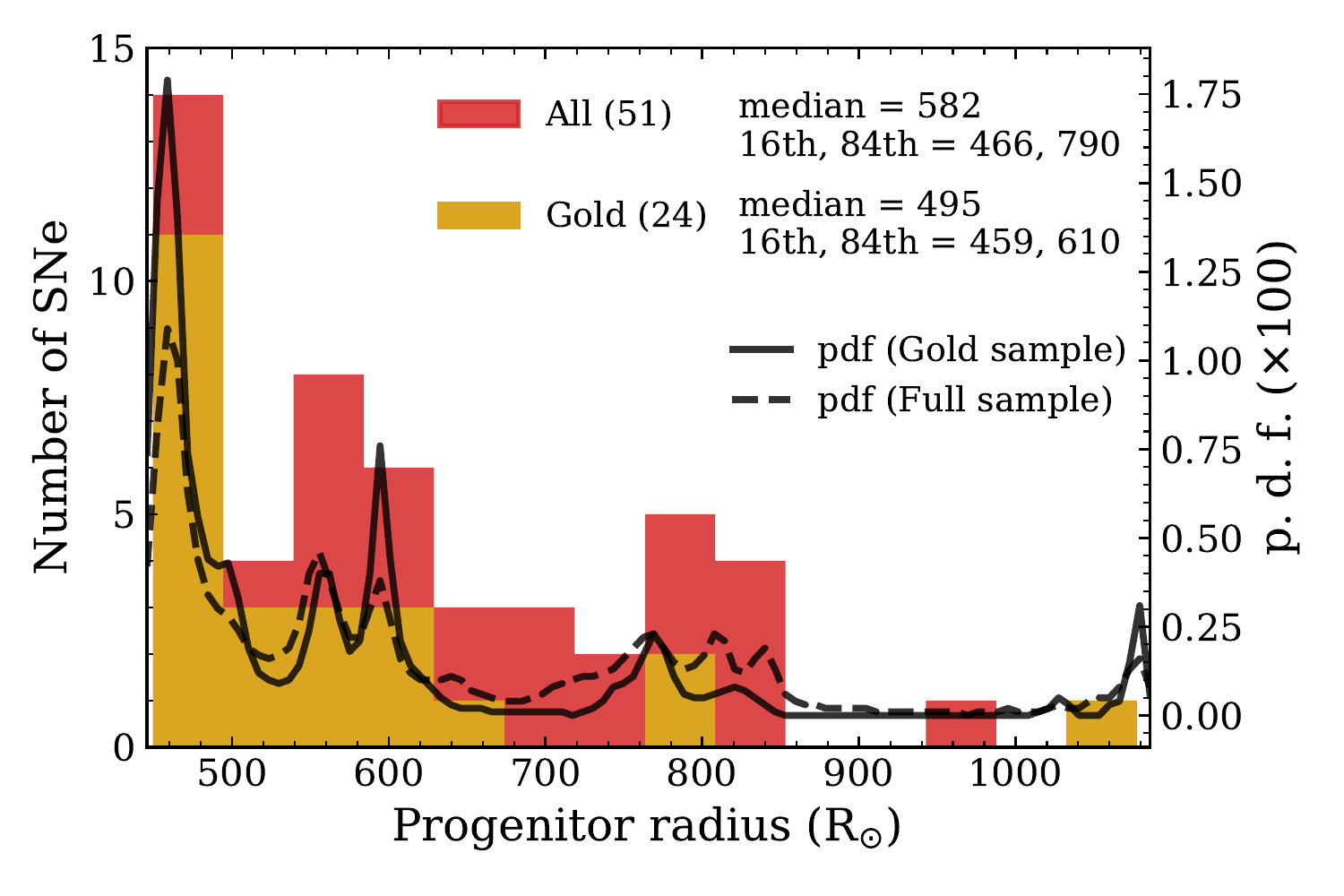}
\caption{Histograms of three progenitor parameters: \mzams\ (top panel), \mej\ (middle panel), and \ra\ (bottom panel). The gold sample is represented by yellow bars, and the red bars are the histograms for the full sample. In each panel, the number of \sneii\ is listed, together with the median and the 16th and 84th percentiles. Probability density functions of the physical parameters for the gold and full samples are represented by solid and dashed lines, respectively.}
\label{fig:hist1}
\end{figure}

\begin{figure}
\centering
\includegraphics[width=0.45\textwidth]{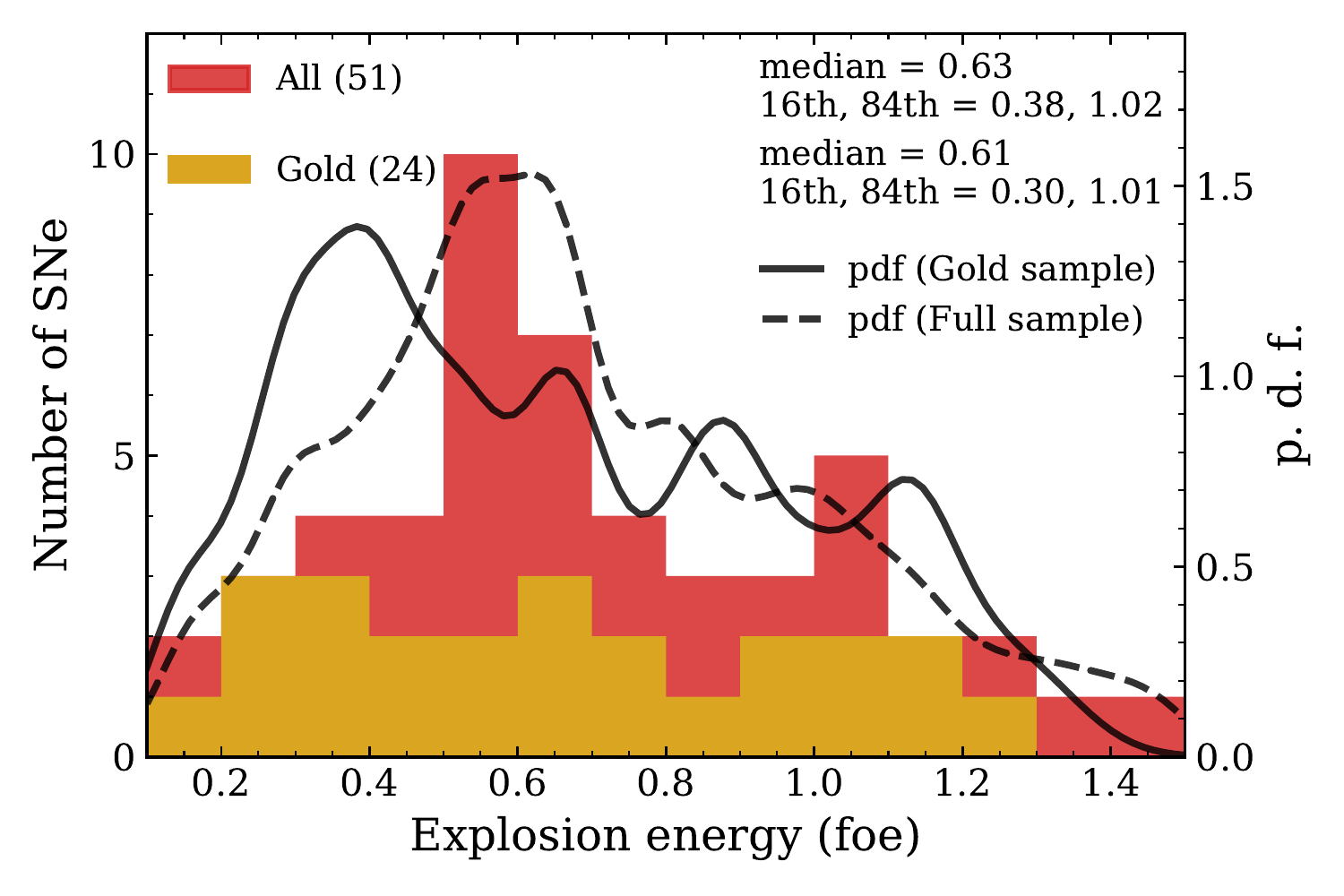} 

\includegraphics[width=0.45\textwidth]{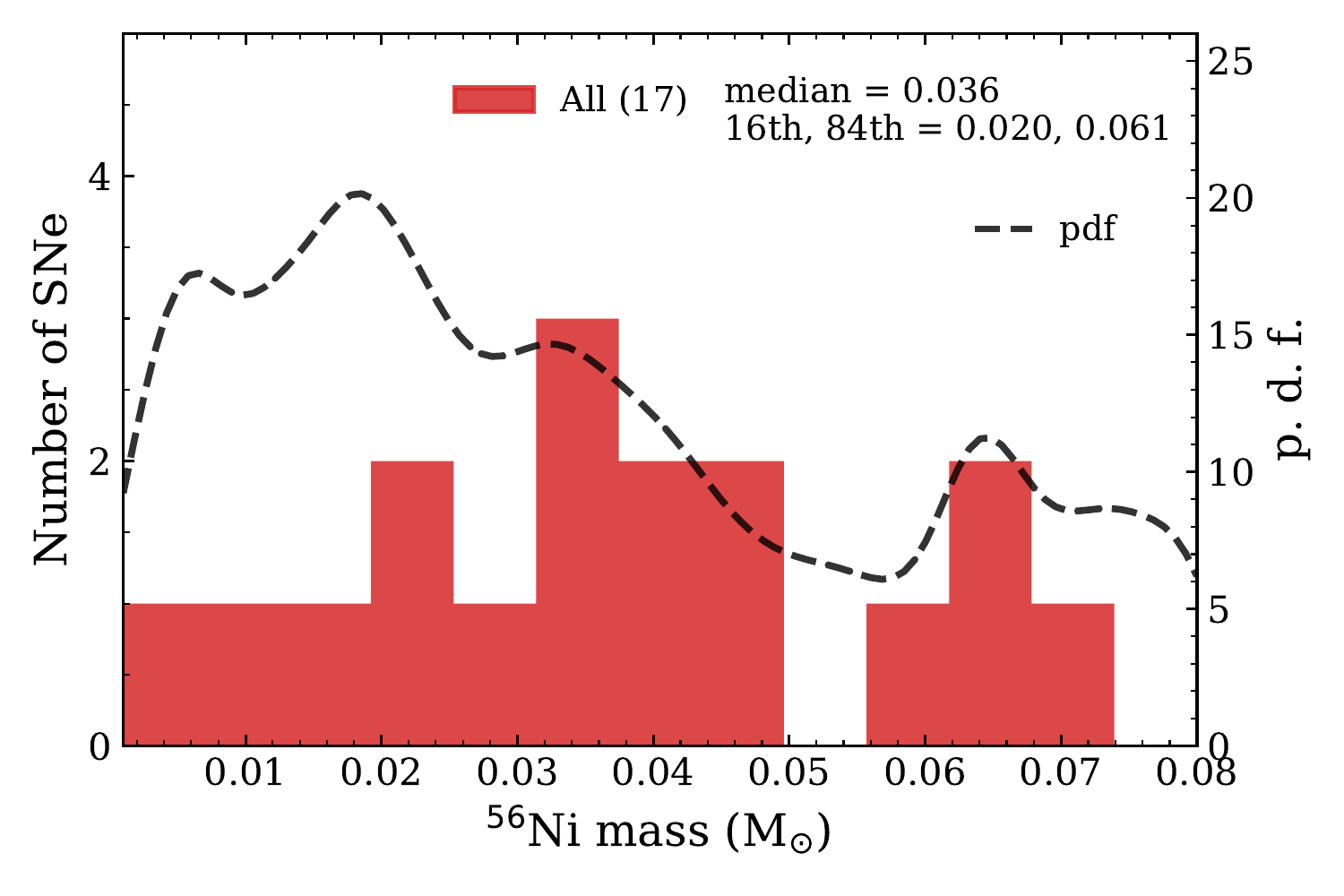} 

\includegraphics[width=0.45\textwidth]{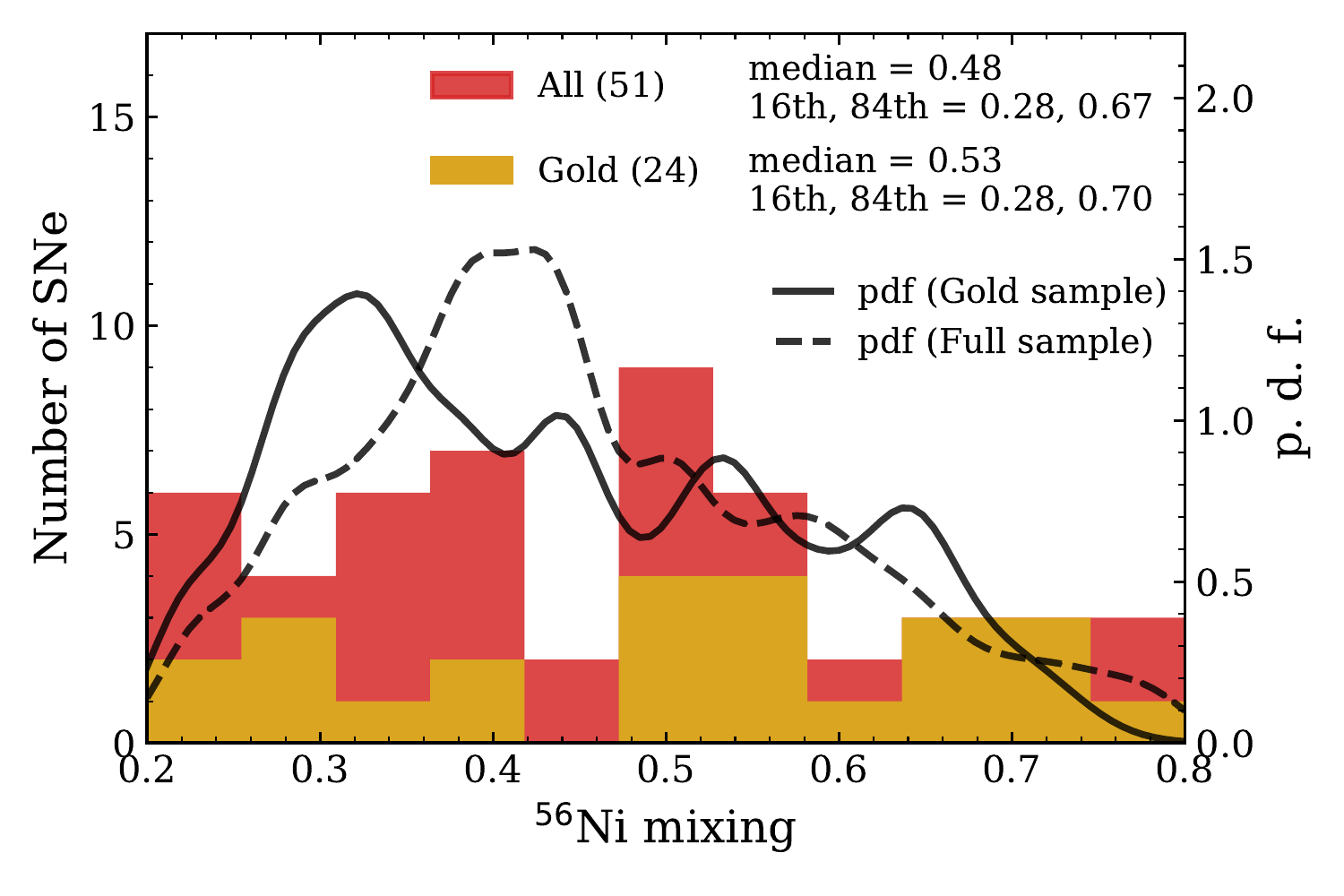}
\caption{Same as in Fig.~\ref{fig:hist1} but for \e\ (top panel), \mni\ (middle panel), and \mix\ (bottom panel).}
\label{fig:hist2}
\end{figure}

We searched for the probability distributions of the parameters for each \snii\ in the sample employing the set of explosion models and the MCMC procedure presented in Sect.~\ref{sec:methods}. 
We define a `gold sample' of our \sneii, selecting those events that followed our selection criteria:
(a) bolometric LCs covering the photospheric phase and at least the beginning of the transition to the radioactive tail phase, since the latter is crucial to constraining principally the \mej\ and \e;
(b) at least two \ion{Fe}{ii} velocity measurements during the photospheric phase and separated in time by more than ten days; and (c) the maximum a posteriori model reproduces the observations\footnote{Sometimes none of the models in our grid can reproduce the LC phases or the evolution of the photospheric velocity.}.
The assessment of the quality of the fits to the models was achieved visually and independently by the first three authors of this paper (LM, MB, and JA).
A small number of contentious cases were discussed; however, their inclusion or exclusion in the gold sample does not affect our results or conclusions (see Appendix~\ref{app:gold_sample}).
In total, 24~\sneii\ fall into the gold sample (see Table~\ref{table:results}).

The full sample of \sneii\ comprises 53 events and is formed by the SNe in the gold sample plus 29 additional objects for which progenitor and/or explosion parameters were determined, but insufficient data coverage or fitting quality prevent these \sneii\ from being considered gold events.
SN~2008bk was considered a gold event even though the conditions were not fulfilled with our dataset since there are no observations during the transition from the plateau to the tail phase. SN~2008bk is a well-studied SN and for this reason, we used already published data to restrict the end of the plateau \citep{vandyk+12a}.

Unfortunately, the following SNe do not fall into either of the previous groups and were excluded from the rest of this work.
The models in our set of explosions cannot reproduce the observed behaviour of five \sneii\footnote{We note that five \sneii\ are not a negligible fraction of the sample ($\sim$10\% of the full sample). However, if our models had reproduced these five SNe, only one would be part of the gold sample (SN~2006Y), which would not alter significantly the results. The other SNe do not show enough observations to belong to the gold sample. At the same time, this relatively high percentage shows that there is significant diversity within the SNe~II population beyond that which can be produced by the explosion of `standard' progenitors.}: 2006Y, 2008bu, 2008bm, 2009aj, and 2009au. The first two SNe show atypical optically thick phase durations (`optd'\footnote{optd corresponds to the time between explosion and the mid-point of the transition from plateau to the radioactive tail.}) of only 64~$\pm$~4~days and 52~$\pm$~7~days for SN~2006Y and SN~2008bu, respectively \citepalias{martinez+21}. No model within our grid presents such a short optd (see also Sect.~\ref{sec:nonstd}).
We are able to model short-plateau SNe~II by increasing the mass loss during the evolution of their progenitors, which reduces the extent of the hydrogen-rich envelope at the time of collapse. These results are presented in \citetalias{paper3_submitted} \citep[see also][for short-plateau SN~II modelling]{hiramatsu+21}.
SNe 2008bm, 2009aj, and 2009au were already analysed in \citet{rodriguez+20}. Those authors showed that these events share the following common characteristics: low expansion velocities, absolute \emph{V-}band LCs much brighter compared to normal \sneii\ with such velocities, and signs of interaction of the ejecta with CSM, among others. Moreover, based on hydrodynamic simulations, \citet{rodriguez+20} found that a massive CSM of $\sim$3.6~\ms\ was needed to reproduce the entire LCs and expansion velocities of SN~2009aj. Such a large CSM mass is expected to influence SNe~II properties at epochs much later than the 30 day limit we assumed for the rest of the sample.
As we mentioned in Sect.~\ref{sec:modelling_background}, our explosion models were calculated without including any CSM. Therefore, it is understandable that we are not able to find a set of parameters that can represent the full observations of these three events.

The explosion epoch was not estimated for SN~2005gk, SN~2005hd, or SN~2005kh, which makes it impossible to infer reliable results.
Finally, a number of \sneii\ have insufficient data for constraining their physical properties from LC modelling (SNe~2004dy, 2005K, 2005es, 2006bc, 2006it, 2006ms, 2008F, 2008bh, 2008bp, 2008hg, 2008ho, 2008il, and 2009A) and were also excluded from this work. 

Figure~\ref{fig:2006ai} gives an example of our fits for one specific SN (SN~2006ai), where models drawn from the posterior distribution of the parameters are shown in comparison with the observed bolometric LC (top panel) and \ion{Fe}{ii} line velocities (bottom panel).
Additionally, Appendix~\ref{app:physical_pars} compares models and observations for the entire CSP-I \snii\ sample, and shows an example corner plot with the posterior probability distribution of the parameters of SN~2006ai.
We characterise the results of the sample by the median of the marginal distributions. The 16th and 84th percentiles measure the width of the distribution of the sample, which are adopted as the lower and upper uncertainties. The results are reported in Table~\ref{table:results} where we also include estimates of \mej, \mhenv, the total mass of hydrogen at the time of collapse (\mh), and \ra. 
We emphasise that these progenitor properties are not model parameters and, therefore, they were not fitted. These values were linearly interpolated from the \mzams\ derived from the fitting.
We note that most of our estimations have small error bars. This arises from the high cadence and quality of the observations. However, these errors do not take into account systematics such as the uncertainties in the hydrodynamic simulations and stellar evolution modelling. In the latter, we used `standard' values for various parameters (wind efficiency, convection, metallicity). Changes in these parameters give different pre-SN configurations, and therefore, different results. As a consequence, the errors on the physical parameters are likely to be underestimated.
We reiterate that the results presented in the following sections are achieved by using standard pre-SN models (see details in Sect.~\ref{sec:hydro_models}), similar to those used by the studies that determine initial masses from progenitor detection in pre-explosion images. 
While this brings several caveats (that are discussed in later sections), it affords a consistent comparison with various other literature results.

Having inferred the progenitor and explosion properties for a large sample of \sneii\ (24 SNe in the gold sample and 53 in total), we now analyse the distributions of the physical parameters. This is the largest set of physical properties of SNe~II derived to date. For two of the 53 \sneii\ (SN~2005af and SN~2007it) only \mni\ was derived. SN~2005af was only observed at late times, from the late recombination phase to the radioactive tail phase. SN~2007it was also observed at late phases and, additionally, it has early observations (<\,30 days). These data are insufficient to determine all the physical parameters, and only \mni\ was inferred.

Figures~\ref{fig:hist1} and \ref{fig:hist2} display histograms of progenitor (\mzams, \mej, and \ra) and explosion parameters (\e, \mni, and \mix), respectively. In each panel, the yellow bars represent \sneii\ in the gold sample, while the histogram of the full sample is indicated in red bars. 
We characterise the distributions by the 16th, 50th, and 84th percentiles.
Additionally, Figs.~\ref{fig:hist1} and \ref{fig:hist2} show the probability density function of the parameters. These functions were computed using a kernel density estimation of the sum of the posterior distributions marginalised over the parameters for the \sneii\ in the gold (solid lines) and full (dashed lines) samples. 
We note that some values exceed unity. This is not erroneous since the figure shows the probability density function. The integral of the probability density function along the entire range of values equals unity.

The \mzams\ distribution for the \sneii\ in the gold sample (Fig.~\ref{fig:hist1}, top panel) is characterised by a median value of 10.4~\ms, with most progenitors (21 of 24) being less massive than 13~\ms. When the full sample is considered the median is slightly larger, \mzams~=~11.7~\ms.
For both the gold and full samples, the lowest and highest \mzams\ values are found to be 9.2 and 20.9~\ms\ for SN~2009N and SN~2008ag, respectively.
SN~2008ag presents one of the longest and brightest optically thick phases in the sample \citepalias{martinez+21}, and thus it is not surprising to find a massive progenitor with large pre-SN mass and radius to reproduce its observations.
None of the \sneii\ in the sample is consistent with explosion models for stars more massive than 21~\ms. This is in accordance with several studies that analyse \snii\ progenitors in pre-explosion images, that also find a lack of high-mass RSG progenitors \citep{smartt15,davies+18}.
On the other hand, the lower boundary of our \mzams\ distribution is consistent with the lowest-\mzams\ progenitor model.
In Sect.~\ref{sec:imf}, a full analysis of the \mzams\ distribution is given.

The histograms of \mej\ (Fig.~\ref{fig:hist1}, middle panel) and \ra\ (Fig.~\ref{fig:hist1}, bottom panel) can be explained from the \mzams\ distributions. The \mej\ and \ra\ are not model parameters, and therefore these values were directly inferred from the \mzams\ value derived from the fitting. As more low-\mzams\ progenitors are recovered, it is expected that most of the \sneii\ have low \mej\ and \ra\ (under the assumption of standard single-star evolution used in this study). 
The median \mej\ is 8.4~\ms\ for the gold sample, and 9.2~\ms\ when all \sneii\ are analysed. As for the \mzams, the lowest and highest \mej\ and \ra\ are obtained for SN~2009N and SN~2008ag, respectively (for both gold and full samples). The values of \mej\ range from 7.9 to 14.8~\ms, and from 450 to 1077~\rs\ for the pre-explosion radius. The median \ra\ is 495~\rs\ for the gold sample, and 582~\rs\ for the full sample.

The top panel of Fig.~\ref{fig:hist2} shows the histograms for the explosion energy. The gold sample ranges from 0.15~foe (SN~2008bk) to 1.25~foe (SN~2006ai) with a median value of 0.61~foe. The full sample spans a larger range of \e\ from 0.15~foe (SN~2008bk) to 1.40~foe (SN~2006bl), with a median value of 0.63~foe.
The \mni\ distribution is displayed in the middle panel of Fig.~\ref{fig:hist2}. We could determine the \Ni\ mass for 17~\sneii\ as only these present observations during the radioactive tail. 
Uncertainties in the explosion epoch, host-galaxy extinction, and distance impact on the determination of \mni. While some \sneii\ have lower uncertainties in the parameters mentioned above, these SNe are not necessarily the same as those in the gold sample above. Thus, to avoid confusion with the previously defined sub-samples, we did not separate SNe~II into different samples for the derivation of their \Ni\ masses.
The median of the distribution is 0.036~\ms, and it ranges from 0.006~\ms\ for SN~2008bk, to 0.069~\ms\ for SN~2007X.
In \citetalias{martinez+21}, we found that SN~2008bk is the lowest-luminosity event at the radioactive tail phase, and therefore it is unsurprising that SN~2008bk has the lowest estimated \mni.
The bottom panel of Fig.~\ref{fig:hist2} shows the \mix\ distributions. \mix\ values range from the inner ejecta (0.2) to the outer ejecta (0.8) with a median of 0.53 for the gold sample, similar to that for the full sample. Only a few \snii\ explosions are consistent with extended \mix\ ($\sim$0.8) into the ejecta.

\section{Analysis}
\label{sec:analysis}

\subsection{The mass function of \snii\ progenitors}
\label{sec:imf}

\begin{table}
\caption{Parameters of the cumulative distributions.}
\label{tab:imf_results}
\centering
\begin{tabular}{cccc}
\hline\hline\noalign{\smallskip}
 & \mmin\ (\ms)& \mmax\ (\ms) & $\Gamma$ \\
\hline\noalign{\smallskip}
Gold sample & 9.3$^{+0.1}_{-0.1}$ & 21.3$^{+3.8}_{-0.4}$\tablefootmark{a} & $-$6.35$^{+0.52}_{-0.57}$ \\
\noalign{\smallskip}
Full sample & 9.3$^{+0.1}_{-0.1}$ & 21.5$^{+1.2}_{-0.8}$ & $-$4.07$^{+0.29}_{-0.29}$ \\
\noalign{\smallskip}
\hline\noalign{\smallskip}
Gold sample & 8.5$^{+0.1}_{-0.1}$ & 19.1$^{+0.2}_{-0.2}$ & $-$2.35 \\
\noalign{\smallskip}
Full sample & 9.0$^{+0.1}_{-0.1}$ & 19.0$^{+0.2}_{-0.2}$ & $-$2.35 \\
\noalign{\smallskip}
\hline
\end{tabular}
\tablefoot{
We show cases where the power-law slope ($\Gamma$) is unconstrained (top panel) and where it is constrained to $-$2.35 (bottom panel).
The parameters are characterised by the median of the marginal distribution adopting the 16th and 84th percentiles as the lower and upper uncertainties.
\tablefoottext{a}{Using the mode of the distribution (see text).}
}
\end{table}

\begin{figure}
\centering
\includegraphics[width=0.49\textwidth]{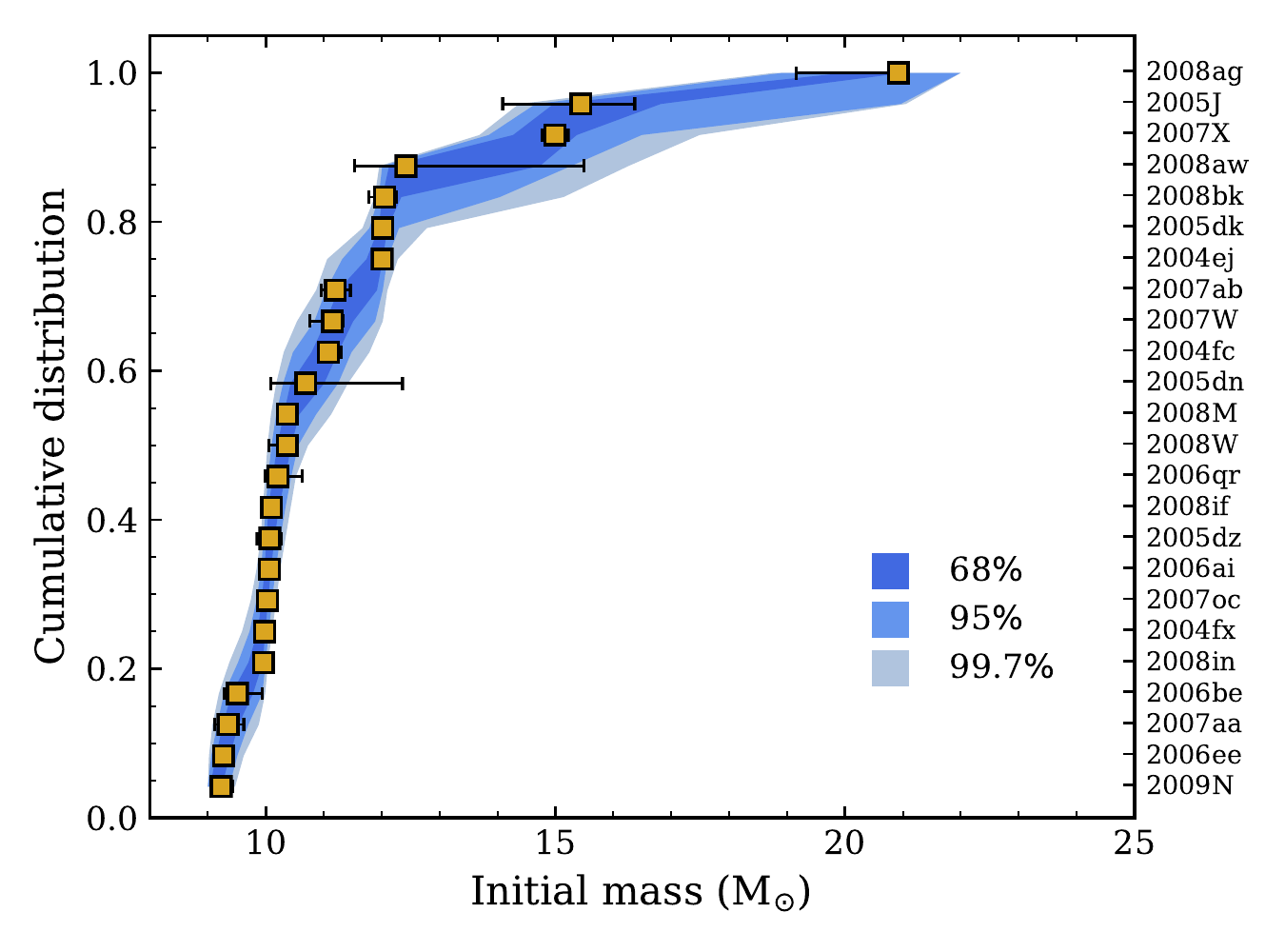}
\caption{Cumulative distribution of \mzams\ for the \snii\ progenitors in the gold sample. The derived masses are shown in yellow squares. The shaded contours show the confidence regions of the CD.}
\label{fig:cmd_gold}
\end{figure}

\begin{figure}
\centering
\includegraphics[width=0.49\textwidth]{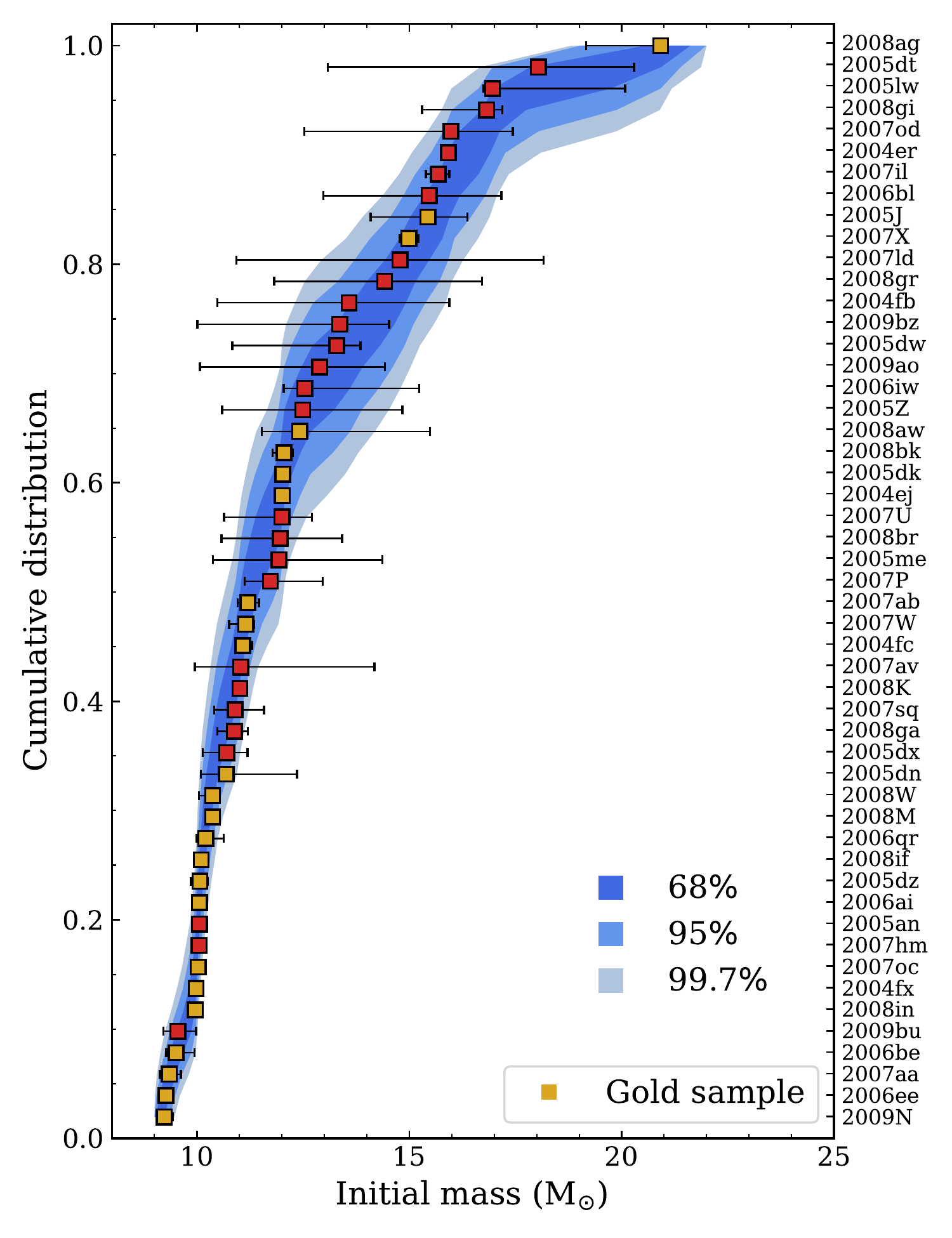}
\caption{Same as in Fig.~\ref{fig:cmd_gold} but for the \snii\ progenitors in the full sample. Yellow squares represent the \sneii\ in the gold sample.}
\label{fig:cmd_full}
\end{figure}

\begin{figure*}
\centering
\includegraphics[width=0.47\textwidth]{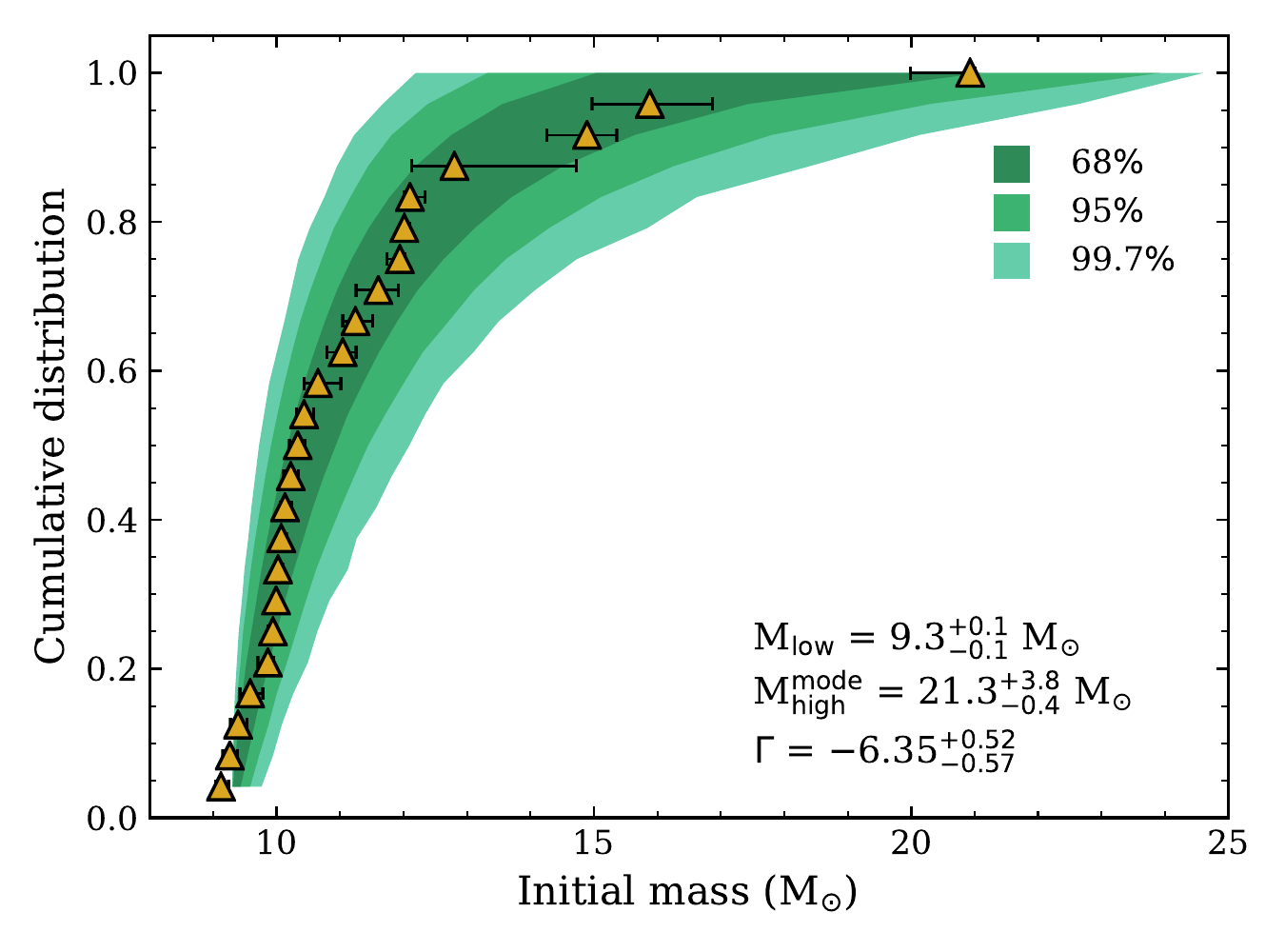}
\includegraphics[width=0.35\textwidth]{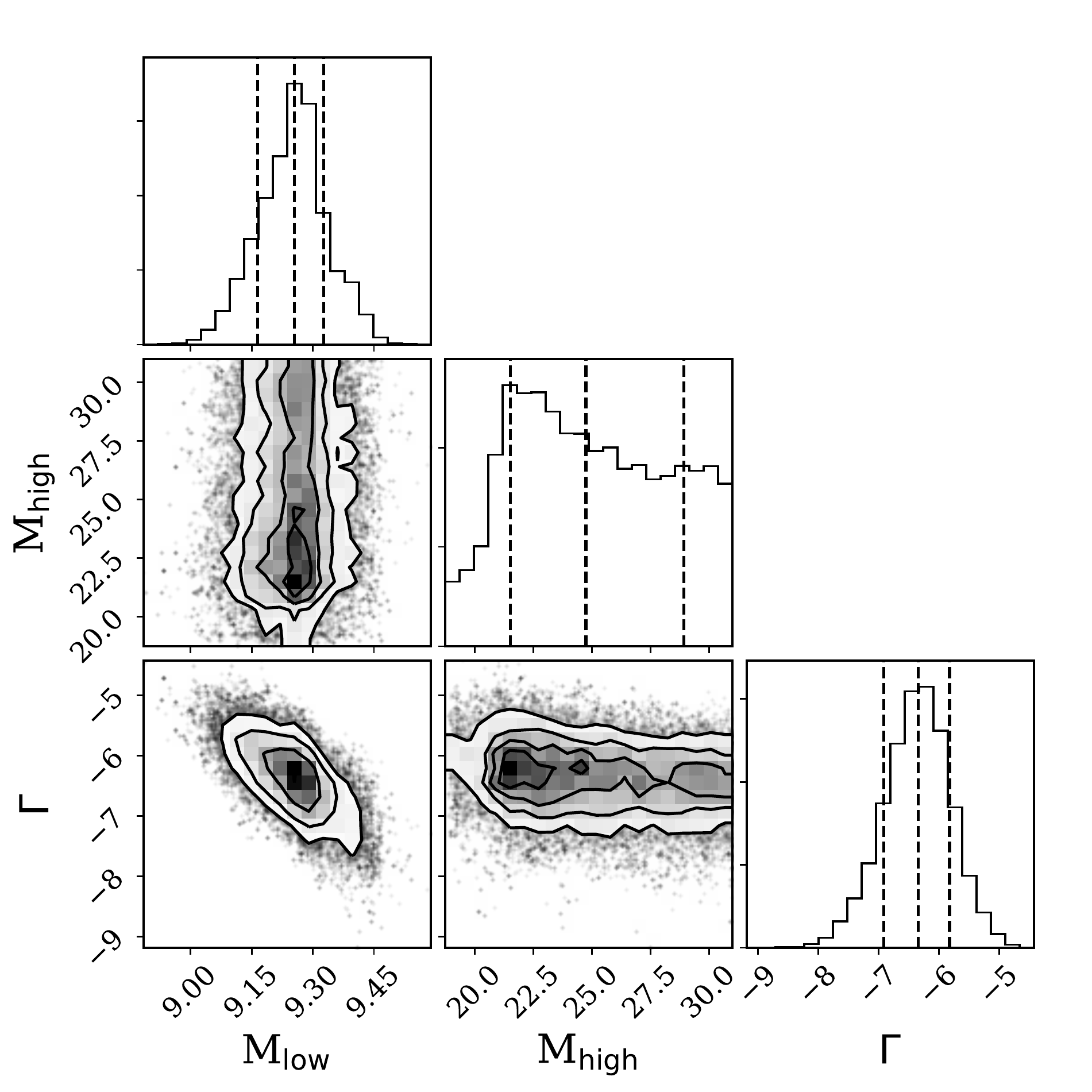}
\caption{Fits to the CD of \mzams\ of \snii\ progenitors in the gold sample. \textit{Left panel:} CD of \mzams\ derived in our work for the gold sample (yellow triangles and error bars), corresponding to the median and the 68\% confidence limit of the distribution presented in Fig.~\ref{fig:cmd_gold} as shaded contours, in comparison with the model CD constructed with the median values of the marginal distributions of the parameters (shaded contours).
\textit{Right panel:} Corner plot of the joint posterior probability distribution of the parameters. Dashed lines indicate the 16th, 50th, and 84th percentiles of the distributions.}
\label{fig:cmd_teo_gold}
\end{figure*}

\begin{figure*}
\centering
\includegraphics[width=0.47\textwidth]{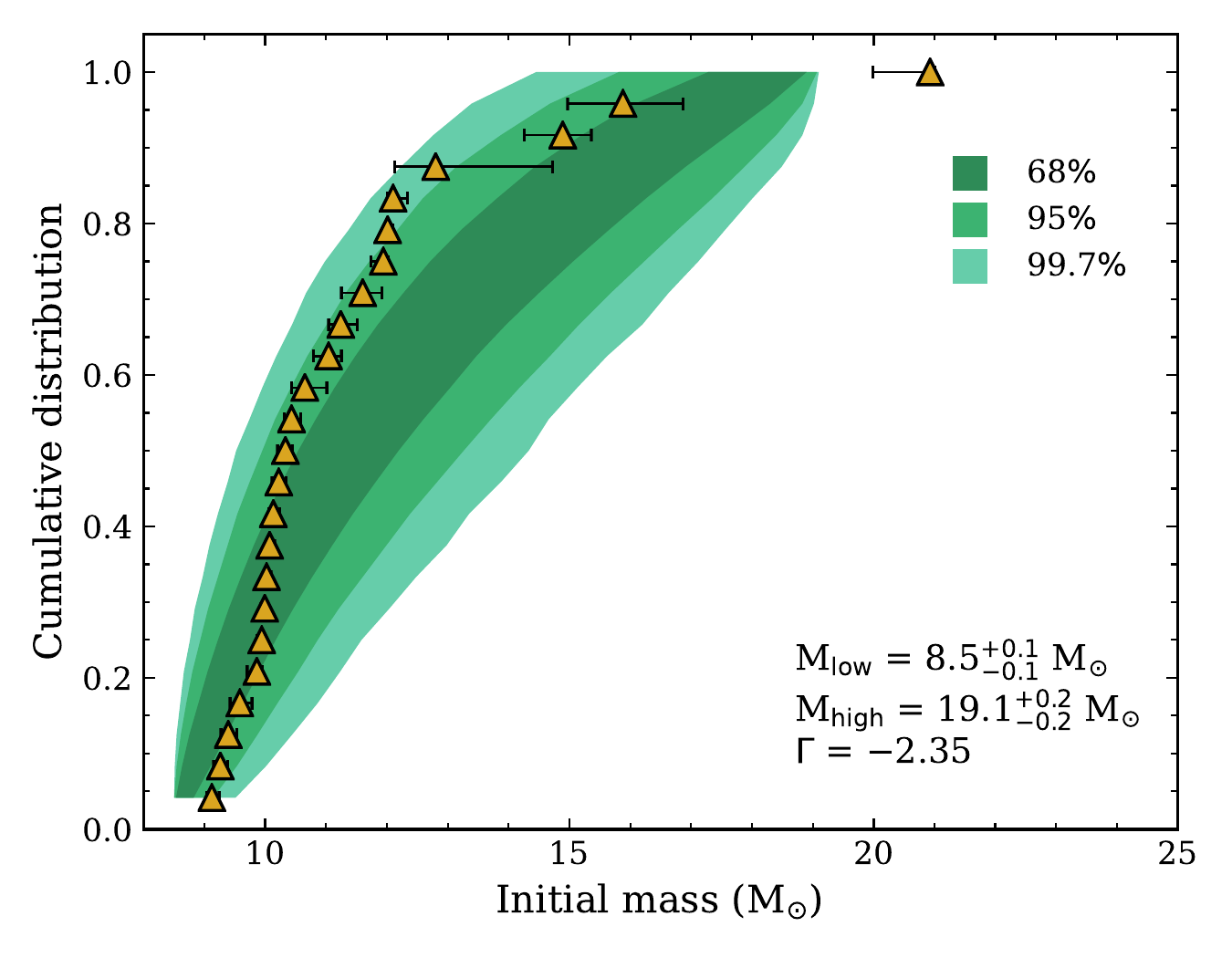}
\includegraphics[width=0.32\textwidth]{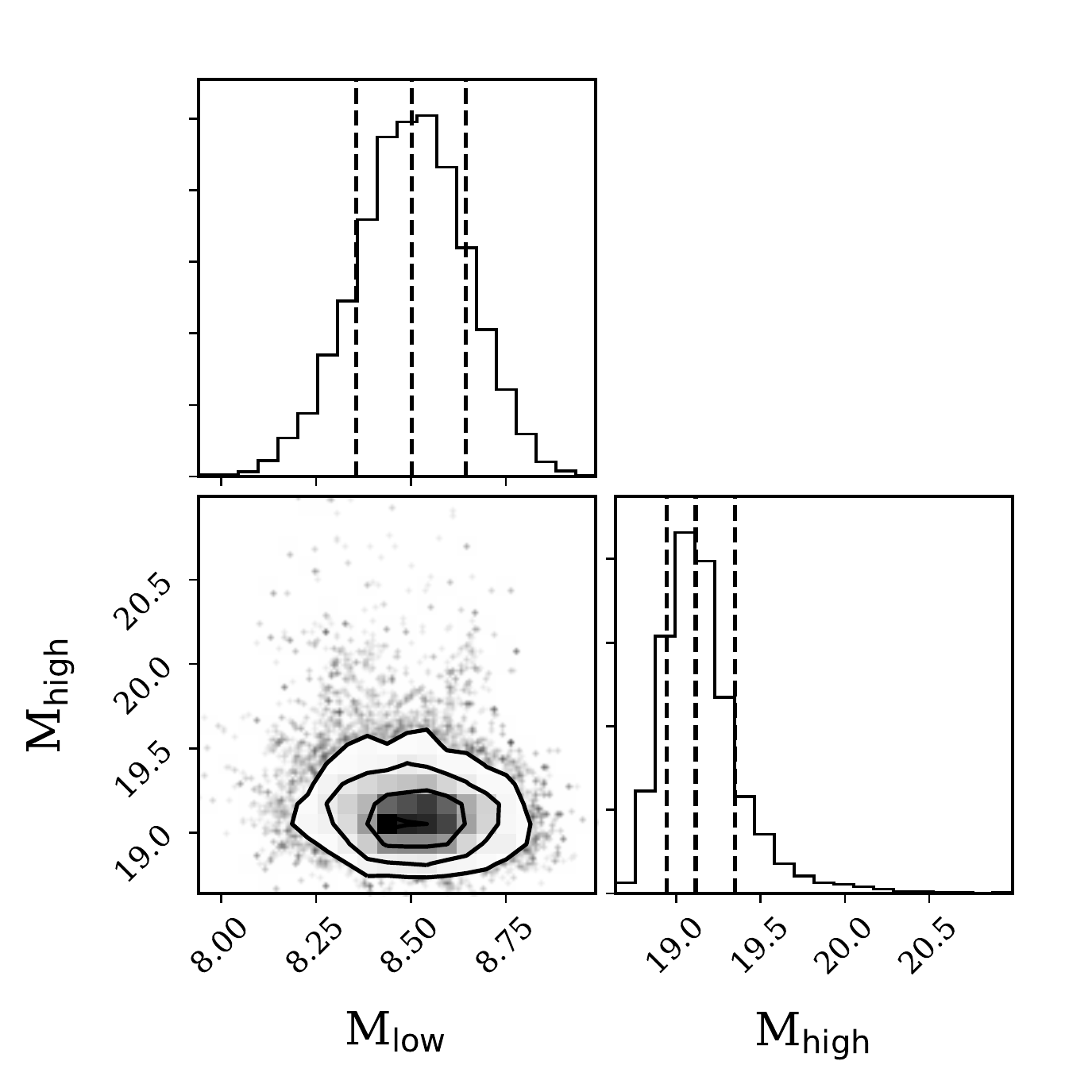}
\caption{Same as in Fig.~\ref{fig:cmd_teo_gold} but when the power-law slope is fixed to $\Gamma$~=~$-$2.35, that is, the slope of a Salpeter massive-star IMF.}
\label{fig:cmd_teo_gold_2.35}
\end{figure*}

\begin{figure*}
\centering
\includegraphics[width=0.45\textwidth]{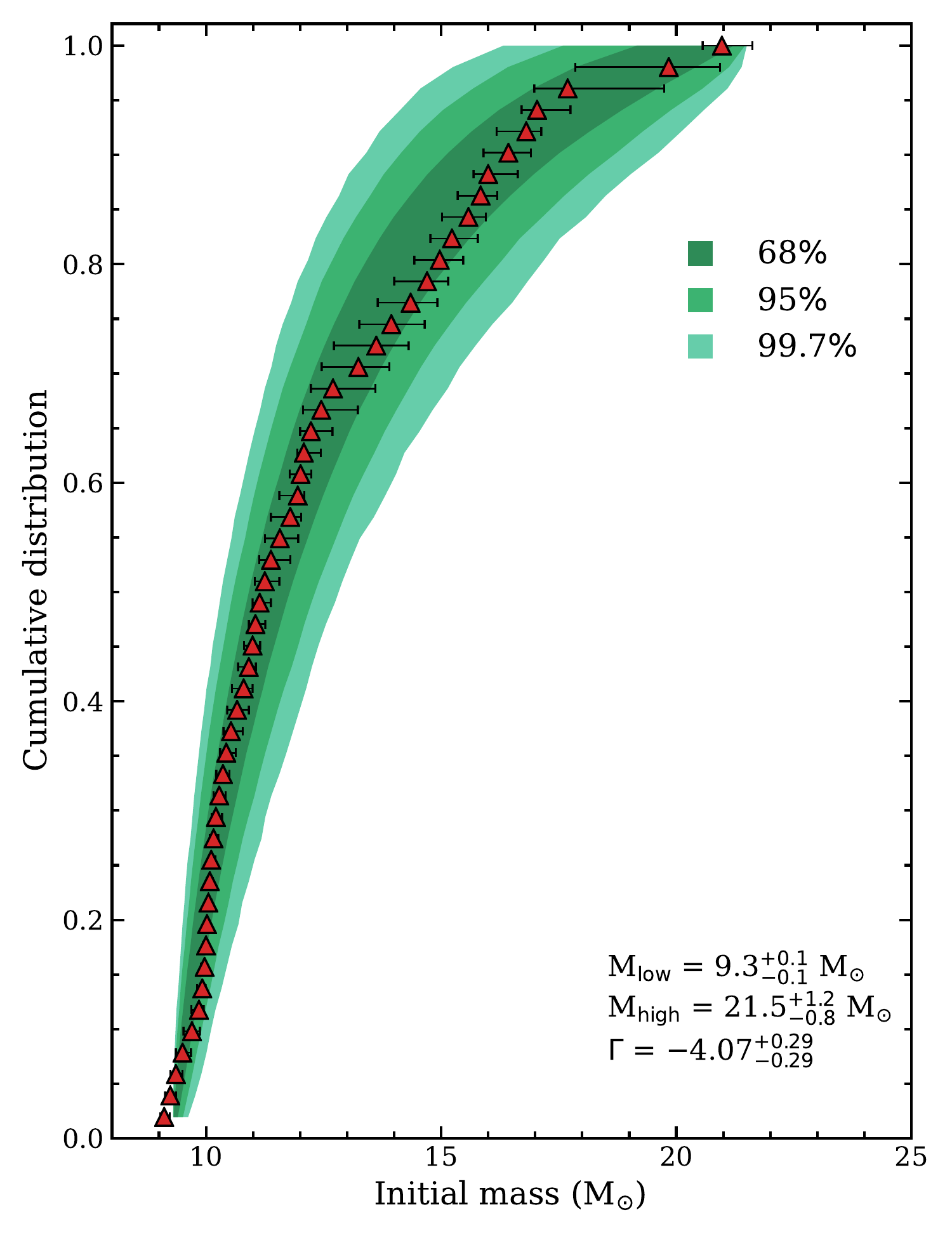}
\includegraphics[width=0.45\textwidth]{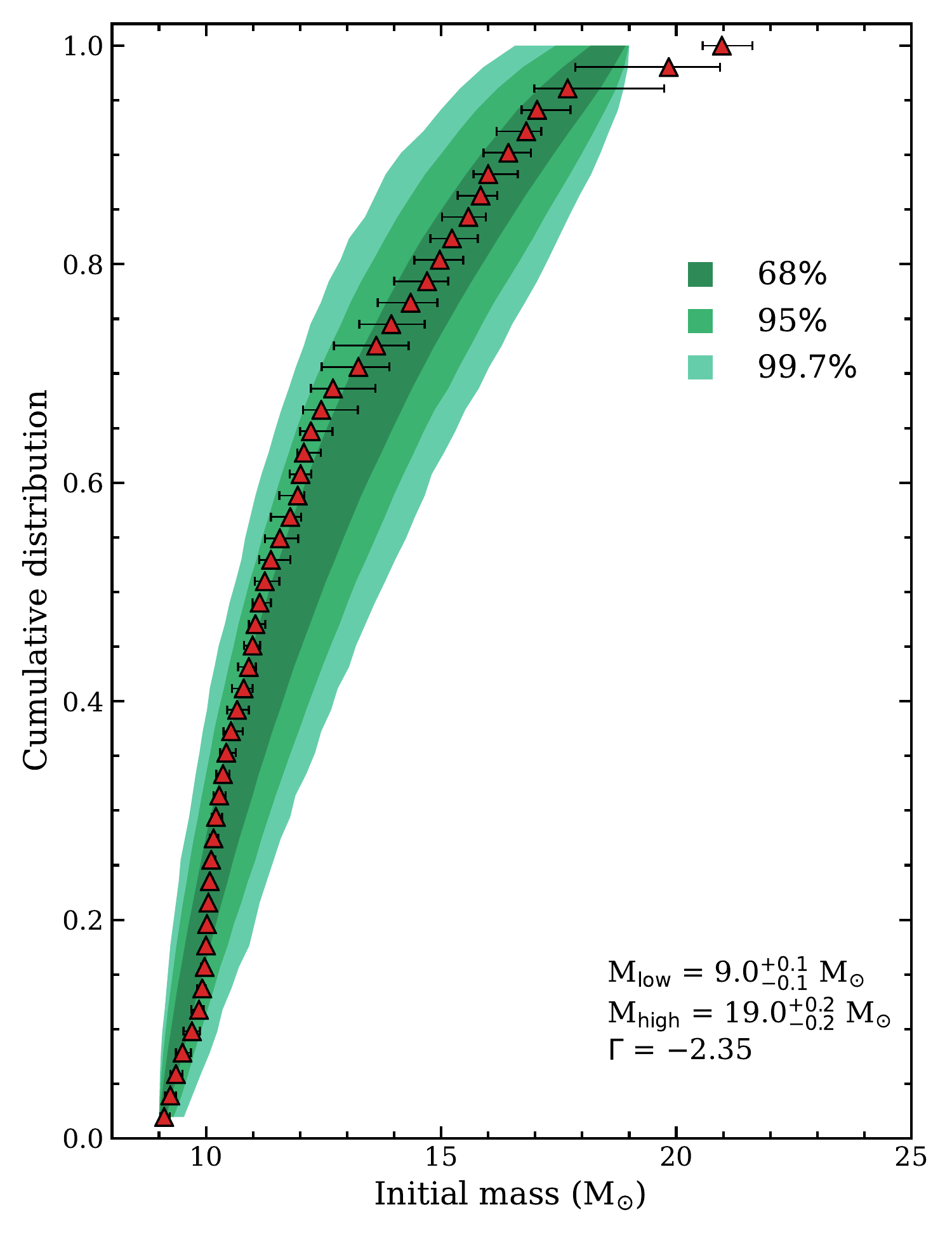} 

\includegraphics[width=0.37\textwidth]{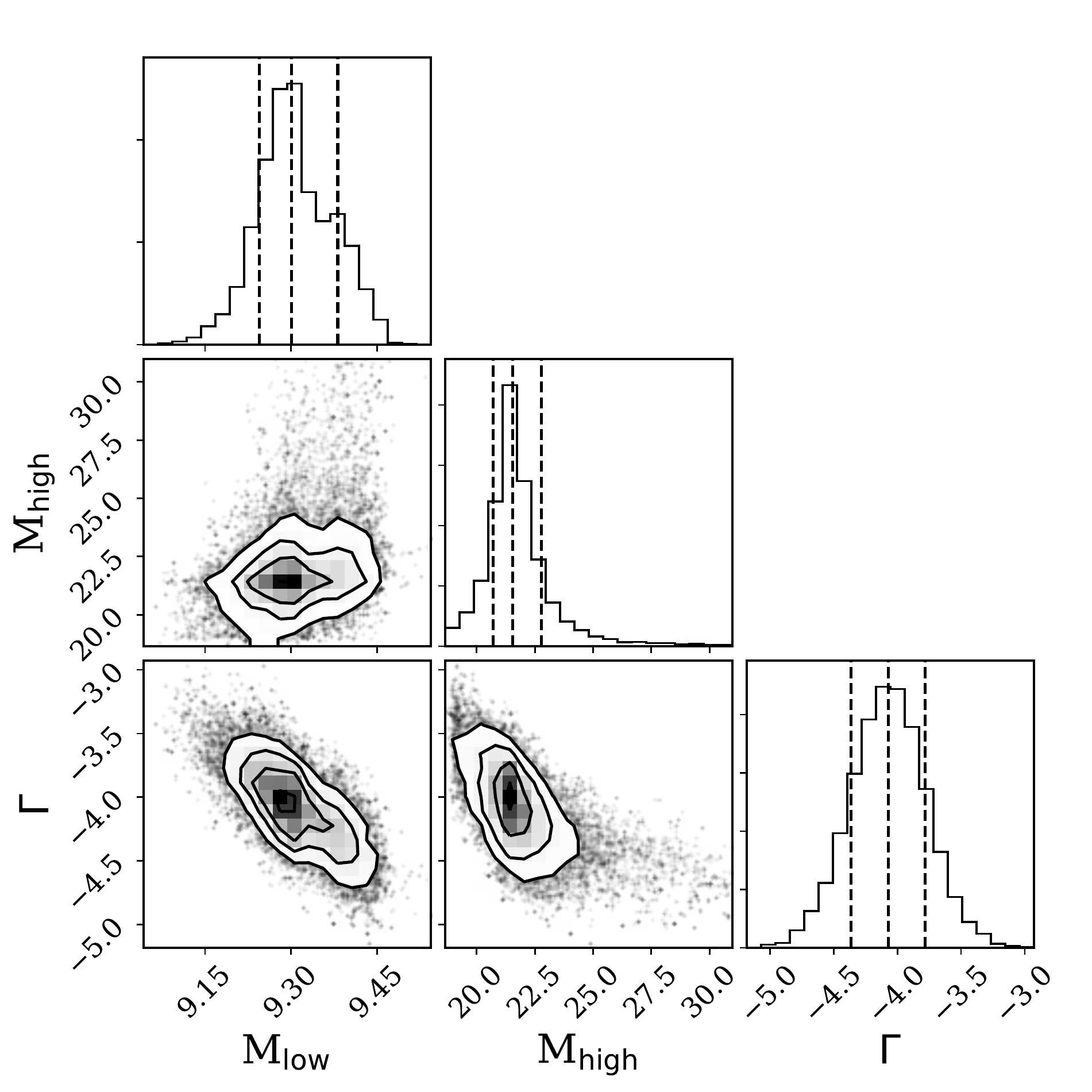}
\includegraphics[width=0.275\textwidth]{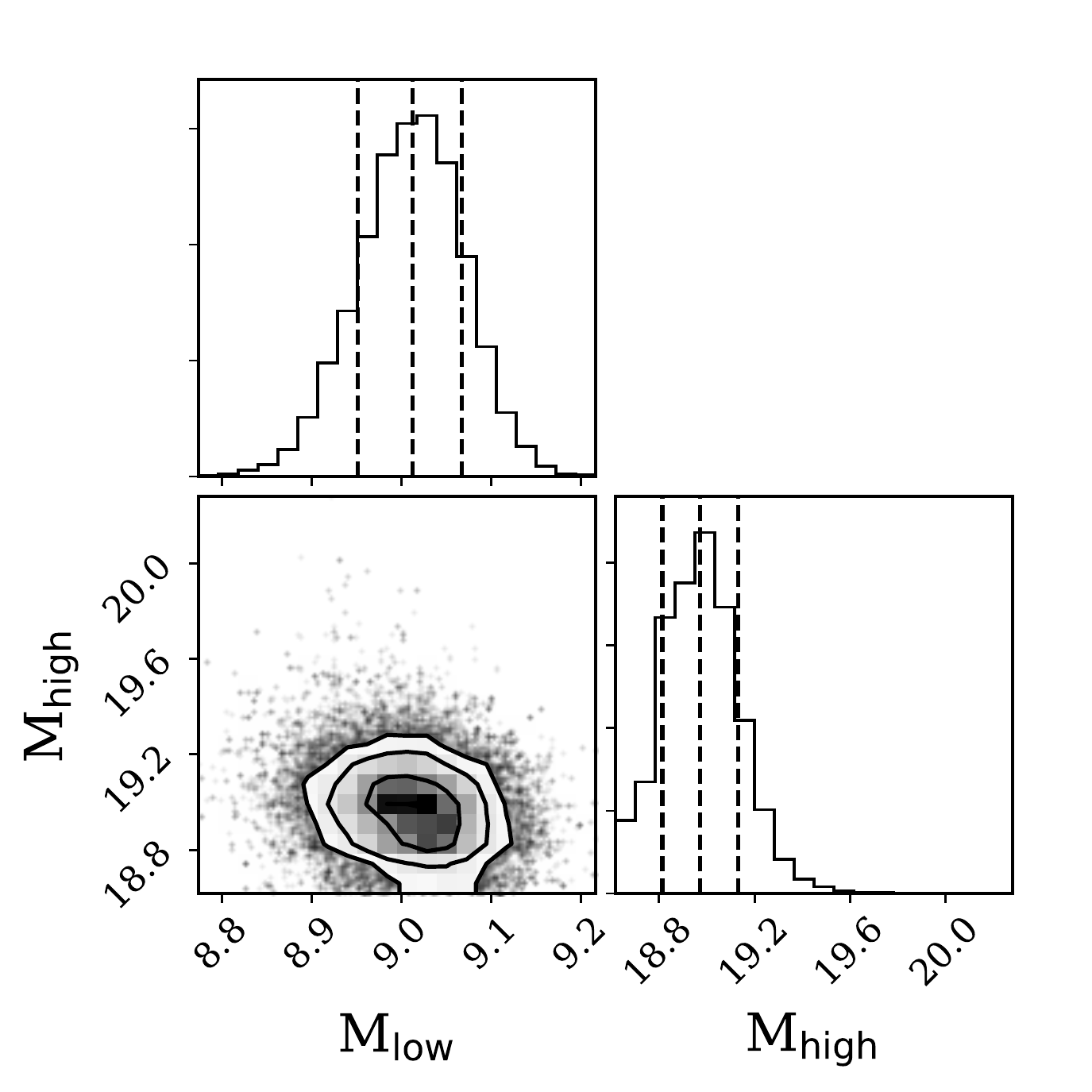}
\caption{Fits to the CD of \mzams\ of \snii\ progenitors in the full sample. \textit{Top left:} CD of \mzams\ derived in our work for the full sample (red triangles and the error bars), corresponding to the median and the 68\% confidence limit of the distribution presented in Fig.~\ref{fig:cmd_full} as shaded contours, in comparison with the model CD constructed with the median values of the marginal distributions of the parameters (shaded contours). \textit{Bottom left:} Corner plot of the joint posterior probability distribution of the parameters. Dashed lines indicate the 16th, 50th, and 84th percentiles of the distributions. The right panels show the same as the left panels but when the power-law slope is fixed to $\Gamma$~=~$-$2.35.}
\label{fig:cmd_teo_full}
\end{figure*}

Most studies of the initial mass function (IMF) of \snii\ progenitors use information from pre-explosion images \citep[e.g.][although see \citealt{morozova+18}]{smartt+09,smartt15,davies+18}. A discussion of their results is presented in Sect.~\ref{sec:rsg_problem}.
In the current work, we study the \mzams\ distribution of \snii\ progenitors using the values derived for the CSP-I sample. 
Restricting ourselves to the gold sample we have 24 \mzams\ estimations. This number increases to 51 if we consider the full sample. To date, this is the most homogeneous and largest sample of progenitors of \sneii\ analysed from hydrodynamical modelling.
Our analysis is based on fitting a power law to the inferred \mzams\ distribution employing a Monte Carlo (MC) method to randomly sample the masses from a master population. This technique is similar to that presented by \citet{davies+20}, where the reader is referred for details. 

We first constructed the cumulative distribution (CD) of the \mzams\ for the progenitors. We performed 10000 MC simulations to determine the probability distribution of the CD. For each simulation, \mzams\ was randomly sampled from the probability distribution of each progenitor, which was obtained from our MCMC procedure (Sect.~\ref{sec:mcmc}), and then the progenitors were ordered in increasing \mzams.
As noted in \citet{davies+20}, when sampling from the posterior distribution, the \mzams\ of a progenitor can take different values producing changes in the order of the progenitors. For this reason, it is necessary to re-order the progenitor masses at each MC trial.
Figures~\ref{fig:cmd_gold} and \ref{fig:cmd_full} show the progenitor masses in increasing order (filled squares) and the confidence regions of the CD of the masses (filled contours) for the gold and full samples, respectively.

Then, the CDs derived above (i.e. the shaded contours in Figs.~\ref{fig:cmd_gold} and \ref{fig:cmd_full}) were fitted with models of CD based on power laws.
These models have the following input parameters: the lower and upper \mzams\ limits of the distribution (\mmin\ and \mmax, respectively), and the slope of the power law ($\Gamma$). 
For each set of parameters, we employed 10000 MC trials to determine the posterior probability distribution for the model CD of the masses. We found the probability distributions of the parameters via MCMC methods.
Results are characterised by the median of the marginal distributions, adopting the 16th and 84th percentiles as the lower and upper uncertainties, with one exception (see below).

In the following, we focus on the analysis of the gold sample. The full sample is analysed later.
Figure~\ref{fig:cmd_teo_gold} shows the CD of \mzams\ obtained for the gold sample in comparison with the theoretical CD based on power laws.
The yellow triangles and error bars represent the median and the 68\% confidence level of the distribution presented in Fig.~\ref{fig:cmd_gold} as shaded contours.
The shaded regions in Fig.~\ref{fig:cmd_teo_gold} correspond to the theoretical CD constructed with the median values of the marginal distributions of the parameters.
As previously indicated, 10000 MC simulations were performed to construct this theoretical CD.
A corner plot of the joint posterior probability distribution of the parameters is presented in the right panel of Fig.~\ref{fig:cmd_teo_gold}.
We find \mmin\,=\,9.3$^{+0.1}_{-0.1}$~\ms, \mmax~=~24.7$^{+4.2}_{-3.2}$~\ms, and $\Gamma$~=~$-$6.35$^{+0.52}_{-0.57}$. These values are also listed in Table~\ref{tab:imf_results}.
We note that the median value of \mmax\ is offset to larger masses from the peak of the distribution, that is, the distribution is skewed to the right. For this reason, we used the mode to characterise the \mmax\ distribution and the 68\% confidence interval for the uncertainties, finding $M_{\rm high}^{\rm mode}$~=~21.3$^{+3.8}_{-0.4}$~\ms. 
The right-skewed distribution is because of the steepness of the CD of the progenitors. The mass function of \snii\ progenitors we derive is much steeper than that for a Salpeter massive-star IMF with $\Gamma$~=~$-$2.35 \citep{salpeter55}.
With a steep power law of $\Gamma$~=~$-$6.35 we do not expect high-mass progenitors to be within our sample. The vast majority of the stars are found near the value of \mmin, and changes in \mmax\ to larger values do not significantly alter the distribution.
For a standard Salpeter IMF, 90\% of the stars between 9--25~\ms\ (the range of our pre-SN models) are in the 9--21~\ms\ range. Then, in our sample of 24 progenitors, we would expect to find two progenitors with masses above 21~\ms. This is not the case for our defined $\Gamma$. 
Not only we do not have progenitors with \mzams\ above 21~\ms, we also find a large number of low-mass progenitors. In the gold sample, 87\% of the progenitors have \mzams\,<\,13~\ms, while the expected value for a Salpeter IMF is $\sim$50\%.

Following the above results when leaving the power law as a free parameter, we now compare the CD of \mzams\ to models with the power-law slope set to $-$2.35 (i.e. a standard Salpeter IMF). In this present case, we find \mmin\,=\,8.5$^{+0.1}_{-0.1}$~\ms\ and
\mmax\,=\,19.1$^{+0.2}_{-0.2}$~\ms\ (Fig.~\ref{fig:cmd_teo_gold_2.35}, right panel). 
The CD model drawn from the posterior distribution of the parameters shows discrepancies with the CD of progenitor masses derived from our modelling (Fig.~\ref{fig:cmd_teo_gold_2.35}, left panel).
The differences are due to the large number of low-\mzams\ progenitors found.

The differences are due to the large number of low-\mzams\ progenitors found.

The same analysis was performed for the full sample of \snii\ progenitors. Here, we find the following parameter values: \mmin\,=\,9.3$^{+0.1}_{-0.1}$~\ms, \mmax\,=\,21.5$^{+1.2}_{-0.8}$~\ms, and $\Gamma$\,=\,$-$4.07$^{+0.29}_{-0.29}$.
In Fig.~\ref{fig:cmd_teo_full}, we show the CD of \mzams\ obtained for the full sample (red triangles and error bars corresponding to the median and 68\% confidence limit of the distribution presented in Fig.~\ref{fig:cmd_full} as shaded contours) in comparison with the model CD constructed with the median values of the marginal distributions (shaded contours). In addition, Fig.~\ref{fig:cmd_teo_full} shows a corner plot of the posterior probability distribution of the parameters. 
The values of \mmin\ and \mmax\ are similar for both the gold and full samples. The largest differences are found for the power-law slope. This is shallower than that obtained for the gold sample, although it is still steeper than a Salpeter IMF.
In Sect.~\ref{sec:sample_size} we analyse whether the different sample sizes influence the power-law slope obtained.

Additionally, we computed the lower and upper mass limits if we constrain $\Gamma$\,=\,$-$2.35 (top-right and bottom-right panels of Fig.~\ref{fig:cmd_teo_full}). We find \mmin\,=\,9.0$^{+0.1}_{-0.1}$~\ms\ and \mmax\,=\,19.0$^{+0.2}_{-0.2}$~\ms. Here, the model drawn from the median values of the parameter distributions can mostly reproduce the behaviour of the \mzams\ distribution.
At first glance, \mmax\ is smaller than the upper mass boundary of the derived cumulative mass distribution (red triangle in the top-right panel of Fig.~\ref{fig:cmd_teo_full}). However, the upper mass boundary has a value of \mzams\,=\,21.0$^{+1.0}_{-2.2}$~\ms\ (99.7\% confidence), that is, they are coincident at the 99.7\% significance level. 
In both the gold and full samples, we find steeper power laws than that of a Salpeter massive-star IMF. In Sect.~\ref{sec:discussion} we discuss possible reasons of this discrepancy, which we term `the IMF incompatibility'.

\subsection{Ejecta masses and explosion energies}
\label{sec:mej_e}

\begin{figure}
\centering
\includegraphics[width=0.47\textwidth]{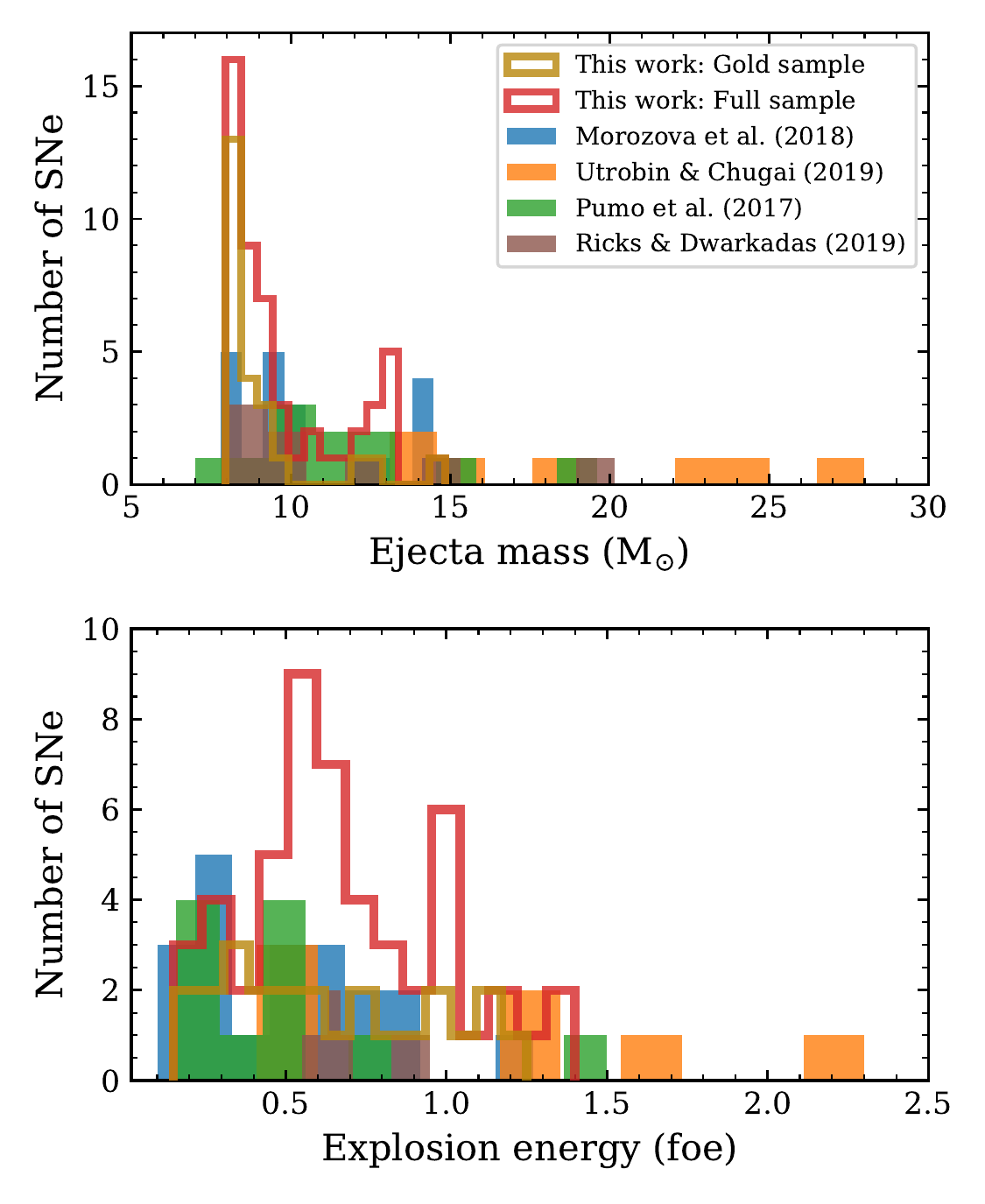}
\caption{Comparison of ejecta masses (top panel) and explosion energies (bottom panel) derived in the current work to those from previous studies containing large samples of \sneii.}
\label{fig:comp_mej_e}
\end{figure}

Here we discuss the range of ejecta masses and explosion energies obtained in our work. Figure~\ref{fig:comp_mej_e} compares our results to those from other studies involving large samples of \sneii. For this purpose, we used the results published in \citet{utrobin+19}, which summarises previous estimates from a series of papers by those authors and the results of \citet{pumo+17} on under-luminous \sneii. \citet{pumo+17} also complies the results of several `normal' \sneii\ obtained with the same code.
We also compare with the ejecta masses and explosion energies from \citet{ricks+19}.
All the above-mentioned studies present detailed modelling of LCs and expansion velocities for several \sneii. 
\citet{morozova+18} presented LC modelling for 20~\sneii. Given that our distributions were derived from the modelling of the LC and expansion velocities simultaneously, a direct comparison is difficult. However, we include the results from \citet{morozova+18} in the comparison.
Unfortunately, we cannot compare with the findings from \citet{eldridge+19b} as only the initial masses were published.

We observe that our \mej\ and \e\ are generally consistent with those from previous studies as our results fall within the range of previous estimates. However, we note a spike at $\sim$8$-$9~\ms\ that is not found in the other studies\footnote{A direct comparison with previous works is difficult for several reasons. Firstly, most of them present hydrodynamical modelling of small sample sizes, except for \citet{morozova+18}. However, these authors do not consider the expansion velocities in their fitting procedure. In addition, previous studies used different techniques to create pre-SN models (polytropic models or stellar evolution calculations) and different fitting methods (choosing the best-fitting models by eye or through statistical techniques).}.
This spike is the reason why we found so many low-\mzams\ progenitors, causing the IMF incompatibility (see Sect.~\ref{sec:imf_incompatibility}).

The highest estimated \mej\ and \e\ for the CSP-I \snii\ sample are 14.8~\ms\ and 1.25~foe (1.40~foe for the full sample), respectively. A number of higher \mej\ and \e\ are found in the other works, especially in those from \citet{utrobin+19}. This could be because these authors use non-evolutionary models as pre-SN configurations that are capable of producing a large variety of pre-explosion structures, each with different mass, radius, chemical composition, and density profile.
In the present work, we used standard single-star evolution calculations as pre-SN models. Therefore, the initial density profiles were not freely chosen, and the range of pre-SN parameters is confined. In our set of progenitor models, the highest \mej\ is 15.7~\ms, coming from a 24~\ms\ progenitor in the ZAMS \citepalias[see][their Fig.~2]{martinez+20}. Therefore, it is impossible for us to find larger \mej\ given that we only consider \mzams\ up to 25~\ms. 
\citet{ricks+19} obtained a \mej\ of $\sim$20~\ms\ for SN~2015ba using evolutionary models to initialise the explosion. However, for this particular progenitor model, the authors used a lower scaling factor for the efficiency of mass loss via winds during the stellar evolution producing a more massive star at explosion time.
We again stress that our pre-SN models were calculated using the standard assumptions and parameters during the evolution. Different stellar evolution assumptions can lead to different results.

\citet{pumo+17} found, on average, lower \mej\ and \e\ than other studies probably because most of their sample corresponds to low-luminosity \sneii.
While the range of parameters found is in good agreement with that from \citet{morozova+18}, we find some differences in the distributions. \citet{morozova+18} found more \sneii\ with lower energies, while at the same time, a spike at \mej~$\sim$14~\ms. 

Finally, we compare the explosion energies in this work with those expected from neutrino-driven explosion models.
1D core-collapse models found explosion energies ranging from 0.1 to 2.0~foe \citep{ugliano+12,ertl+16,sukhbold+16}.
The \sneii\ modelled in this work do not cover the entire range of explosion energies predicted by the explosion models, particularly because our maximum value is somewhat lower (1.25~foe for the gold sample and 1.40~foe for the full sample).
More energetic explosions produce more luminous \sneii, which are not found in our sample. For example, the highest explosion energies found in \citet{sukhbold+16} produce plateau luminosities higher than 10$^{42.6}$~erg\,s$^{-1}$ (see their Fig.~31), while in the CSP-I \snii\ sample, only one object -- SN~2009aj -- has plateau luminosities above this value. In addition, as previously discussed in Sect.~\ref{sec:results}, SN~2009aj may be powered by interaction between the ejecta and a massive CSM.
More recently, \citet{burrows+20} conducted 3D core-collapse simulations in a wide range of progenitor masses. The lowest-\mzams\ model (9~\ms) in their grid achieves an explosion energy of $\sim$0.1~foe, which is compatible with the lowest explosion energy found in our study. Unfortunately, the other core-collapse models have not reached the asymptotic explosion energy by the end of the simulations. However, these authors do find that progenitors with higher \mzams\ attain higher explosion energies. We find the same trend in \citetalias{paper3_submitted}.

\subsection{Progenitor radii}
\label{sec:radii}

\citet{levesque+05,levesque+06} studied Galactic and Magellanic Cloud RSGs and obtained their radii using the Stefan-Boltzman law from effective temperatures and bolometric luminosities. These studies found that most RSGs have radii between 100 and 1500~\rs. 
In Sect.~\ref{sec:results}, most progenitor radii inferred from our LC and velocity modelling are found within the range of 450$-$600~\rs.
However, we emphasise that progenitor radius is not an independent parameter in our modelling. Our fitting routine derives \mzams, which relates to pre-SN structures with different characteristics (e.g. ejecta mass, radius, among other physical parameters).
Additionally, while early-time optical LCs are most sensitive to progenitor radius, in the current work fits were performed only to observations later than 30~days from explosion (see Sect.~\ref{sec:modelling_background}). 

\subsection{\Ni\ masses}
\label{sec:nickel}

The \Ni\ masses for the CSP-I \snii\ sample range from 0.006~\ms\ through 0.069~\ms\ with a median of 0.036~\ms. 
Our values were determined for 17~\sneii\ which do not cover the entire \Ni\ mass range found in the literature. Some \sneii\ have been found with more and less \Ni\ \citep[e.g.][]{hamuy03,pastorello+04,pumo+11}.
\mni\ was estimated for all CSP-I \sneii\ with bolometric data during the radioactive tail phase\footnote{With the exception of SN~2005hd and SN~2005kh, since they do not have any estimate for the explosion epoch.}, suggesting that the \mni\ range of our models (see Table~\ref{tab:models}) was sufficient and the inclusion of models with \mni\ larger than 0.08~\ms\ into our grid of simulations was not necessary. 

\citet{pejcha+15b} estimated \mni\ for 21 \sneii\ and found a minimum of 0.0045~$\pm$~0.0008~\ms\ for SN~2001dc\footnote{The minimum \mni\ in \citet{pejcha+15b} is 0.0015\ms\ for SN~2006bp. However, the authors note that this \Ni\ mass estimate is uncertain given that there is only photometry in one band during the transition and the tail phase, which poorly constrained the temperature term in their modelling. For this reason, we exclude SN~2006bp from the comparison.} and a median of 0.030~\ms. Both the median and minimum values are consistent with our determination.
However, a large discrepancy is found for the maximum value. \citet{pejcha+15b} inferred a \Ni\ mass yield of 0.28~\ms\ for SN~1992H, which is significantly higher than the 0.069~\ms\ estimated in our study for SN~2007X. \citet{pejcha+15b} inferred a larger distance to SN~1992H than previous estimates \citep{schmidt+94,clocchiatti+96} making this SN more luminous compared to normal \sneii, and therefore, with a larger \mni.

\citet{muller+17} determined \mni\ for 19~\sneii\ and combined their sample with that from \citet{pejcha+15b} making a larger sample of 38 objects. Their \mni\ distribution is described by a median value of 0.031~\ms\ -- consistent with ours -- with values between 0.005 and 0.280~\ms. The latter value corresponds to SN~1992H as described above. 
More recently, from a compilation of 115 \snii\ \Ni\ mass estimates from the literature, \citet{anderson19} estimated a median value of 0.032~\ms, in agreement with our results. 
The minimum and maximum values in \citet{anderson19} are 0.001 and 0.360~\ms, respectively. This range is larger than that determined in our study, particularly because the high \Ni\ mass estimate of 0.360~\ms\ \citep[for SN~1992am;][]{nadyozhin03} is significantly larger than the maximum value estimated for the CSP-I sample. 
The larger sample in \citet{anderson19} may indicate that values above the adopted maximum in our models of 0.08~\ms\ are exceptional.
Around 10\% of the \sneii\ in the literature have estimated \Ni\ masses above 0.08~\ms\ \citep[see Fig.~1 from][]{anderson19}.
However, some of these \sneii\ are classified as 1987A-like events, which have relatively high \Ni\ masses compared to the `normal' \sneii\ studied in the current work.
Additionally, a few \sneii\ have more than one \Ni\ mass estimate in the literature, and some of these are consistent with our adopted maximum \Ni\ mass. Differences are attributed to distinct distance and reddening values used in the literature for the same SN.

\citet{sukhbold+16} report \Ni\ masses based on neutrino-powered explosions for numerous pre-SN models within a large range of initial masses. The minimum \mni\ values in \citet{sukhbold+16} are 0.003 and 0.006~\ms\ for the 9.25 and 9.0~\ms\ models, respectively. The different values correspond to the different codes used for their calculation. These values are similar to the lowest \mni\ derived for our \snii\ sample.
For stars between 10 and 25~\ms\ (the maximum initial mass in our pre-SN models), most \mni\ values are found within the range of 0.01$-$0.08~\ms, which is consistent with our findings. Only a few explosion models produce larger \mni\ in \citet{sukhbold+16}, with a maximum value of $\sim$0.10~\ms.

\subsection{\Ni\ mixing}
\label{sec:ni_mixing}

During a SN explosion, a shock wave emerges heating the stellar material. If the shock temperature is sufficiently high, explosive nucleosynthesis takes place. 
Explosive nucleosynthesis efficiently produces heavy elements from Si to Zn \citep{umeda+02}.
\Ni\ dominates the production of nuclear species in the inner regions of the star, but Co, Zn, and additional Ni isotopes are also mostly produced here \citep{woosley+95,thielemann+96,umeda+02}.
Once the shock reaches the composition interfaces between the carbon-oxygen core and the helium core, and the helium core and the hydrogen-rich envelope, RT instabilities appear causing a large-scale spatial mixing. This is the mechanism by which the heavy elements synthesised during explosive burning can reach the outer regions of the envelope while hydrogen can be mixed inwards.

The first evidence of chemical mixing during the explosion was based on the observations of SN~1987A. A substantial amount of \Ni\ mixed into the hydrogen-rich envelope is required to explain the rise time to peak of SN~1987A \citep{shigeyema+88,shigeyama+90}, and the LC and spectral evolution of X-rays and $\gamma$-rays \citep{kumagai+89}. 
Furthermore, the presence of significant Fe emission at high velocities during the nebular phase demonstrates that a significant amount of \Ni\ was mixed into the hydrogen-rich envelope during the explosion \citep{haas+90}.
While the progenitor of SN~1987A is a blue supergiant (BSG) star, \snii\ progenitors are known to be RSGs with significantly distinct pre-SN structures. Recent 3D hydrodynamical simulations of the evolution of the SN shock for both BSG and RSG explosions have shown that RSG models achieve higher maximum velocities for the \Ni\ into the ejecta \citep{wongwathanarat+15} implying more extended mixing of \Ni. 

In Sect.~\ref{sec:results} we showed that the \Ni\ distribution in the ejecta covers the range of initial values, from 0.2 (inner ejecta) to 0.8 (outer ejecta).
However, in both the gold and full samples, the vast majority of the events are found within low (0.2$-$0.4) and moderate (0.4$-$0.6) values and only a low number of \sneii\ are found with a large extent of \mix\ (Fig~\ref{fig:hist2}, bottom panel). 
An extreme mixing of \Ni\ is hard to obtain in RSGs because of their large envelopes. Our results are consistent with this concept given that an extended \Ni\ mixing was only found for one \snii\ in the gold sample (SN~2008ag, \mix\,=\,0.8). 
We note that the \mix\ is characterised as a fraction of the pre-SN mass. This means that the same value of \mix\ for different progenitors may represent different extents of \Ni\ in mass and radial coordinates.

We stress that in this study the \mix\ is treated as a free parameter. Recently, \citet{wongwathanarat+15} computed 3D hydrodynamical simulations and found that the extent of the mixing of the metal-rich ejecta depends on the early-time asymmetries generated by the neutrino-driven mechanism, explosion energy, and progenitor structure. In the progenitor structure, the most important features are the width of the carbon-oxygen core, the density structure of the helium core, and the density gradient at the composition interface between the helium core and hydrogen-rich envelope.
However, only a few hydrodynamical simulations for RSG models have been carried out in the literature. Further studies are crucial for better understanding of the \mix\ process in \sneii.

\subsection{Explosion epochs}

We compared the explosion epochs derived in our analysis with those from \citet{gutierrez+17I} for each SN. Our estimations are always inside the range of values derived by \citet{gutierrez+17I}, which is by design since we did not allow our fitting procedure to sample values outside that range (see Sect.~\ref{sec:mcmc}).
The mean difference between both estimates is 0.1~days with a standard deviation of 4.8~days.
There are a few cases where our explosion epoch estimates are very close to the observational limits, possibly suggesting explosion dates beyond the uncertainties of \citet{gutierrez+17I}.
In \citetalias{martinez+20}, we tested this by relaxing the prior for the explosion epoch and obtained a similar marginal distribution of the physical parameters, indicating no changes in our results.

That the mean offset is so close to zero gives significant support to both methodologies of estimating the explosion epoch. While we did not allow our modelling estimates to be outside the range of the observational errors, the modelling explosion epochs could have been biased towards -- on average -- higher or lower values than the observational values. 
In the case of observational explosion epochs, the constraints from the non-detection that the SN has not exploded at that date are only as strong as the survey depth providing the non detection. One could therefore imagine that `true' explosion epochs could be biased towards values close to the non detection. However, the above analysis shows this not to be the case, and therefore gives support to the observation epochs as estimated by \citet{gutierrez+17I}.
At the same time, one could imagine that a bias in the modelling technique could mean that the fitting could attempt to use the prior on the explosion epoch to systematically offset the plateau duration -- and therefore the ejecta mass -- to lower or higher values by constraining the explosion epoch to be later or earlier, respectively (with respect to the observational value). Again, this is not seen in our data, giving support to the robustness of our model fitting methods.

\subsection{Non-standard \sneii}
\label{sec:nonstd}

In this study, we utilise a large grid of explosion models assuming standard values for several stellar evolution parameters (metallicity, wind efficiency, mixing length, and overshooting) and explosion parameters \citepalias{martinez+20}.
Although most of the well-sampled \sneii\ in the CSP-I sample can be adequately reproduced by the models in this grid, some are not.

SN~2006Y and SN~2008bu reside in the lower end of the optically thick phase duration distribution with 64\,$\pm$\,4~days and 52\,$\pm$\,7~days, respectively \citepalias{martinez+21}.
Unfortunately, no model in our grid has such a short optd.
In the case of SN~2006Y, the closest fitting models can reproduce the velocities and the plateau luminosity reasonably well. However, the optd is overestimated by $\sim$10~days. At first sight, this difference is small, but the models also largely underestimate the radioactive tail luminosity by $\sim$0.6~dex. Additional \Ni\ can easily fit the tail, but it is known that more \Ni\ also produces a longer plateau phase \citep[e.g.][]{kasen+09,bersten13phd}, which increases the discrepancy with the observations.
On the other hand, for SN~2008bu, the situation is even worse as the optd lasts $\sim$10~days less than that of SN~2006Y.

Theoretical studies suggest that progenitors with smaller hydrogen-rich envelope masses at time of collapse produce shorter optd \citep[e.g.][]{litvinova+83,kasen+09,bersten13phd}.
Following the standard mass-loss rate in the stellar modelling of single stars (as we assume in this study) it is not possible to find such stripped progenitors with \mzams\ smaller than 25~\ms, which is the highest-mass progenitor in our grid.
More massive stars can finish with smaller envelopes but they are more difficult to explode \citep{sukhbold+20}.
Therefore, lower ejecta masses can be achieved by increasing the mass-loss rate via stellar winds, interacting binaries, or rotation, among other possibilities (as discussed in Sect.~\ref{sec:discussion}).
We calculated new pre-SN models with enhanced mass loss providing much better fits to SN~2006Y and SN~2008bu.
These results are presented in \citetalias{paper3_submitted}. Our results agree with those from \citet{hiramatsu+21}. These authors presented models for SN~2006Y, as well as other two short-plateau \sneii, from high-mass progenitors that experience more mass loss than standard stellar evolution.

SN~2004er also shows some discrepancies between models and observations, especially with the velocities.
SN~2004er shows the longest optd with 146~$\pm$~2~days \citepalias{martinez+21}. Its bolometric LC is mostly well-reproduced with the exception that the fitting models show shorter optd by $\sim$10~days.
The largest discrepancy is found in the velocity evolution, where the observed \ion{Fe}{ii} velocities are greater than the models by $\sim$2000~km~s$^{-1}$. 
In fact, SN~2004er displays higher \ion{Fe}{ii} velocities than most \sneii. For example, the \ion{Fe}{ii} velocity at 55~days post-explosion is 5033~$\pm$~751~km~s$^{-1}$, while the mean \ion{Fe}{ii} velocity of \sneii\ at 53~days is 3537~$\pm$~851~km~s$^{-1}$ \citep{gutierrez+17I}.
Such large velocities can be achieved through higher explosion energies, although this would produce more luminous and shorter plateau phases. However, larger ejecta masses have the opposite effect, that is, dimmer \sneii\ with longer plateaus.
Therefore, SN~2004er was probably a high-energy explosion of a star with a massive pre-SN envelope -- a model that is beyond the parameter space sampled in the current study (and therefore beyond stellar evolution with standard assumptions and/or values of the input parameters).
Something similar happens for SN~2007sq. Its bolometric LC is well fit by our standard models, but our model velocities fall below observations.
The \sneii\ above analysed (SNe~2004er, 2006Y, 2007sq, and 2008bu) may be examples of high-mass RSG progenitors that could explain the lack of more massive progenitors in the sample, although a detailed analysis is necessary.

Finally, we briefly describe the fitting models of SN~2008K. These models show offsets both in the bolometric LC and velocities with respect to the observations.
The bolometric LC is well fit in general, but when looking at the details, it is seen that the models do not resemble the linear behaviour of SN~2008K, which has a plateau decline rate of 1.70~$\pm$~0.31~mag per 100 days \citepalias{martinez+21}. SN~2008K also has higher \ion{Fe}{ii} velocities than most \sneii. 
In principle, more energy is needed to reproduce the velocities. This implies more mass to conserve the plateau duration (more energy reduces the plateau length). However, we compared SN~2008K with higher-energy and more-massive explosion models and found even larger discrepancies given that the new parameters produce LCs similar to typical SNe~IIP. The detailed pre-SN and explosion properties of SN~2008K are still uncertain.
The examples above reflect that some of the standard assumptions in stellar evolution cannot cover all the physical properties of \snii\ progenitors. Non-standard stellar evolution is required to explain the properties of some individual SNe~II.

\section{Discussion}
\label{sec:discussion}

\subsection{The RSG problem}
\label{sec:rsg_problem}

During recent years, the number of works studying the \mzams\ distribution of \snii\ progenitors has increased.
\citet{smartt+09} were the first to analyse the \mzams\ distribution from observed progenitors in pre-explosion images, 
founding a maximum \mzams\ of 16.5\,$\pm$\,1.5~\ms.
This is in contrast with model predictions and that more massive RSGs are observed in the Local Group \citep{levesque+06,neugent+20}. The lack of higher-mass progenitors is called the `RSG problem'. 
Since then, special attention has been given to the upper mass boundary of the \mzams\ distribution. This limit is crucial to understanding the evolutionary pathways of massive stars. 
A larger set of \snii\ progenitors consisting in 13 detections and the same number of upper limits was studied by \citet{smartt15}, but no progenitors were found with \mzams\ above 18~\ms.

Inferring the progenitor \mzams\ from a direct detection requires a previous estimation of the progenitor luminosity. 
This is used to compare  with the luminosity predicted by stellar evolution theory. Thus, the physical models provide an estimate of the mass from a luminosity measurement.
In this context, \citet{straniero+19} studied the role of convection, rotation, and binarity in \snii\ progenitor evolution finding that neither of these uncertain processes in stellar modelling can mitigate the RSG problem. 

The progenitor luminosity can be achieved by fits to the spectral energy distribution if the progenitor is detected in several bands, or by using bolometric corrections to convert single-band flux into luminosity. The uncertainties in the latter case may be large if only single-band or two-band detections are available.
\cite{davies+18} investigated this source of systematic error and derived new bolometric corrections and extinction values. Those authors found an increased upper mass limit of \mmax\,=\,19$^{+2.5}_{-1.3}$\ms\ (68\% confidence), with a 95\% upper confidence limit of <\,27~\ms.
Additionally, \citet{davies+18} analysed the effects of a small sample size on \mmax, concluding that this causes a systematic error of $\sim$2~\ms\ that shifts the upper mass limit to larger values.
Recently, \citet{davies+20} inferred the properties of the progenitor distribution but working directly with the observed luminosities, thus eliminating the uncertainties introduced in stellar modelling when converting the observed luminosities into \mzams. 
Finally, they compared their maximum luminosity with stellar models and found a \mmax\ of 18$-$20~\ms. However, they remark that the sample size should be at least doubled to enable a reduction in the large uncertainties on \mmax.

The direct detection of progenitors in pre-explosion images is the most powerful tool to determine the nature of progenitor stars as it can be directly linked to a progenitor system. However, the analysis of direct detections can only be applied -- when pre-SN images exist -- to the nearest SNe ($d$\,$\lesssim$\,30~Mpc) due to the lack of resolution for more distant objects.
The study of the stellar populations in the immediate SN environments is an alternative indirect method that is also useful for nearby objects where individual stars or clusters are resolved. This information can yield constraints on the ages, and therefore, on the progenitor masses. These measurements are mostly consistent with initial masses of \snii\ progenitors lower than 20~\ms, although they present larger uncertainties than the direct identification of the progenitor star \citep[e.g.][]{vandyk+99,williams+14,maund17}. 

Nebular spectra can also be used to constrain the progenitor mass. After the optically thick phase, when all the hydrogen is recombined, the core of the progenitor star becomes visible and nucleosynthesis yields can be analysed \citep{jerkstrand+12,dessart+20}.
Using this method, most of the mass estimates agree with the lack of high-mass progenitors \citep[see e.g.][]{jerkstrand+14,valenti+16}. However, \citet{anderson+18} infer an initial mass of 17--25~\ms\ for SN~2015bs through comparison to nebular-phase spectral models suggesting an explosion of a higher-mass progenitor than previously observed for a \snii.

Different alternatives have been proposed to solve the RSG problem.
One possibility is that high-mass progenitors do not explode and instead collapse into black holes. Massive stars in the range of the missing RSG progenitors could have pre-SN core structures that are generally more difficult to explode \citep[see e.g.][]{oconnor+11,sukhbold+20}. The search for failed SNe has found possible candidates corresponding to high-mass stars \citep{adams+17a,allan+20,neustadt+21}, lending support to theoretical predictions. 
In addition, it has been suggested that more massive RSGs produce weak explosions leading to faint \sneii\ \citep{fryer99,heger+03}.
\citet{zampieri+03} investigated the possibility that two low-luminosity \sneii\ (SN~1997D and SN~1999br) could be produced in low-energy explosions of massive progenitors.
An alternative solution posits that stellar evolution models could be modified in order to increase the mass loss and reduce the maximum theoretical mass of stars exploding as a RSG. With this assumption, massive stars in the range of the missing RSG progenitors will explode as stripped-envelope SNe instead of \snii\ events \citep[see e.g.][]{yoon+10,ekstrom+12}. 

Unfortunately, most of the previously mentioned methods to constrain progenitor parameters of SNe are restricted to nearby events because they require either images of high enough resolution and sensitivity or spectra at late-time epochs when the SN is significantly dim. 
Hydrodynamical modelling of LCs (and sometimes spectra) is one of the most commonly applied indirect methods to infer physical properties of \sneii\ as it can be applied to large distances, and therefore statistically significant samples. 
\citet{morozova+18} inferred progenitor and explosion properties of a sample of 20 \sneii\ from LC modelling and used the derived \mzams\ to explore the mass distribution. They found an upper mass cut-off of \mmax\,=\,22.9~\ms\ ($\sim$30~\ms\ at 95\% confidence limit). However, \citet{morozova+18} did not consider expansion velocities in their modelling, the exclusion of which has been shown to bias progenitor masses to larger values \citepalias[see][]{martinez+20}.

In Sect.~\ref{sec:imf} we determined the IMF parameters (\mmin, \mmax, and $\Gamma$) for the gold and full samples. For each sample, we also fitted the IMF leaving $\Gamma$ as a free parameter or fixing it to $-$2.35. Thus, in total, we have four different estimations of the IMF parameters. 
All these estimations find \mmax\,$\lesssim$\,21.5~\ms. In principle, this is consistent with the RSG problem. However, different levels of significance are found for each estimation, which are examined below.
We first analyse the cases when $\Gamma$ is left as a free parameter. 
We find \mmax\,=\,21.3$^{+3.8}_{-0.4}$~\ms\ for the gold sample, that is, \mmax~=~25~\ms\ is within the 68\% confidence limit.
The study of the full sample indicates \mmax\,=\,21.5$^{+1.2}_{-0.8}$~\ms\ (68\% confidence) with a 95\% upper confidence limit of <\,27.1~\ms.
While this increases the significance of our results with respect to the RSG problem, we note that the results obtained from the full sample are not as robust as those from the gold sample.
When $\Gamma$ is fixed to the value of a Salpeter massive-star IMF, we find \mmax\,<\,21~\ms\ for both the gold and full samples. In these two cases, the RSG problem is recovered with a significance above the 99\% significance level.

\subsection{The IMF incompatibility}
\label{sec:imf_incompatibility}

The \snii\ progenitor IMF derived in Sect.~\ref{sec:imf} is inconsistent with a Salpeter IMF due to the large difference found in the slope of the distributions -- which we named the IMF incompatibility.
However, the massive-star IMF is very well established. Several studies of OB associations and clusters suggest that the majority of the stars above a few solar masses are drawn from a Salpeter power-law \citep{bastian+10}.
In addition, progenitor mass constraints from the stellar populations surrounding a SN event and \snii\ progenitor mass estimations from pre-SN imaging are also drawn from a Salpeter IMF \citep{smartt15,koplitz+21}. 
The apparent incompatibility of our results with a standard IMF is driven by the large number of low-\mzams\ progenitors found with respect to more massive progenitors.
In Sect.~\ref{sec:sample} we pointed out that CSP-I built a magnitude-limited sample of \sneii.
The luminosity function of a magnitude-limited distribution could be biased against lower-luminosity events that generally come from low-\mzams\ progenitors and low-energy explosions \citepalias[see][]{paper3_submitted}.
This is the opposite of the volume-limited sample of \sneii\ with progenitor detection in pre-explosion images \citep{smartt+09}.
Given that most \sneii\ in our sample are consistent with such low-mass events, there does not seem to be significant evidence that this bias is affecting our results or conclusions.
We therefore believe that this incompatibility reflects a lack of understanding of some physical ingredients in our study or the completeness of our sample. Here we discuss possible reasons, but a deep and systematic analysis should be carried out in order to identify the ultimate reason for the IMF incompatibility. 

\subsubsection{Sample size}
\label{sec:sample_size}

\begin{figure}
\centering
\includegraphics[width=0.46\textwidth]{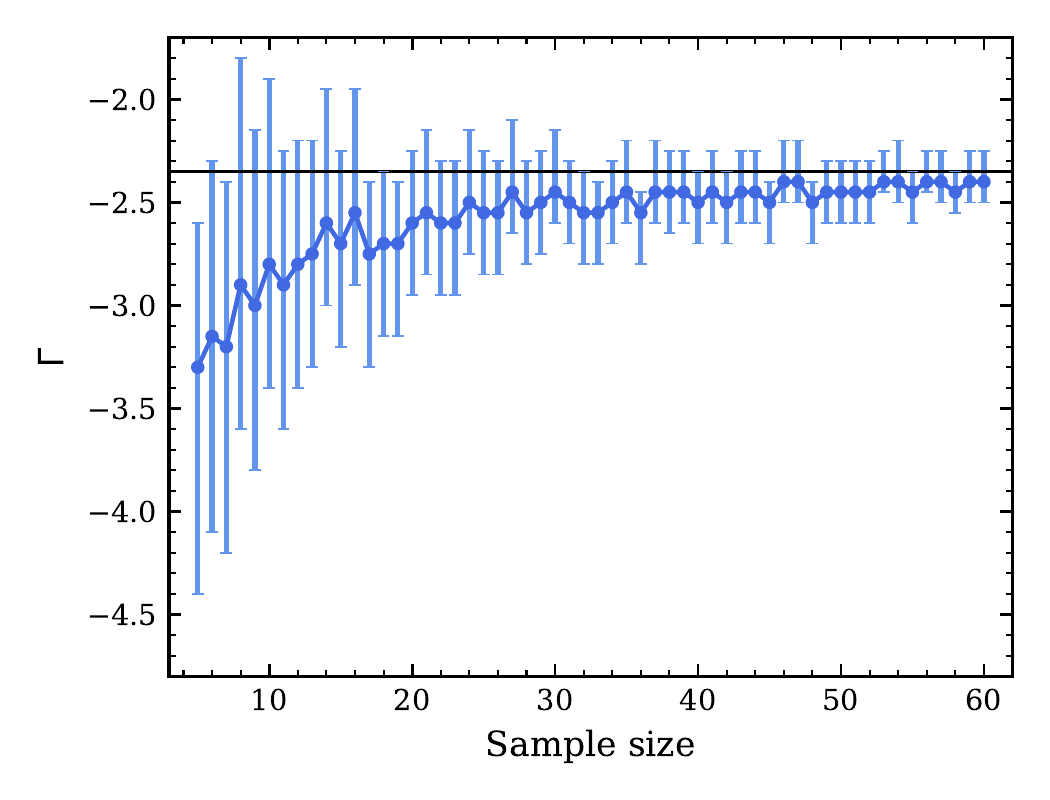}
\caption{Best-fitting power-law slope of the CD of the progenitor masses for a range of sample sizes. The error bars correspond to the 68\% confidence limit. The larger the sample, the higher the probability that the best-fitting power-law slope is that of a Salpeter IMF. The solid black line indicates the Salpeter IMF slope for massive stars.}
\label{fig:sample_size}
\end{figure}

First, we explore if the IMF incompatibility is consequence of the sample size. 
We performed a test similar to that presented in Sect.~\ref{sec:imf} for different sample sizes ($N$).
We randomly sampled the initial masses of $N$ progenitors assuming a Salpeter massive-star IMF with \mmin~=~9~\ms\ and \mmax~=~25~\ms\ (the \mzams\ range of our pre-SN models), and ordered the progenitors in increasing mass.
We performed 10000 MC simulations to determine the CD of the masses for each value of $N$, and then we fit power-law models to these CDs. Given that we are interested in testing the sample size effect in the estimation of $\Gamma$, we constrained \mmin\ and \mmax\ to 9 and 25~\ms, respectively, which means that $\Gamma$ was the only parameter to determine.
The posterior distribution of the $\Gamma$ parameter was found via an MCMC procedure.
We employed 10000 MC trials to determine the posterior probability distribution for the progenitors at each $\Gamma$ value. We performed the same experiment for a range of sample sizes. 

The results of this test are presented in Fig.~\ref{fig:sample_size}. The larger the sample size, the best-fitting $\Gamma$ tends to the value of the Salpeter IMF slope ($-$2.35, black solid line).
For 24 objects (i.e. the number of \sneii\ in the gold sample), the best-fitting $\Gamma$ is $-$2.50$^{+0.25}_{-0.35}$, which means that $\Gamma$~=~$-$2.35 is within the 68\% confidence interval. 
Something similar happens for $N$~=~51 (i.e. the number of \sneii\ in the full sample with \mzams\ estimates), where we find $-$2.45$^{+0.15}_{-0.15}$.
This indicates that our sample size do not produce the bias we found in the IMF. 
As the IMF incompatibility cannot be explained by the sample size, in the following sections we discuss some possible physical reasons for this discrepancy.

From Fig.~\ref{fig:sample_size}, we also note the best-fitting $\Gamma$ differs significantly from that of a Salpeter IMF for samples smaller than 20 objects. This restricts future studies to work with larger samples for a reliable estimate of the power-law slope.

\subsubsection{Pre-SN models}

The IMF incompatibility is produced by the large relative number of low- to high-\mzams\ progenitors. 
In Sect.~\ref{sec:results} we found a large number of explosions with progenitor structures compatible with \mej\ in the range of 8$-$10~\ms\ and \ra\ between 450 and 650~\rs\ (Fig.~\ref{fig:hist1}, middle and bottom panels). In the context of standard single-star evolution adopted in the current study, these pre-SN structures are consistent with low-\mzams\ progenitors.
However, if the standard assumptions in stellar evolution change (which is possible given the uncertainties in massive-star modelling), different pre-SN configurations could be obtained for the same \mzams\ value.
Below we discuss how different factors such as the envelope stripping processes, metallicity, mixing length, and overshooting can affect the pre-SN structure, and therefore our results. 

\vspace{0.2cm}
\noindent
\textsf{Mass-loss rate.}
Similar \mej\ (for higher \mzams) can be achieved by increasing the mass-loss rate via stellar winds, eruptive mass-loss events, rotation, binary interaction, or a combination of these phenomena.
If more mass were lost by progenitors, then the estimated \mej\ could be reduced for more massive \mzams, and thus this would reduce the inconsistency of our results with a standard IMF.
While higher mass loss may change the \mzams\ distribution derived from hydrodynamical modelling, additional mass stripping does not change the progenitor \mzams\ estimations from pre-SN imaging. This is because the progenitor final luminosity is closely related to its helium core mass and it is almost independent of the final envelope mass (direct progenitor observations do not provide ejecta mass estimates). The sensitivity of the initial mass--final luminosity relationship on the adopted mass-loss rate is small \citep{straniero+19,farrell+20}.

Recent studies infer that most massive stars have rotational velocities in the range of 0$-$300~km~s$^{-1}$ \citep{ramirez+13,dufton+13}. Since rotation increases mass loss, the final mass of rotating stars is smaller than that of non-rotating stars \citep{hirschi+04}.
Following this, if we consider pre-SN models of rotating massive stars we could find lower-\mej\ for higher-\mzams\ progenitors, that is, in the direction to explain the IMF incompatibility.

\citet{beasor+20} and \citet{humphreys+20} derived new mass-loss prescriptions based on direct measurements and did not find larger mass-loss rates in the RSG phase than previous estimations.
In addition, \citet{beasor+20} conclude that quiescent mass-loss during the RSG phase (when most of the mass loss occurs) has little effect in stripping the envelope.
In principle, this excludes the RSG winds as a mechanism to strip a larger part of the hydrogen-rich envelope \citep[see also][]{beasor+21}.
However, \citet{ekstrom+12} proposed new mass-loss prescriptions for RSGs considering that they may undergo episodic mass-loss events (i.e. more mass is lost).
\citet{neugent+20} compared the observed luminosity function of RSGs in M31 to that predicted by the stellar models from \citet{ekstrom+12} and concluded that the new prescription for RSG mass-loss rate reproduces the observations.
Given the inconsistencies in previous estimations, it is not clear whether mass-loss processes on isolated RSGs are responsible for stripping a significant part of the envelope.

Very early spectra have shown the presence of CSM around \snii\ progenitors \citep[e.g.][]{khazov+16,yaron+17,bruch+21} indicating that enhanced mass loss might occur prior to core collapse. This is also inferred from early-time LC modelling \citep[e.g.][]{morozova+18,forster+18}.
Such high mass-loss rates are not consistent with direct measurements of RSG winds possibly implying another mechanism to induce mass loss.
\citet{fuller17} suggests that pre-SN outbursts via waves driven by convection during the late stages of evolution may reproduce the CSM properties.
However, despite early-time observations of many \sneii\ indicate very high mass-loss rates shortly before the explosion, the amount of mass involved is small, which means that this process cannot explain the low pre-SN masses.

An efficient way to remove a large fraction of (if not all) the outer envelope of massive stars is through mass transfer to a companion star in a binary system \citep[e.g.][]{podsiadlowski+92,benvenuto+13}.
Therefore, this mechanism can produce \snii\ events with lower ejecta masses than predicted by single-star evolution.
According to recent studies, most massive stars belong to binary systems \citep[or higher-order multiples;][]{moe+17}. Moreover, it has been shown that a large fraction of massive stars will exchange mass with a companion prior to the RSG phase \citep{sana+12,sana+13}.
Recently, from analytical estimations and population synthesis simulations, \citet{zapartas+19} showed that a large fraction (from one-third to half) of \snii\ progenitors will interact with a companion star during its evolution.
Additionally, \citet{eldridge+18} showed that the diversity of CCSN LCs can be interpreted as the result of binary interaction. 
Previously, it has generally been assumed that SN~II arise from progenitors that explode with most of their hydrogen envelopes intact. SN~II modelling assuming such progenitors has generally been consistent with observations and thus it has not been necessary to invoke binarity. However, this does not mean that the effects of binary evolution are unimportant in explaining SN~II observations and diversity. Our results and the above mentioned studies support the hypothesis that many SN~II progenitors may experience larger mass loss than expected assuming standard single-star evolution. 
In addition, a different treatment of mass loss is required in the cases of SN~2006Y and SN~2008bu (see Sect.~\ref{sec:nonstd} and \citetalias{paper3_submitted}). Thus, it may well be that binarity plays a larger role in shaping SN~II behaviour than previously assumed.
We now discuss whether other stellar parameters can explain the discrepancy.

\vspace{0.2cm}
\noindent
\textsf{Metallicity.}
The influence of metallicity on the winds of hot-massive stars is well known \citep{vink+01}. However, for single-massive stars, most of the mass lost happens in the RSG phase, where the rate of mass loss is assumed to be nearly independent of metallicity \citep{vanloon06,goldman+17} at least within the metallicity range of the CSP-I \snii\ sample (see below). 
The influence of metallicity on the winds can change the pre-SN structure at the time of collapse. Lower metallicities produce more compact and more massive progenitors \citep{dessart+13}. 
However, we do not expect large variations in the metallicity of the CSP-I sample. \citet{anderson+16} inferred environment metallicities for the CSP-I \snii\ sample and found a narrow range in metallicities, which is not expected to significantly alter the stellar evolution.
In addition, \citet{anderson+16} found that there is no relationship between progenitor metallicity and \snii\ LC and spectral properties (except for the metal line equivalent widths).
\citet{gutierrez+18} also concluded that the hydrogen-rich envelope mass and the explosion energy are not correlated with the metallicity. 
Therefore, we do not expect that metallicity significantly modifies our results obtained from solar-metallicity stellar models.

\vspace{0.2cm}
\noindent
\textsf{Mixing-length parameter.}
Pre-SN models with different mixing-length parameters differ primarily in the final radius (larger mixing-length parameters produce smaller radii).
In our study, we find that most \snii\ progenitors are consistent with RSG radii between 450 and 600~\rs\ (Fig.~\ref{fig:hist1}, bottom panel).
The minimum progenitor radius in our grid of pre-SN models is 450~\rs\ (using a mixing-length parameter of two). Smaller radii can be obtained with a mixing-length parameter of three, as previously suggested by \citet{dessart+13} based on the analysis on SN~1999em.
Adopting a larger mixing-length parameter, we would obtain pre-SN structures with similar pre-SN masses but smaller radii.
A smaller pre-SN radius for the same initial mass might help to solve the IMF incompatibility as more massive progenitors are needed to derive our radius distribution. However, a direct comparison is not possible given that larger \mzams\ progenitors also imply larger ejecta masses and the LC models will dramatically change.
Therefore, it would be necessary to carry out a series of simulations before being able to conclude whether or not changes in this parameter could be a determining factor in the IMF incompatibility.

\vspace{0.2cm}
\noindent
\textsf{Overshooting.} A more extended convective zone produces a more massive helium core, and therefore, higher luminosities.
Because of the higher luminosity, the mass-loss rate is larger producing smaller final hydrogen-rich envelope masses.
In principle, this can help to solve the discrepancy, but larger overshooting parameters produce much more extended RSGs with dense envelopes \citep{wagle+20}, acting in the opposite direction to our results. A detailed future analysis should be carried out to fully understand these effects. 

\subsubsection{Failed \sneii}

In the previous sections, we give possible explanations for the IMF incompatibility, but always assuming that the mass function of \snii\ progenitors should be consistent with a Salpeter massive-star IMF. However, these two distributions may be intrinsically different.
Here, we test whether the lack of massive progenitors might be caused by high-\mzams\ progenitors failing to produce a SN explosion, where their collapse leads to an implosion into a black hole.
If the shock produced by the core collapse fails to explode the RSG star, it may produce a low-luminosity red transient lasting for about a year \citep{lovegrove+13}.
Such weak explosions would be difficult to detect and may contribute to the lack of high-\mzams\ progenitors.

We constructed a \snii\ progenitor IMF using the results from \citet{sukhbold+16}, where the authors calculated explosion models for a large and fine grid of pre-SN models and distinguish between exploding and non-exploding cases.
The procedure is similar to that presented in Sect.~\ref{sec:imf}, but with one exception. Here, at each MC simulation, we randomly sampled a certain number of stars in the range from 9 to 25~\ms\ (the lower and upper \mzams\ of our pre-SN models) assuming a Salpeter massive-star IMF with the condition that the sample has 24 objects (i.e. the number of \sneii\ in the gold sample) after removing the non-exploding stars.
We determined the probability distribution of the CD of the \mzams\ for the \snii\ progenitors.
We fitted this CD in the same manner as in Sect.~\ref{sec:imf} and found the following parameters: \mmin~=~9.1$^{+0.1}_{-0.2}$~\ms, \mmax~=~20.8$^{+3.0}_{-2.3}$~\ms, and $\Gamma$~=~$-$3.32$^{+0.71}_{-0.68}$.
These results imply that if failed \sneii\ are taken into account, the upper \mzams\ limit and the slope of the \snii\ progenitor IMF are modified.
A steeper power-law than that of a Salpeter IMF is found, although not as steep as the slope inferred for the gold sample.
In addition, for the case of \mmax, a lower value than that assumed for the Salpeter IMF is obtained.

\subsubsection{Modelling}

Here we briefly discuss if the differences found in the IMF can be attributed to the simplifications in our hydrodynamical simulations, or to a bias in our fitting procedure.

\vspace{0.2cm}
\noindent
\textsf{Hydrodynamical simulations.} While the details of the explosion mechanism are still under debate \citep[see][for a recent review]{burrows+20}, observations (polarisation, SN remnants, and nebular spectroscopy) show that the explosion is asymmetric in the inner regions \citep[e.g.][]{hoflich91,delaney+10,taubenberger+09,tanaka+12}.
However, our hydrodynamic code assumes spherically symmetric explosions, which might produce systematic effects (that would be inherent to all 1D codes).

In our explosion models, the explosion energy is deposited at the inner layers by artificially adding internal energy in form of a thermal bomb. The explosion energy is used to overcome gravity and lifts the envelope out of its potential.
The high initial explosion energy produces a strong shock front and pushes the material above resulting in the ejection of the envelope.
Inner low-velocity and low-density regions expanding with a certain minimum velocity appear. However, such regions are artefacts of the 1D nature of the modelling.
In a spherical explosion, a hot low-density and fast region below a cold and denser stellar structure is formed. Thus, this region is highly RT unstable, with instabilities growing on timescales of seconds to minutes \citep{janka+93}.
In multi-dimensional calculations, RT instabilities mix material filling the inner low-density region.
The inner low-density and low-velocity region present in 1D codes might produce systematic errors in the physical parameter estimations, although a detailed analysis of this effect in our results is beyond the scope of this work.

In addition, RT instabilities cause a large-scale mixing of heavy elements into the hydrogen-rich envelope when the shock wave reaches the composition interfaces between the carbon-oxygen core and the helium core, and the helium core and the hydrogen-rich envelope \citep{muller+91,herant+91,kifonidis+03}, and also the mixing of hydrogen inside the helium core \citep[see][and references therein]{utrobin+08}. Such mixing of hydrogen inwards is not taken into account in our simulations. However, as shown in previous studies, this causes a minor effect on the LC morphology \citep{utrobin+08}. 
Additionally, it has been proposed that RT instabilities could also modify the pre-SN density profile \citep{utrobin+08}, leading to a smoother transition between zones of different chemical abundances. 
As this process happens before the arrival of the shock wave to the surface, it could be taken into account simply by modifying the pre-SN model assumed in the 1D hydrodynamic simulation.
The above discussion is the reason why some authors have decided not to use stellar evolution models as initial configurations in 1D simulations. 
However, it is difficult to conclude whether this effect could be the reason for the IMF incompatibility. 

Another important factor that can impact our results is found in the opacity calculation. Our hydrodynamical code adopts a minimum value of the opacity sometimes referred to as the `opacity floor' to account for the actual contribution of the line opacity and the non-thermal excitation or ionisation of electrons that are not included in our opacity calculation \citep[see more details in][]{bersten+11}. 
Different values of the opacity floor produce differences in our LC and photospheric velocity models that could modify our results. This could be a systematic source of error that needs to be studied. However, the adopted value for the opacity floor was selected based on several tests \citep[see][]{bersten+11,bersten13phd} with the STELLA code \citep{blinnikov+98} that uses multi-group radiative transfer and includes the effect of line opacities \citep{sorokina+02}.

\vspace{0.2cm}
\noindent
\textsf{Fitting technique.} Another possibility is that our fitting procedure could bias our results to low-\mzams\ progenitors because of the degeneracy between physical parameters, meaning that explosions with different physical properties can produce similar photometric and spectroscopic properties \citep{dessart+19,goldberg+19,martinez+19}. However, in \citetalias{martinez+20}, we derived the progenitor and explosion parameters for eight~\sneii\ using the same fitting method and found very good agreement with the progenitor initial mass estimations from pre-SN imaging and nebular spectral modelling.
Furthermore, in \citetalias{martinez+20}, we compared \sneii\ observations with several high-mass models, finding that no such model could reproduce the observations \citepalias[see][their Fig.~11]{martinez+20}.
We note that the eight \sneii\ from \citetalias{martinez+20} have progenitor and observed properties within the ranges derived in this study and \citetalias{martinez+21}, respectively.
We also propose several high-mass models as observations to test our procedure, recovering the high-mass models in each case (see Appendix~\ref{app:high_mass}). We conclude that our fitting procedure is not biased against high-mass solutions.

\vspace{0.3cm}

In this section we have discussed several possible explanations for the IMF incompatibility.
It is clear that the uncertainties in stellar evolution can have a significant impact on the final structure of progenitors, and therefore, on LCs and spectra \citep[see e.g.][]{dessart+13}. 
From the present discussion, we conclude that higher mass-loss rates have a clear effect going in the direction needed to explain the IMF incompatibility, although we cannot rule out other possible factors. 
Additional mass stripping during evolution may be required to obtain similar ejecta masses as those from this study, but for higher-\mzams\ progenitors.
Lower masses at the time of explosion can be reached by a higher wind efficiency \citep[although see][]{beasor+20}, unstable and episodic mass-loss events, stellar rotation, and/or mass transfer in binary systems.
Such mass transfer may be the main mechanism for stripping larger amounts of envelope masses since most massive stars are found in interacting binary systems \citep{sana+12}. 
Higher \mzams\ also imply larger progenitor radii, which modifies the pre-SN structure.
Therefore, a larger mixing-length parameter may be more appropriate to use, as suggested by \citet{dessart+13} and \citet{gonzalez+15}.
To test this conclusion, we carried out a smaller number of additional simulations of high-\mzams\ progenitors evolved with a larger mixing-length parameter that experienced an increased mass loss. The analysis of these models is presented in Appendix~\ref{app:nonstd}, and suggests that indeed such models may also provide reasonable fits to some of our SNe~II.
A detailed study in this direction is clearly warranted in the future.

\section{Summary and conclusions}
\label{sec:conclusions}

In this, the second paper of a series, we have derived the physical properties (\mzams, \mej, \ra, \e, \mni, and \mix) from the hydrodynamical modelling of bolometric LCs and expansion velocities for a sample of 53~\sneii.
This is the largest sample of \sneii\ analysed so far that comes from the same follow-up programme where the SNe were all observed and processed in the same manner.
We used a grid of explosion models applied to stellar evolution progenitors and compared them to SN observations using a quantified fitting technique based on MCMC methods. The pre-SN models were constructed assuming non-rotating solar-metallicity single-star evolution with standard values for various evolutionary parameters (efficiency of mass loss through winds, mixing length, overshooting, and semiconvection). 
Therefore, our conclusions are valid for standard pre-SN models, and variations in stellar modelling may provide different results.

Using progenitors produced by standard stellar evolution, we were able to model the majority of the objects in the sample, with the exception of a few \sneii\ that will be analysed in \citetalias{paper3_submitted}.
The following ranges of physical parameters were estimated, \mej~=~7.9$-$14.8~\ms, \ra~=~450$-$1077~\rs, \e~=~0.15$-$1.40~foe, and \mni~=~0.006$-$0.069~\ms, with different degrees of \mix\ within the ejecta, from a more centrally concentrated \Ni\ to a more extended spatial distribution (with a median value of $\sim$0.5 in mass fraction). An extreme mixing of \Ni\ was only found for one SN in the gold sample: SN~2008ag. Values of
\mzams\ are found between 9.2 and 20.9~\ms. 
In summary, we find ranges of progenitor and explosion properties of \sneii\ that qualitatively agree with previous observational and theoretical estimates.
Five \sneii\ in the CSP-I sample with well-sampled bolometric LCs (and sometimes well-sampled velocity curves) could not be fully reproduced with our grid of hydrodynamical simulations applied to standard single-star models, implying that non-standard stellar evolution and/or binary evolution models are required\footnote{Two of these objects (SN~2006Y and SN~2008bu) are modelled in \citetalias{paper3_submitted} assuming higher mass loss.}.

We analysed the distributions for all the physical parameters and found that most \sneii\ in our sample are consistent with ejecta masses in the range of 8$-$10~\ms\ and RSG radii between 450 and 600~\rs, which are the result of the evolution of low-\mzams\ progenitors between 9 and 12~\ms.
The study of the \mzams\ distribution for both the gold and full samples, assuming a Salpeter massive-star IMF or a free power-law slope, suggests a lack of \snii\ progenitors more massive than 21.5~\ms, consistent with the RSG problem at different levels of significance.
This is the first time that the RSG problem is recovered from LC and photospheric velocity modelling.
The gold sample gives an upper mass boundary of \mmax~=~21.3$^{+3.8}_{-0.4}$~\ms\ and a power-law slope of $\Gamma$~=~$-$6.35$^{+0.52}_{-0.57}$. This power law is not consistent with previous IMF estimations for massive stars.
The steeper IMF found for the \snii\ progenitors is caused by the large relative number of low- to high-ejecta mass that biases the \mzams\ distribution to lower masses.
We discussed several possible explanations for this discrepancy, suggesting that the bias in our results towards low \mzams\ may be caused by the pre-SN structures utilised. Our analysis indicates that similar pre-SN structures, for higher \mzams, may be obtained by considering a higher rate of mass loss during massive-star evolution, probably as a result of binary interaction prior to the RSG phase, and larger mixing-length parameters.
However, a systematic analysis of the different factors needs to be carried out to properly identify the reason for the IMF incompatibility.

The progenitor and explosion properties inferred in this work are used in \citetalias{paper3_submitted} to search for correlations between physical parameters and observed bolometric and spectroscopic properties for the same sample of \sneii. This provides an extensive analysis of \snii\ diversity dependence on the physics of the progenitors and their explosions.

\begin{acknowledgements}
We thank the referee for the useful comments that improved the manuscript.
The work of the Carnegie Supernova Project was supported by the National Science Foundation under grants AST-0306969, AST-0607438, AST-1008343, AST-1613426, AST-1613472, and AST-1613455.
L.M. acknowledges support from a CONICET fellowship.
L.M. and M.O. acknowledge support from UNRN~PI2018~40B885 grant.
M.H. acknowledges support from the Hagler Institute of Advanced Study at Texas A\&M University.
S.G.G. acknowledges support by FCT under Project CRISP PTDC/FIS-AST-31546/2017 and~Project~No.~UIDB/00099/2020.
M.S. is supported by grants from the VILLUM FONDEN (grant number 28021) and the Independent Research Fund Denmark (IRFD; 8021-00170B).
F.F. acknowledges support from the National Agency for Research and Development (ANID) grants: BASAL Center of Mathematical Modelling AFB-170001, Ministry of Economy, Development, and Tourism’s Millennium Science Initiative through grant IC12009, awarded to the Millennium Institute of Astrophysics, and FONDECYT Regular \#1200710.
L.G. acknowledges financial support from the Spanish Ministry of Science, Innovation and Universities (MICIU) under the 2019 Ram\'on y Cajal program RYC2019-027683 and from the Spanish MICIU project PID2020-115253GA-I00.
P.H. acknowledges the support by National Science Foundation (NSF) grant AST-1715133.
\\
\textit{Software:} \texttt{corner.py} \citep{corner}, \texttt{emcee} \citep{emcee}, \texttt{NumPy} \citep{numpyguide2006,numpy2011}, \texttt{matplotlib} \citep{matplotlib}, \texttt{MESA} \citep{paxton+11,paxton+13,paxton+15,paxton+18,paxton+19}, \texttt{SciPy} \citep{scipy2020}, \texttt{Pandas} \citep{pandas}, \texttt{ipython/jupyter} \citep{jupyter}.
\end{acknowledgements}

\bibliographystyle{aa}
\bibliography{biblio}

\newcommand{\noop}[1]{}
\begin{thebibliography}{172}
\expandafter\ifx\csname natexlab\endcsname\relax\def\natexlab#1{#1}\fi

\bibitem[{{Adams} {et~al.}(2017){Adams}, {Kochanek}, {Gerke}, {Stanek}, \&
  {Dai}}]{adams+17a}
{Adams}, S.~M., {Kochanek}, C.~S., {Gerke}, J.~R., {Stanek}, K.~Z., \& {Dai},
  X. 2017, \mnras, 468, 4968

\bibitem[{{Allan} {et~al.}(2020){Allan}, {Groh}, {Mehner}, {Smith}, {Boian},
  {Farrell}, \& {Andrews}}]{allan+20}
{Allan}, A.~P., {Groh}, J.~H., {Mehner}, A., {et~al.} 2020, \mnras, 496, 1902

\bibitem[{{Anderson}(2019)}]{anderson19}
{Anderson}, J.~P. 2019, \aap, 628, A7

\bibitem[{{Anderson} {et~al.}(2014{\natexlab{a}}){Anderson}, {Dessart},
  {Gutierrez}, {Hamuy}, {Morrell}, {Phillips}, {Folatelli}, {Stritzinger},
  {Freedman}, {Gonz{\'a}lez-Gait{\'a}n}, {McCarthy}, {Suntzeff}, \&
  {Thomas-Osip}}]{anderson+14_ha}
{Anderson}, J.~P., {Dessart}, L., {Gutierrez}, C.~P., {et~al.}
  2014{\natexlab{a}}, \mnras, 441, 671

\bibitem[{{Anderson} {et~al.}(2018){Anderson}, {Dessart}, {Guti{\'e}rrez},
  {Kr{\"u}hler}, {Galbany}, {Jerkstrand}, {Smartt}, {Contreras}, {Morrell},
  {Phillips}, {Stritzinger}, {Hsiao}, {Gonz{\'a}lez-Gait{\'a}n}, {Agliozzo},
  {Castell{\'o}n}, {Chambers}, {Chen}, {Flewelling}, {Gonzalez},
  {Hosseinzadeh}, {Huber}, {Fraser}, {Inserra}, {Kankare}, {Mattila},
  {Magnier}, {Maguire}, {Lowe}, {Sollerman}, {Sullivan}, {Young}, \&
  {Valenti}}]{anderson+18}
{Anderson}, J.~P., {Dessart}, L., {Guti{\'e}rrez}, C.~P., {et~al.} 2018, Nature
  Astronomy, 2, 574

\bibitem[{{Anderson} {et~al.}(2014{\natexlab{b}}){Anderson},
  {Gonz{\'a}lez-Gait{\'a}n}, {Hamuy}, {Guti{\'e}rrez}, {Stritzinger}, {Olivares
  E.}, {Phillips}, {Schulze}, {Antezana}, {Bolt}, {Campillay}, {Castell{\'o}n},
  {Contreras}, {de Jaeger}, {Folatelli}, {F{\"o}rster}, {Freedman},
  {Gonz{\'a}lez}, {Hsiao}, {Krzemi{\'n}ski}, {Krisciunas}, {Maza}, {McCarthy},
  {Morrell}, {Persson}, {Roth}, {Salgado}, {Suntzeff}, \&
  {Thomas-Osip}}]{anderson+14_lc}
{Anderson}, J.~P., {Gonz{\'a}lez-Gait{\'a}n}, S., {Hamuy}, M., {et~al.}
  2014{\natexlab{b}}, \apj, 786, 67

\bibitem[{{Anderson} {et~al.}(2016){Anderson}, {Guti{\'e}rrez}, {Dessart},
  {Hamuy}, {Galbany}, {Morrell}, {Stritzinger}, {Phillips}, {Folatelli},
  {Boffin}, {de Jaeger}, {Kuncarayakti}, \& {Prieto}}]{anderson+16}
{Anderson}, J.~P., {Guti{\'e}rrez}, C.~P., {Dessart}, L., {et~al.} 2016, \aap,
  589, A110

\bibitem[{{Arcavi} {et~al.}(2012){Arcavi}, {Gal-Yam}, {Cenko}, {Fox},
  {Leonard}, {Moon}, {Sand }, {Soderberg}, {Kiewe}, {Yaron}, {Becker},
  {Scheps}, {Birenbaum}, {Chamudot}, \& {Zhou}}]{arcavi+12}
{Arcavi}, I., {Gal-Yam}, A., {Cenko}, S.~B., {et~al.} 2012, \apjl, 756, L30

\bibitem[{{Arnett}(1996)}]{arnett96book}
{Arnett}, D. 1996, {Supernovae and Nucleosynthesis: An Investigation of the
  History of Matter from the Big Bang to the Present}

\bibitem[{{Arnett}(1980)}]{arnett80}
{Arnett}, W.~D. 1980, \apj, 237, 541

\bibitem[{{Barbon} {et~al.}(1979){Barbon}, {Ciatti}, \& {Rosino}}]{barbon+79}
{Barbon}, R., {Ciatti}, F., \& {Rosino}, L. 1979, \aap, 72, 287

\bibitem[{{Barker} {et~al.}(2021){Barker}, {Harris}, {Warren}, {O'Connor}, \&
  {Couch}}]{barker+21}
{Barker}, B.~L., {Harris}, C.~E., {Warren}, M.~L., {O'Connor}, E.~P., \&
  {Couch}, S.~M. 2021, arXiv e-prints, arXiv:2102.01118

\bibitem[{{Bastian} {et~al.}(2010){Bastian}, {Covey}, \& {Meyer}}]{bastian+10}
{Bastian}, N., {Covey}, K.~R., \& {Meyer}, M.~R. 2010, \araa, 48, 339

\bibitem[{{Beasor} {et~al.}(2021){Beasor}, {Davies}, \& {Smith}}]{beasor+21}
{Beasor}, E.~R., {Davies}, B., \& {Smith}, N. 2021, \apj, 922, 55

\bibitem[{{Beasor} {et~al.}(2020){Beasor}, {Davies}, {Smith}, {van Loon},
  {Gehrz}, \& {Figer}}]{beasor+20}
{Beasor}, E.~R., {Davies}, B., {Smith}, N., {et~al.} 2020, \mnras, 492, 5994

\bibitem[{{Benvenuto} {et~al.}(2013){Benvenuto}, {Bersten}, \&
  {Nomoto}}]{benvenuto+13}
{Benvenuto}, O.~G., {Bersten}, M.~C., \& {Nomoto}, K. 2013, \apj, 762, 74

\bibitem[{{Bersten}(2013)}]{bersten13phd}
{Bersten}, M.~C. 2013, arXiv e-prints, arXiv:1303.0639

\bibitem[{{Bersten} {et~al.}(2011){Bersten}, {Benvenuto}, \&
  {Hamuy}}]{bersten+11}
{Bersten}, M.~C., {Benvenuto}, O., \& {Hamuy}, M. 2011, \apj, 729, 61

\bibitem[{{Bersten} \& {Hamuy}(2009)}]{bersten+09}
{Bersten}, M.~C. \& {Hamuy}, M. 2009, \apj, 701, 200

\bibitem[{{Blinnikov} {et~al.}(1998){Blinnikov}, {Eastman}, {Bartunov},
  {Popolitov}, \& {Woosley}}]{blinnikov+98}
{Blinnikov}, S.~I., {Eastman}, R., {Bartunov}, O.~S., {Popolitov}, V.~A., \&
  {Woosley}, S.~E. 1998, \apj, 496, 454

\bibitem[{{Bruch} {et~al.}(2021){Bruch}, {Gal-Yam}, {Schulze}, {Yaron}, {Yang},
  {Soumagnac}, {Rigault}, {Strotjohann}, {Ofek}, {Sollerman}, {Masci},
  {Barbarino}, {Ho}, {Fremling}, {Perley}, {Nordin}, {Cenko}, {Adams},
  {Adreoni}, {Bellm}, {Blagorodnova}, {Bulla}, {Burdge}, {De}, {Dhawan},
  {Drake}, {Duev}, {Dugas}, {Graham}, {Graham}, {Irani}, {Jencson},
  {Karamehmetoglu}, {Kasliwal}, {Kim}, {Kulkarni}, {Kupfer}, {Liang},
  {Mahabal}, {Miller}, {Prince}, {Riddle}, {Sharma}, {Smith}, {Taddia},
  {Taggart}, {Walters}, \& {Yan}}]{bruch+21}
{Bruch}, R.~J., {Gal-Yam}, A., {Schulze}, S., {et~al.} 2021, \apj, 912, 46

\bibitem[{{Burrows} {et~al.}(2020){Burrows}, {Radice}, {Vartanyan}, {Nagakura},
  {Skinner}, \& {Dolence}}]{burrows+20}
{Burrows}, A., {Radice}, D., {Vartanyan}, D., {et~al.} 2020, \mnras, 491, 2715

\bibitem[{{Burrows} \& {Vartanyan}(2021)}]{burrows+21}
{Burrows}, A. \& {Vartanyan}, D. 2021, \nat, 589, 29

\bibitem[{{Chandrasekhar}(1939)}]{chandrasekhar1939}
{Chandrasekhar}, S. 1939, {An introduction to the study of stellar structure}

\bibitem[{{Chevalier}(1976)}]{chevalier76}
{Chevalier}, R.~A. 1976, \apj, 207, 872

\bibitem[{{Clocchiatti} {et~al.}(1996){Clocchiatti}, {Benetti}, {Wheeler},
  {Wren}, {Boisseau}, {Cappellaro}, {Turatto}, {Patat}, {Swartz}, {Harkness},
  {Brotherton}, {Wills}, {Hemenway}, {Cornell}, {Frueh}, \&
  {Kaiser}}]{clocchiatti+96}
{Clocchiatti}, A., {Benetti}, S., {Wheeler}, J.~C., {et~al.} 1996, \aj, 111,
  1286

\bibitem[{{Contreras} {et~al.}(2010){Contreras}, {Hamuy}, {Phillips},
  {Folatelli}, {Suntzeff}, {Persson}, {Stritzinger}, {Boldt}, {Gonz{\'a}lez},
  {Krzeminski}, {Morrell}, {Roth}, {Salgado}, {Maureira}, {Burns}, {Freedman},
  {Madore}, {Murphy}, {Wyatt}, {Li}, \& {Filippenko}}]{contreras+10}
{Contreras}, C., {Hamuy}, M., {Phillips}, M.~M., {et~al.} 2010, \aj, 139, 519

\bibitem[{{Curtis} {et~al.}(2021){Curtis}, {Wolfe}, {Fr{\"o}hlich}, {Miller},
  {Wollaeger}, \& {Ebinger}}]{curtis+21}
{Curtis}, S., {Wolfe}, N., {Fr{\"o}hlich}, C., {et~al.} 2021, \apj, 921, 143

\bibitem[{{Davies} \& {Beasor}(2018)}]{davies+18}
{Davies}, B. \& {Beasor}, E.~R. 2018, \mnras, 474, 2116

\bibitem[{{Davies} \& {Beasor}(2020)}]{davies+20}
{Davies}, B. \& {Beasor}, E.~R. 2020, \mnras, 493, 468

\bibitem[{{Davis} {et~al.}(2019){Davis}, {Hsiao}, {Ashall}, {Hoeflich},
  {Phillips}, {Marion}, {Kirshner}, {Morrell}, {Sand}, {Burns}, {Contreras},
  {Stritzinger}, {Anderson}, {Baron}, {Diamond}, {Guti{\'e}rrez}, {Hamuy},
  {Holmbo}, {Kasliwal}, {Krisciunas}, {Kumar}, {Lu}, {Pessi}, {Piro}, {Prieto},
  {Shahbandeh}, \& {Suntzeff}}]{davis+19}
{Davis}, S., {Hsiao}, E.~Y., {Ashall}, C., {et~al.} 2019, \apj, 887, 4

\bibitem[{{de Jaeger} {et~al.}(2018){de Jaeger}, {Anderson}, {Galbany},
  {Gonz{\'a}lez-Gait{\'a}n}, {Hamuy}, {Phillips}, {Stritzinger}, {Contreras},
  {Folatelli}, {Guti{\'e}rrez}, {Hsiao}, {Morrell}, {Suntzeff}, {Dessart}, \&
  {Filippenko}}]{dejaeger+18}
{de Jaeger}, T., {Anderson}, J.~P., {Galbany}, L., {et~al.} 2018, \mnras, 476,
  4592

\bibitem[{{de Jaeger} {et~al.}(2019){de Jaeger}, {Zheng}, {Stahl},
  {Filippenko}, {Brink}, {Bigley}, {Blanchard}, {Blanchard}, {Bradley},
  {Cargill}, {Casper}, {Cenko}, {Channa}, {Choi}, {Clubb}, {Cobb}, {Cohen}, {de
  Kouchkovsky}, {Ellison}, {Falcon}, {Fox}, {Fuller}, {Ganeshalingam}, {Gould},
  {Graham}, {Halevi}, {Hayakawa}, {Hestenes}, {Hyland}, {Jeffers}, {Joubert},
  {Kandrashoff}, {Kelly}, {Kim}, {Kim}, {Kumar}, {Leonard}, {Li}, {Lowe}, {Lu},
  {Mason}, {McAllister}, {Mauerhan}, {Modjaz}, {Molloy}, {Perley}, {Pina},
  {Poznanski}, {Ross}, {Shivvers}, {Silverman}, {Soler}, {Stegman}, {Taylor},
  {Tang}, {Wilkins}, {Wang}, {Wang}, {Yuk}, {Yunus}, \& {Zhang}}]{dejaeger+19}
{de Jaeger}, T., {Zheng}, W., {Stahl}, B.~E., {et~al.} 2019, \mnras, 490, 2799

\bibitem[{{DeLaney} {et~al.}(2010){DeLaney}, {Rudnick}, {Stage}, {Smith},
  {Isensee}, {Rho}, {Allen}, {Gomez}, {Kozasa}, {Reach}, {Davis}, \&
  {Houck}}]{delaney+10}
{DeLaney}, T., {Rudnick}, L., {Stage}, M.~D., {et~al.} 2010, \apj, 725, 2038

\bibitem[{{Dessart} \& {Hillier}(2005)}]{dessart+05}
{Dessart}, L. \& {Hillier}, D.~J. 2005, \aap, 439, 671

\bibitem[{{Dessart} \& {Hillier}(2019)}]{dessart+19}
{Dessart}, L. \& {Hillier}, D.~J. 2019, \aap, 625, A9

\bibitem[{{Dessart} \& {Hillier}(2020)}]{dessart+20}
{Dessart}, L. \& {Hillier}, D.~J. 2020, \aap, 642, A33

\bibitem[{{Dessart} {et~al.}(2013){Dessart}, {Hillier}, {Waldman}, \&
  {Livne}}]{dessart+13}
{Dessart}, L., {Hillier}, D.~J., {Waldman}, R., \& {Livne}, E. 2013, \mnras,
  433, 1745

\bibitem[{{Dufton} {et~al.}(2013){Dufton}, {Langer}, {Dunstall}, {Evans},
  {Brott}, {de Mink}, {Howarth}, {Kennedy}, {McEvoy}, {Potter},
  {Ram{\'\i}rez-Agudelo}, {Sana}, {Sim{\'o}n-D{\'\i}az}, {Taylor}, \&
  {Vink}}]{dufton+13}
{Dufton}, P.~L., {Langer}, N., {Dunstall}, P.~R., {et~al.} 2013, \aap, 550,
  A109

\bibitem[{{Ekstr{\"o}m} {et~al.}(2012){Ekstr{\"o}m}, {Georgy}, {Eggenberger},
  {Meynet}, {Mowlavi}, {Wyttenbach}, {Granada}, {Decressin}, {Hirschi},
  {Frischknecht}, {Charbonnel}, \& {Maeder}}]{ekstrom+12}
{Ekstr{\"o}m}, S., {Georgy}, C., {Eggenberger}, P., {et~al.} 2012, \aap, 537,
  A146

\bibitem[{{Eldridge} {et~al.}(2019){Eldridge}, {Guo}, {Rodrigues}, {Stanway},
  \& {Xiao}}]{eldridge+19b}
{Eldridge}, J.~J., {Guo}, N.~Y., {Rodrigues}, N., {Stanway}, E.~R., \& {Xiao},
  L. 2019, \pasa, 36, e041

\bibitem[{{Eldridge} {et~al.}(2018){Eldridge}, {Xiao}, {Stanway}, {Rodrigues},
  \& {Guo}}]{eldridge+18}
{Eldridge}, J.~J., {Xiao}, L., {Stanway}, E.~R., {Rodrigues}, N., \& {Guo},
  N.~Y. 2018, \pasa, 35, 49

\bibitem[{{Englert Urrutia} {et~al.}(2020){Englert Urrutia}, {Bersten}, \&
  {Cidale}}]{englert+20}
{Englert Urrutia}, B.~N., {Bersten}, M.~C., \& {Cidale}, L.~S. 2020, Boletin de
  la Asociacion Argentina de Astronomia La Plata Argentina, 61B, 51

\bibitem[{{Ertl} {et~al.}(2016){Ertl}, {Janka}, {Woosley}, {Sukhbold}, \&
  {Ugliano}}]{ertl+16}
{Ertl}, T., {Janka}, H.~T., {Woosley}, S.~E., {Sukhbold}, T., \& {Ugliano}, M.
  2016, \apj, 818, 124

\bibitem[{{Faran} {et~al.}(2014{\natexlab{a}}){Faran}, {Poznanski},
  {Filippenko}, {Chornock}, {Foley}, {Ganeshalingam}, {Leonard}, {Li},
  {Modjaz}, {Nakar}, {Serduke}, \& {Silverman}}]{faran+14a}
{Faran}, T., {Poznanski}, D., {Filippenko}, A.~V., {et~al.} 2014{\natexlab{a}},
  \mnras, 442, 844

\bibitem[{{Faran} {et~al.}(2014{\natexlab{b}}){Faran}, {Poznanski},
  {Filippenko}, {Chornock}, {Foley}, {Ganeshalingam}, {Leonard}, {Li},
  {Modjaz}, {Serduke}, \& {Silverman}}]{faran+14b}
{Faran}, T., {Poznanski}, D., {Filippenko}, A.~V., {et~al.} 2014{\natexlab{b}},
  \mnras, 445, 554

\bibitem[{{Farmer} {et~al.}(2016){Farmer}, {Fields}, {Petermann}, {Dessart},
  {Cantiello}, {Paxton}, \& {Timmes}}]{farmer+16}
{Farmer}, R., {Fields}, C.~E., {Petermann}, I., {et~al.} 2016, \apjs, 227, 22

\bibitem[{{Farrell} {et~al.}(2020){Farrell}, {Groh}, {Meynet}, \&
  {Eldridge}}]{farrell+20}
{Farrell}, E.~J., {Groh}, J.~H., {Meynet}, G., \& {Eldridge}, J.~J. 2020,
  \mnras, 494, L53

\bibitem[{{Folatelli} {et~al.}(2013){Folatelli}, {Morrell}, {Phillips},
  {Hsiao}, {Campillay}, {Contreras}, {Castell{\'o}n}, {Hamuy}, {Krzeminski},
  {Roth}, {Stritzinger}, {Burns}, {Freedman}, {Madore}, {Murphy}, {Persson},
  {Prieto}, {Suntzeff}, {Krisciunas}, {Anderson}, {F{\"o}rster}, {Maza},
  {Pignata}, {Rojas}, {Boldt}, {Salgado}, {Wyatt}, {Olivares E.}, {Gal-Yam}, \&
  {Sako}}]{folatelli+13}
{Folatelli}, G., {Morrell}, N., {Phillips}, M.~M., {et~al.} 2013, \apj, 773, 53

\bibitem[{Foreman-Mackey(2016)}]{corner}
Foreman-Mackey, D. 2016, The Journal of Open Source Software, 24

\bibitem[{{Foreman-Mackey} {et~al.}(2013){Foreman-Mackey}, {Hogg}, {Lang}, \&
  {Goodman}}]{emcee}
{Foreman-Mackey}, D., {Hogg}, D.~W., {Lang}, D., \& {Goodman}, J. 2013, \pasp,
  125, 306

\bibitem[{{F{\"o}rster} {et~al.}(2018){F{\"o}rster}, {Moriya}, {Maureira},
  {Anderson}, {Blinnikov}, {Bufano}, {Cabrera-Vives}, {Clocchiatti}, {de
  Jaeger}, {Est{\'e}vez}, {Galbany}, {Gonz{\'a}lez-Gait{\'a}n}, {Gr{\"a}fener},
  {Hamuy}, {Hsiao}, {Huentelemu}, {Huijse}, {Kuncarayakti}, {Mart{\'\i}nez},
  {Medina}, {Olivares E.}, {Pignata}, {Razza}, {Reyes}, {San Mart{\'\i}n},
  {Smith}, {Vera}, {Vivas}, {de Ugarte Postigo}, {Yoon}, {Ashall}, {Fraser},
  {Gal-Yam}, {Kankare}, {Le Guillou}, {Mazzali}, {Walton}, \&
  {Young}}]{forster+18}
{F{\"o}rster}, F., {Moriya}, T.~J., {Maureira}, J.~C., {et~al.} 2018, Nature
  Astronomy, 2, 808

\bibitem[{{Fryer}(1999)}]{fryer99}
{Fryer}, C.~L. 1999, \apj, 522, 413

\bibitem[{{Fuller}(2017)}]{fuller17}
{Fuller}, J. 2017, \mnras, 470, 1642

\bibitem[{{Galbany} {et~al.}(2016){Galbany}, {Hamuy}, {Phillips}, {Suntzeff},
  {Maza}, {de Jaeger}, {Moraga}, {Gonz{\'a}lez-Gait{\'a}n}, {Krisciunas},
  {Morrell}, {Thomas-Osip}, {Krzeminski}, {Gonz{\'a}lez}, {Antezana},
  {Wishnjewski}, {McCarthy}, {Anderson}, {Guti{\'e}rrez}, {Stritzinger},
  {Folatelli}, {Anguita}, {Galaz}, {Green}, {Impey}, {Kim}, {Kirhakos},
  {Malkan}, {Mulchaey}, {Phillips}, {Pizzella}, {Prosser}, {Schmidt},
  {Schommer}, {Sherry}, {Strolger}, {Wells}, \& {Williger}}]{galbany+16}
{Galbany}, L., {Hamuy}, M., {Phillips}, M.~M., {et~al.} 2016, \aj, 151, 33

\bibitem[{{Goldberg} {et~al.}(2019){Goldberg}, {Bildsten}, \&
  {Paxton}}]{goldberg+19}
{Goldberg}, J.~A., {Bildsten}, L., \& {Paxton}, B. 2019, \apj, 879, 3

\bibitem[{{Goldman} {et~al.}(2017){Goldman}, {van Loon}, {Zijlstra}, {Green},
  {Wood}, {Nanni}, {Imai}, {Whitelock}, {Matsuura}, {Groenewegen}, \&
  {G{\'o}mez}}]{goldman+17}
{Goldman}, S.~R., {van Loon}, J.~T., {Zijlstra}, A.~A., {et~al.} 2017, \mnras,
  465, 403

\bibitem[{{Gonz{\'a}lez-Gait{\'a}n} {et~al.}(2015){Gonz{\'a}lez-Gait{\'a}n},
  {Tominaga}, {Molina}, {Galbany}, {Bufano}, {Anderson}, {Gutierrez},
  {F{\"o}rster}, {Pignata}, {Bersten}, {Howell}, {Sullivan}, {Carlberg}, {de
  Jaeger}, {Hamuy}, {Baklanov}, \& {Blinnikov}}]{gonzalez+15}
{Gonz{\'a}lez-Gait{\'a}n}, S., {Tominaga}, N., {Molina}, J., {et~al.} 2015,
  \mnras, 451, 2212

\bibitem[{{Grassberg} {et~al.}(1971){Grassberg}, {Imshennik}, \&
  {Nadyozhin}}]{grassberg+71}
{Grassberg}, E.~K., {Imshennik}, V.~S., \& {Nadyozhin}, D.~K. 1971, \apss, 10,
  28

\bibitem[{{Guti{\'e}rrez} {et~al.}(2014){Guti{\'e}rrez}, {Anderson}, {Hamuy},
  {Gonz{\'a}lez-Gait{\'a}n}, {Folatelli}, {Morrell}, {Stritzinger}, {Phillips},
  {McCarthy}, {Suntzeff}, \& {Thomas-Osip}}]{gutierrez+14}
{Guti{\'e}rrez}, C.~P., {Anderson}, J.~P., {Hamuy}, M., {et~al.} 2014, \apjl,
  786, L15

\bibitem[{{Guti{\'e}rrez} {et~al.}(2017{\natexlab{a}}){Guti{\'e}rrez},
  {Anderson}, {Hamuy}, {Gonz{\'a}lez-Gaitan}, {Galbany}, {Dessart},
  {Stritzinger}, {Phillips}, {Morrell}, \& {Folatelli}}]{gutierrez+17II}
{Guti{\'e}rrez}, C.~P., {Anderson}, J.~P., {Hamuy}, M., {et~al.}
  2017{\natexlab{a}}, \apj, 850, 90

\bibitem[{{Guti{\'e}rrez} {et~al.}(2017{\natexlab{b}}){Guti{\'e}rrez},
  {Anderson}, {Hamuy}, {Morrell}, {Gonz{\'a}lez-Gaitan}, {Stritzinger},
  {Phillips}, {Galbany}, {Folatelli}, {Dessart}, {Contreras}, {Della Valle},
  {Freedman}, {Hsiao}, {Krisciunas}, {Madore}, {Maza}, {Suntzeff}, {Prieto},
  {Gonz{\'a}lez}, {Cappellaro}, {Navarrete}, {Pizzella}, {Ruiz}, {Smith}, \&
  {Turatto}}]{gutierrez+17I}
{Guti{\'e}rrez}, C.~P., {Anderson}, J.~P., {Hamuy}, M., {et~al.}
  2017{\natexlab{b}}, \apj, 850, 89

\bibitem[{{Guti{\'e}rrez} {et~al.}(2018){Guti{\'e}rrez}, {Anderson},
  {Sullivan}, {Dessart}, {Gonz{\'a}lez-Gaitan}, {Galbany}, {Dimitriadis},
  {Arcavi}, {Bufano}, {Chen}, {Dennefeld}, {Gromadzki}, {Haislip},
  {Hosseinzadeh}, {Howell}, {Inserra}, {Kankare}, {Leloudas}, {Maguire},
  {McCully}, {Morrell}, {Olivares E}, {Pignata}, {Reichart}, {Reynolds},
  {Smartt}, {Sollerman}, {Taddia}, {Tak{\'a}ts}, {Terreran}, {Valenti}, \&
  {Young}}]{gutierrez+18}
{Guti{\'e}rrez}, C.~P., {Anderson}, J.~P., {Sullivan}, M., {et~al.} 2018,
  \mnras, 479, 3232

\bibitem[{{Haas} {et~al.}(1990){Haas}, {Colgan}, {Erickson}, {Lord}, {Burton},
  \& {Hollenbach}}]{haas+90}
{Haas}, M.~R., {Colgan}, S. W.~J., {Erickson}, E.~F., {et~al.} 1990, \apj, 360,
  257

\bibitem[{{Hamuy}(2003)}]{hamuy03}
{Hamuy}, M. 2003, \apj, 582, 905

\bibitem[{{Hamuy} {et~al.}(2006){Hamuy}, {Folatelli}, {Morrell}, {Phillips},
  {Suntzeff}, {Persson}, {Roth}, {Gonzalez}, {Krzeminski}, {Contreras},
  {Freedman}, {Murphy}, {Madore}, {Wyatt}, {Maza}, {Filippenko}, {Li}, \&
  {Pinto}}]{hamuy+06}
{Hamuy}, M., {Folatelli}, G., {Morrell}, N.~I., {et~al.} 2006, \pasp, 118, 2

\bibitem[{{Heger} {et~al.}(2003){Heger}, {Fryer}, {Woosley}, {Langer}, \&
  {Hartmann}}]{heger+03}
{Heger}, A., {Fryer}, C.~L., {Woosley}, S.~E., {Langer}, N., \& {Hartmann},
  D.~H. 2003, \apj, 591, 288

\bibitem[{{Herant} \& {Benz}(1991)}]{herant+91}
{Herant}, M. \& {Benz}, W. 1991, \apjl, 370, L81

\bibitem[{{Hillier} \& {Dessart}(2019)}]{hillier+19}
{Hillier}, D.~J. \& {Dessart}, L. 2019, \aap, 631, A8

\bibitem[{{Hiramatsu} {et~al.}(2021){Hiramatsu}, {Howell}, {Moriya},
  {Goldberg}, {Hosseinzadeh}, {Arcavi}, {Anderson}, {Guti{\'e}rrez}, {Burke},
  {McCully}, {Valenti}, {Galbany}, {Fang}, {Maeda}, {Folatelli}, {Hsiao},
  {Morrell}, {Phillips}, {Stritzinger}, {Suntzeff}, {Gromadzki}, {Maguire},
  {M{\"u}ller-Bravo}, \& {Young}}]{hiramatsu+21}
{Hiramatsu}, D., {Howell}, D.~A., {Moriya}, T.~J., {et~al.} 2021, \apj, 913, 55

\bibitem[{{Hirschi} {et~al.}(2004){Hirschi}, {Meynet}, \&
  {Maeder}}]{hirschi+04}
{Hirschi}, R., {Meynet}, G., \& {Maeder}, A. 2004, \aap, 425, 649

\bibitem[{{Hoflich}(1991)}]{hoflich91}
{Hoflich}, P. 1991, \aap, 246, 481

\bibitem[{{Humphreys} {et~al.}(2020){Humphreys}, {Helmel}, {Jones}, \&
  {Gordon}}]{humphreys+20}
{Humphreys}, R.~M., {Helmel}, G., {Jones}, T.~J., \& {Gordon}, M.~S. 2020, \aj,
  160, 145

\bibitem[{Hunter(2007)}]{matplotlib}
Hunter, J.~D. 2007, Computing In Science \& Engineering, 9, 90

\bibitem[{{Janka} \& {M{\"u}ller}(1993)}]{janka+93}
{Janka}, H.~T. \& {M{\"u}ller}, E. 1993, in Frontiers of Neutrino Astrophysics,
  203--217

\bibitem[{{Jerkstrand} {et~al.}(2012){Jerkstrand}, {Fransson}, {Maguire},
  {Smartt}, {Ergon}, \& {Spyromilio}}]{jerkstrand+12}
{Jerkstrand}, A., {Fransson}, C., {Maguire}, K., {et~al.} 2012, \aap, 546, A28

\bibitem[{{Jerkstrand} {et~al.}(2014){Jerkstrand}, {Smartt}, {Fraser},
  {Fransson}, {Sollerman}, {Taddia}, \& {Kotak}}]{jerkstrand+14}
{Jerkstrand}, A., {Smartt}, S.~J., {Fraser}, M., {et~al.} 2014, \mnras, 439,
  3694

\bibitem[{{Kasen} {et~al.}(2006){Kasen}, {Thomas}, \& {Nugent}}]{kasen+06}
{Kasen}, D., {Thomas}, R.~C., \& {Nugent}, P. 2006, \apj, 651, 366

\bibitem[{{Kasen} \& {Woosley}(2009)}]{kasen+09}
{Kasen}, D. \& {Woosley}, S.~E. 2009, \apj, 703, 2205

\bibitem[{{Khazov} {et~al.}(2016){Khazov}, {Yaron}, {Gal-Yam}, {Manulis},
  {Rubin}, {Kulkarni}, {Arcavi}, {Kasliwal}, {Ofek}, {Cao}, {Perley},
  {Sollerman}, {Horesh}, {Sullivan}, {Filippenko}, {Nugent}, {Howell}, {Cenko},
  {Silverman}, {Ebeling}, {Taddia}, {Johansson}, {Laher}, {Surace},
  {Rebbapragada}, {Wozniak}, \& {Matheson}}]{khazov+16}
{Khazov}, D., {Yaron}, O., {Gal-Yam}, A., {et~al.} 2016, \apj, 818, 3

\bibitem[{{Kifonidis} {et~al.}(2003){Kifonidis}, {Plewa}, {Janka}, \&
  {M{\"u}ller}}]{kifonidis+03}
{Kifonidis}, K., {Plewa}, T., {Janka}, H.~T., \& {M{\"u}ller}, E. 2003, \aap,
  408, 621

\bibitem[{{Koplitz} {et~al.}(2021){Koplitz}, {Johnson}, {Williams}, {Long},
  {Blair}, {Murphy}, {Dolphin}, \& {Hillis}}]{koplitz+21}
{Koplitz}, B., {Johnson}, J., {Williams}, B.~F., {et~al.} 2021, \apj, 916, 58

\bibitem[{{Kozyreva} {et~al.}(2019){Kozyreva}, {Nakar}, \&
  {Waldman}}]{kozyreva+19}
{Kozyreva}, A., {Nakar}, E., \& {Waldman}, R. 2019, \mnras, 483, 1211

\bibitem[{{Krisciunas} {et~al.}(2017){Krisciunas}, {Contreras}, {Burns},
  {Phillips}, {Stritzinger}, {Morrell}, {Hamuy}, {Anais}, {Boldt}, {Busta},
  {Campillay}, {Castell{\'o}n}, {Folatelli}, {Freedman}, {Gonz{\'a}lez},
  {Hsiao}, {Krzeminski}, {Persson}, {Roth}, {Salgado}, {Ser{\'o}n}, {Suntzeff},
  {Torres}, {Filippenko}, {Li}, {Madore}, {DePoy}, {Marshall}, {Rheault}, \&
  {Villanueva}}]{krisciunas+17}
{Krisciunas}, K., {Contreras}, C., {Burns}, C.~R., {et~al.} 2017, \aj, 154, 211

\bibitem[{{Kumagai} {et~al.}(1989){Kumagai}, {Shigeyama}, {Nomoto}, {Itoh},
  {Nishimura}, \& {Tsuruta}}]{kumagai+89}
{Kumagai}, S., {Shigeyama}, T., {Nomoto}, K., {et~al.} 1989, \apj, 345, 412

\bibitem[{{Leonard} {et~al.}(2002){Leonard}, {Filippenko}, {Gates}, {Li},
  {Eastman}, {Barth}, {Bus}, {Chornock}, {Coil}, {Frink}, {Grady}, {Harris},
  {Malkan}, {Matheson}, {Quirrenbach}, \& {Treffers}}]{leonard+02}
{Leonard}, D.~C., {Filippenko}, A.~V., {Gates}, E.~L., {et~al.} 2002, \pasp,
  114, 35

\bibitem[{{Levesque} {et~al.}(2005){Levesque}, {Massey}, {Olsen}, {Plez},
  {Josselin}, {Maeder}, \& {Meynet}}]{levesque+05}
{Levesque}, E.~M., {Massey}, P., {Olsen}, K.~A.~G., {et~al.} 2005, \apj, 628,
  973

\bibitem[{{Levesque} {et~al.}(2006){Levesque}, {Massey}, {Olsen}, {Plez},
  {Meynet}, \& {Maeder}}]{levesque+06}
{Levesque}, E.~M., {Massey}, P., {Olsen}, K.~A.~G., {et~al.} 2006, \apj, 645,
  1102

\bibitem[{{Li} {et~al.}(2011){Li}, {Leaman}, {Chornock}, {Filippenko},
  {Poznanski}, {Ganeshalingam}, {Wang}, {Modjaz}, {Jha}, {Foley}, \&
  {Smith}}]{li+11a}
{Li}, W., {Leaman}, J., {Chornock}, R., {et~al.} 2011, \mnras, 412, 1441

\bibitem[{{Litvinova} \& {Nadezhin}(1983)}]{litvinova+83}
{Litvinova}, I.~I. \& {Nadezhin}, D.~K. 1983, \apss, 89, 89

\bibitem[{{Litvinova} \& {Nadezhin}(1985)}]{litvinova+85}
{Litvinova}, I.~Y. \& {Nadezhin}, D.~K. 1985, Soviet Astronomy Letters, 11, 145

\bibitem[{{Lovegrove} \& {Woosley}(2013)}]{lovegrove+13}
{Lovegrove}, E. \& {Woosley}, S.~E. 2013, \apj, 769, 109

\bibitem[{{Martinez} {et~al.}(2021{\natexlab{a}}){Martinez}, {Anderson}, \&
  {Bersten}}]{paper3_submitted}
{Martinez}, L., {Anderson}, J.~P., \& {Bersten}, M.~C. 2021{\natexlab{a}},
  \aap, \noop{3001}submitted

\bibitem[{{Martinez} \& {Bersten}(2019)}]{martinez+19}
{Martinez}, L. \& {Bersten}, M.~C. 2019, \aap, 629, A124

\bibitem[{{Martinez} {et~al.}(2020){Martinez}, {Bersten}, {Anderson},
  {Gonz{\'a}lez-Gait{\'a}n}, {F{\"o}rster}, \& {Folatelli}}]{martinez+20}
{Martinez}, L., {Bersten}, M.~C., {Anderson}, J.~P., {et~al.} 2020, \aap, 642,
  A143

\bibitem[{{Martinez} {et~al.}(2021{\natexlab{b}}){Martinez}, {Bersten},
  {Anderson}, {Hamuy}, {Gonz{\'a}lez-Gait{\'a}n}, {Stritzinger}, {Phillips},
  {Guti{\'e}rrez}, {Burns}, {Contreras}, {de Jaeger}, {Ertini}, {Folatelli},
  {F{\"o}rster}, {Galbany}, {Hoeflich}, {Hsiao}, {Morrell}, {Orellana},
  {Pessi}, \& {Suntzeff}}]{martinez+21}
{Martinez}, L., {Bersten}, M.~C., {Anderson}, J.~P., {et~al.}
  2021{\natexlab{b}}, arXiv e-prints, arXiv:2111.06519

\bibitem[{{Maund}(2017)}]{maund17}
{Maund}, J.~R. 2017, \mnras, 469, 2202

\bibitem[{{M}c{K}inney(2010)}]{pandas}
{M}c{K}inney, W. 2010, in {P}roceedings of the 9th {P}ython in {S}cience
  {C}onference, ed. {S}t\'efan van~der {W}alt \& {J}arrod {M}illman, 56 -- 61

\bibitem[{{Minkowski}(1941)}]{minkowski41}
{Minkowski}, R. 1941, \pasp, 53, 224

\bibitem[{{Moe} \& {Di Stefano}(2017)}]{moe+17}
{Moe}, M. \& {Di Stefano}, R. 2017, \apjs, 230, 15

\bibitem[{{Moriya} {et~al.}(2018){Moriya}, {F{\"o}rster}, {Yoon},
  {Gr{\"a}fener}, \& {Blinnikov}}]{moriya+18}
{Moriya}, T.~J., {F{\"o}rster}, F., {Yoon}, S.-C., {Gr{\"a}fener}, G., \&
  {Blinnikov}, S.~I. 2018, \mnras, 476, 2840

\bibitem[{{Morozova} {et~al.}(2015){Morozova}, {Piro}, {Renzo}, {Ott},
  {Clausen}, {Couch}, {Ellis}, \& {Roberts}}]{morozova+15}
{Morozova}, V., {Piro}, A.~L., {Renzo}, M., {et~al.} 2015, \apj, 814, 63

\bibitem[{{Morozova} {et~al.}(2018){Morozova}, {Piro}, \&
  {Valenti}}]{morozova+18}
{Morozova}, V., {Piro}, A.~L., \& {Valenti}, S. 2018, \apj, 858, 15

\bibitem[{{Mueller} {et~al.}(1991){Mueller}, {Fryxell}, \&
  {Arnett}}]{muller+91}
{Mueller}, E., {Fryxell}, B., \& {Arnett}, D. 1991, \aap, 251, 505

\bibitem[{{M{\"u}ller} {et~al.}(2017){M{\"u}ller}, {Prieto}, {Pejcha}, \&
  {Clocchiatti}}]{muller+17}
{M{\"u}ller}, T., {Prieto}, J.~L., {Pejcha}, O., \& {Clocchiatti}, A. 2017,
  \apj, 841, 127

\bibitem[{{Nadyozhin}(2003)}]{nadyozhin03}
{Nadyozhin}, D.~K. 2003, \mnras, 346, 97

\bibitem[{{Neugent} {et~al.}(2020){Neugent}, {Massey}, {Georgy}, {Drout},
  {Mommert}, {Levesque}, {Meynet}, \& {Ekstr{\"o}m}}]{neugent+20}
{Neugent}, K.~F., {Massey}, P., {Georgy}, C., {et~al.} 2020, \apj, 889, 44

\bibitem[{{Neustadt} {et~al.}(2021){Neustadt}, {Kochanek}, {Stanek},
  {Basinger}, {Jayasinghe}, {Garling}, {Adams}, \& {Gerke}}]{neustadt+21}
{Neustadt}, J.~M., {Kochanek}, C.~S., {Stanek}, K.~Z., {et~al.} 2021, in
  American Astronomical Society Meeting Abstracts, Vol.~53, American
  Astronomical Society Meeting Abstracts, 409.03

\bibitem[{{O'Connor} \& {Ott}(2011)}]{oconnor+11}
{O'Connor}, E. \& {Ott}, C.~D. 2011, \apj, 730, 70

\bibitem[{Oliphant(2006)}]{numpyguide2006}
Oliphant, T.~E. 2006, A guide to NumPy, Vol.~1 (Trelgol Publishing USA)

\bibitem[{{Pastorello} {et~al.}(2004){Pastorello}, {Zampieri}, {Turatto},
  {Cappellaro}, {Meikle}, {Benetti}, {Branch}, {Baron}, {Patat}, {Armstrong},
  {Altavilla}, {Salvo}, \& {Riello}}]{pastorello+04}
{Pastorello}, A., {Zampieri}, L., {Turatto}, M., {et~al.} 2004, \mnras, 347, 74

\bibitem[{{Patat} {et~al.}(1994){Patat}, {Barbon}, {Cappellaro}, \&
  {Turatto}}]{patat+94}
{Patat}, F., {Barbon}, R., {Cappellaro}, E., \& {Turatto}, M. 1994, \aap, 282,
  731

\bibitem[{{Paxton} {et~al.}(2011){Paxton}, {Bildsten}, {Dotter}, {Herwig},
  {Lesaffre}, \& {Timmes}}]{paxton+11}
{Paxton}, B., {Bildsten}, L., {Dotter}, A., {et~al.} 2011, \apjs, 192, 3

\bibitem[{{Paxton} {et~al.}(2013){Paxton}, {Cantiello}, {Arras}, {Bildsten},
  {Brown}, {Dotter}, {Mankovich}, {Montgomery}, {Stello}, {Timmes}, \&
  {Townsend}}]{paxton+13}
{Paxton}, B., {Cantiello}, M., {Arras}, P., {et~al.} 2013, \apjs, 208, 4

\bibitem[{{Paxton} {et~al.}(2015){Paxton}, {Marchant}, {Schwab}, {Bauer},
  {Bildsten}, {Cantiello}, {Dessart}, {Farmer}, {Hu}, {Langer}, {Townsend},
  {Townsley}, \& {Timmes}}]{paxton+15}
{Paxton}, B., {Marchant}, P., {Schwab}, J., {et~al.} 2015, \apjs, 220, 15

\bibitem[{{Paxton} {et~al.}(2018){Paxton}, {Schwab}, {Bauer}, {Bildsten},
  {Blinnikov}, {Duffell}, {Farmer}, {Goldberg}, {Marchant}, {Sorokina},
  {Thoul}, {Townsend}, \& {Timmes}}]{paxton+18}
{Paxton}, B., {Schwab}, J., {Bauer}, E.~B., {et~al.} 2018, \apjs, 234, 34

\bibitem[{{Paxton} {et~al.}(2019){Paxton}, {Smolec}, {Schwab}, {Gautschy},
  {Bildsten}, {Cantiello}, {Dotter}, {Farmer}, {Goldberg}, {Jermyn}, {Kanbur},
  {Marchant}, {Thoul}, {Townsend}, {Wolf}, {Zhang}, \& {Timmes}}]{paxton+19}
{Paxton}, B., {Smolec}, R., {Schwab}, J., {et~al.} 2019, \apjs, 243, 10

\bibitem[{{Pejcha} \& {Prieto}(2015)}]{pejcha+15b}
{Pejcha}, O. \& {Prieto}, J.~L. 2015, \apj, 799, 215

\bibitem[{{Perez} \& {Granger}(2007)}]{jupyter}
{Perez}, F. \& {Granger}, B.~E. 2007, Computing in Science and Engineering, 9,
  21

\bibitem[{{Phillips} {et~al.}(2013){Phillips}, {Simon}, {Morrell}, {Burns},
  {Cox}, {Foley}, {Karakas}, {Patat}, {Sternberg}, {Williams}, {Gal-Yam},
  {Hsiao}, {Leonard}, {Persson}, {Stritzinger}, {Thompson}, {Campillay},
  {Contreras}, {Folatelli}, {Freedman}, {Hamuy}, {Roth}, {Shields}, {Suntzeff},
  {Chomiuk}, {Ivans}, {Madore}, {Penprase}, {Perley}, {Pignata}, {Preston}, \&
  {Soderberg}}]{phillips+13}
{Phillips}, M.~M., {Simon}, J.~D., {Morrell}, N., {et~al.} 2013, \apj, 779, 38

\bibitem[{{Podsiadlowski} {et~al.}(1992){Podsiadlowski}, {Joss}, \&
  {Hsu}}]{podsiadlowski+92}
{Podsiadlowski}, P., {Joss}, P.~C., \& {Hsu}, J.~J.~L. 1992, \apj, 391, 246

\bibitem[{{Popov}(1993)}]{popov93}
{Popov}, D.~V. 1993, \apj, 414, 712

\bibitem[{{Poznanski} {et~al.}(2011){Poznanski}, {Ganeshalingam}, {Silverman},
  \& {Filippenko}}]{poznanski+11}
{Poznanski}, D., {Ganeshalingam}, M., {Silverman}, J.~M., \& {Filippenko},
  A.~V. 2011, \mnras, 415, L81

\bibitem[{{Pumo} \& {Zampieri}(2011)}]{pumo+11}
{Pumo}, M.~L. \& {Zampieri}, L. 2011, \apj, 741, 41

\bibitem[{{Pumo} {et~al.}(2017){Pumo}, {Zampieri}, {Spiro}, {Pastorello},
  {Benetti}, {Cappellaro}, {Manic{\`o}}, \& {Turatto}}]{pumo+17}
{Pumo}, M.~L., {Zampieri}, L., {Spiro}, S., {et~al.} 2017, \mnras, 464, 3013

\bibitem[{{Ram{\'\i}rez-Agudelo} {et~al.}(2013){Ram{\'\i}rez-Agudelo},
  {Sim{\'o}n-D{\'\i}az}, {Sana}, {de Koter}, {Sab{\'\i}n-Sanjul{\'\i}an}, {de
  Mink}, {Dufton}, {Gr{\"a}fener}, {Evans}, {Herrero}, {Langer}, {Lennon},
  {Ma{\'\i}z Apell{\'a}niz}, {Markova}, {Najarro}, {Puls}, {Taylor}, \&
  {Vink}}]{ramirez+13}
{Ram{\'\i}rez-Agudelo}, O.~H., {Sim{\'o}n-D{\'\i}az}, S., {Sana}, H., {et~al.}
  2013, \aap, 560, A29

\bibitem[{{Ricks} \& {Dwarkadas}(2019)}]{ricks+19}
{Ricks}, W. \& {Dwarkadas}, V.~V. 2019, \apj, 880, 59

\bibitem[{{Rodr{\'\i}guez} {et~al.}(2020){Rodr{\'\i}guez}, {Pignata},
  {Anderson}, {Moriya}, {Clocchiatti}, {F{\"o}rster}, {Prieto}, {Phillips},
  {Burns}, {Contreras}, {Folatelli}, {Guti{\'e}rrez}, {Hamuy}, {Morrell},
  {Stritzinger}, {Suntzeff}, {Benetti}, {Cappellaro}, {Elias-Rosa},
  {Pastorello}, {Turatto}, {Maza}, {Antezana}, {Cartier}, {Gonz{\'a}lez},
  {Haislip}, {Kouprianov}, {L{\'o}pez}, {Marchi-Lasch}, \&
  {Reichart}}]{rodriguez+20}
{Rodr{\'\i}guez}, {\'O}., {Pignata}, G., {Anderson}, J.~P., {et~al.} 2020,
  \mnras, 494, 5882

\bibitem[{{Rubin} \& {Gal-Yam}(2016)}]{rubin+16}
{Rubin}, A. \& {Gal-Yam}, A. 2016, \apj, 828, 111

\bibitem[{{Salpeter}(1955)}]{salpeter55}
{Salpeter}, E.~E. 1955, \apj, 121, 161

\bibitem[{{Sana} {et~al.}(2013){Sana}, {de Koter}, {de Mink}, {Dunstall},
  {Evans}, {H{\'e}nault-Brunet}, {Ma{\'\i}z Apell{\'a}niz},
  {Ram{\'\i}rez-Agudelo}, {Taylor}, {Walborn}, {Clark}, {Crowther}, {Herrero},
  {Gieles}, {Langer}, {Lennon}, \& {Vink}}]{sana+13}
{Sana}, H., {de Koter}, A., {de Mink}, S.~E., {et~al.} 2013, \aap, 550, A107

\bibitem[{{Sana} {et~al.}(2012){Sana}, {de Mink}, {de Koter}, {Langer},
  {Evans}, {Gieles}, {Gosset}, {Izzard}, {Le Bouquin}, \&
  {Schneider}}]{sana+12}
{Sana}, H., {de Mink}, S.~E., {de Koter}, A., {et~al.} 2012, Science, 337, 444

\bibitem[{{Sanders} {et~al.}(2015){Sanders}, {Soderberg}, {Gezari},
  {Betancourt}, {Chornock}, {Berger}, {Foley}, {Challis}, {Drout}, {Kirshner},
  {Lunnan}, {Marion}, {Margutti}, {McKinnon}, {Milisavljevic}, {Narayan},
  {Rest}, {Kankare}, {Mattila}, {Smartt}, {Huber}, {Burgett}, {Draper},
  {Hodapp}, {Kaiser}, {Kudritzki}, {Magnier}, {Metcalfe}, {Morgan}, {Price},
  {Tonry}, {Wainscoat}, \& {Waters}}]{sanders+15}
{Sanders}, N.~E., {Soderberg}, A.~M., {Gezari}, S., {et~al.} 2015, \apj, 799,
  208

\bibitem[{{Schmidt} {et~al.}(1994){Schmidt}, {Kirshner}, {Eastman}, {Phillips},
  {Suntzeff}, {Hamuy}, {Maza}, \& {Aviles}}]{schmidt+94}
{Schmidt}, B.~P., {Kirshner}, R.~P., {Eastman}, R.~G., {et~al.} 1994, \apj,
  432, 42

\bibitem[{{Shigeyama} \& {Nomoto}(1990)}]{shigeyama+90}
{Shigeyama}, T. \& {Nomoto}, K. 1990, \apj, 360, 242

\bibitem[{{Shigeyama} {et~al.}(1988){Shigeyama}, {Nomoto}, \&
  {Hashimoto}}]{shigeyema+88}
{Shigeyama}, T., {Nomoto}, K., \& {Hashimoto}, M. 1988, \aap, 196, 141

\bibitem[{{Shivvers} {et~al.}(2017){Shivvers}, {Modjaz}, {Zheng}, {Liu},
  {Filippenko}, {Silverman}, {Matheson}, {Pastorello}, {Graur}, {Foley},
  {Chornock}, {Smith}, {Leaman}, \& {Benetti}}]{shivvers+17}
{Shivvers}, I., {Modjaz}, M., {Zheng}, W., {et~al.} 2017, \pasp, 129, 054201

\bibitem[{{Smartt}(2009)}]{smartt09}
{Smartt}, S.~J. 2009, \araa, 47, 63

\bibitem[{{Smartt}(2015)}]{smartt15}
{Smartt}, S.~J. 2015, \pasa, 32, e016

\bibitem[{{Smartt} {et~al.}(2009){Smartt}, {Eldridge}, {Crockett}, \&
  {Maund}}]{smartt+09}
{Smartt}, S.~J., {Eldridge}, J.~J., {Crockett}, R.~M., \& {Maund}, J.~R. 2009,
  \mnras, 395, 1409

\bibitem[{{Sorokina} \& {Blinnikov}(2002)}]{sorokina+02}
{Sorokina}, E.~I. \& {Blinnikov}, S.~I. 2002, in Nuclear Astrophysics, ed.
  W.~{Hillebrandt} \& E.~{M{\"u}ller}, 57--62

\bibitem[{{Spiro} {et~al.}(2014){Spiro}, {Pastorello}, {Pumo}, {Zampieri},
  {Turatto}, {Smartt}, {Benetti}, {Cappellaro}, {Valenti}, {Agnoletto},
  {Altavilla}, {Aoki}, {Brocato}, {Corsini}, {Di Cianno}, {Elias-Rosa},
  {Hamuy}, {Enya}, {Fiaschi}, {Folatelli}, {Desidera}, {Harutyunyan}, {Howell},
  {Kawka}, {Kobayashi}, {Leibundgut}, {Minezaki}, {Navasardyan}, {Nomoto},
  {Mattila}, {Pietrinferni}, {Pignata}, {Raimondo}, {Salvo}, {Schmidt},
  {Sollerman}, {Spyromilio}, {Taubenberger}, {Valentini}, {Vennes}, \&
  {Yoshii}}]{spiro+14}
{Spiro}, S., {Pastorello}, A., {Pumo}, M.~L., {et~al.} 2014, \mnras, 439, 2873

\bibitem[{{Straniero} {et~al.}(2019){Straniero}, {Dominguez}, {Piersanti},
  {Giannotti}, \& {Mirizzi}}]{straniero+19}
{Straniero}, O., {Dominguez}, I., {Piersanti}, L., {Giannotti}, M., \&
  {Mirizzi}, A. 2019, \apj, 881, 158

\bibitem[{{Stritzinger} {et~al.}(2011){Stritzinger}, {Phillips}, {Boldt},
  {Burns}, {Campillay}, {Contreras}, {Gonzalez}, {Folatelli}, {Morrell},
  {Krzeminski}, {Roth}, {Salgado}, {DePoy}, {Hamuy}, {Freedman}, {Madore},
  {Marshall}, {Persson}, {Rheault}, {Suntzeff}, {Villanueva}, {Li}, \&
  {Filippenko}}]{stritzinger+11}
{Stritzinger}, M.~D., {Phillips}, M.~M., {Boldt}, L.~N., {et~al.} 2011, \aj,
  142, 156

\bibitem[{{Sukhbold} \& {Adams}(2020)}]{sukhbold+20}
{Sukhbold}, T. \& {Adams}, S. 2020, \mnras, 492, 2578

\bibitem[{{Sukhbold} {et~al.}(2016){Sukhbold}, {Ertl}, {Woosley}, {Brown}, \&
  {Janka}}]{sukhbold+16}
{Sukhbold}, T., {Ertl}, T., {Woosley}, S.~E., {Brown}, J.~M., \& {Janka}, H.~T.
  2016, \apj, 821, 38

\bibitem[{{Tanaka} {et~al.}(2012){Tanaka}, {Kawabata}, {Hattori}, {Mazzali},
  {Aoki}, {Iye}, {Maeda}, {Nomoto}, {Pian}, {Sasaki}, \&
  {Yamanaka}}]{tanaka+12}
{Tanaka}, M., {Kawabata}, K.~S., {Hattori}, T., {et~al.} 2012, \apj, 754, 63

\bibitem[{{Taubenberger} {et~al.}(2009){Taubenberger}, {Valenti}, {Benetti},
  {Cappellaro}, {Della Valle}, {Elias-Rosa}, {Hachinger}, {Hillebrandt},
  {Maeda}, {Mazzali}, {Pastorello}, {Patat}, {Sim}, \&
  {Turatto}}]{taubenberger+09}
{Taubenberger}, S., {Valenti}, S., {Benetti}, S., {et~al.} 2009, \mnras, 397,
  677

\bibitem[{{Thielemann} {et~al.}(1996){Thielemann}, {Nomoto}, \&
  {Hashimoto}}]{thielemann+96}
{Thielemann}, F.-K., {Nomoto}, K., \& {Hashimoto}, M.-A. 1996, \apj, 460, 408

\bibitem[{{Ugliano} {et~al.}(2012){Ugliano}, {Janka}, {Marek}, \&
  {Arcones}}]{ugliano+12}
{Ugliano}, M., {Janka}, H.-T., {Marek}, A., \& {Arcones}, A. 2012, \apj, 757,
  69

\bibitem[{{Umeda} \& {Nomoto}(2002)}]{umeda+02}
{Umeda}, H. \& {Nomoto}, K. 2002, \apj, 565, 385

\bibitem[{{Utrobin}(2007)}]{utrobin07}
{Utrobin}, V.~P. 2007, \aap, 461, 233

\bibitem[{{Utrobin} \& {Chugai}(2008)}]{utrobin+08}
{Utrobin}, V.~P. \& {Chugai}, N.~N. 2008, \aap, 491, 507

\bibitem[{{Utrobin} \& {Chugai}(2019)}]{utrobin+19}
{Utrobin}, V.~P. \& {Chugai}, N.~N. 2019, \mnras, 490, 2042

\bibitem[{{Valenti} {et~al.}(2016){Valenti}, {Howell}, {Stritzinger}, {Graham},
  {Hosseinzadeh}, {Arcavi}, {Bildsten}, {Jerkstrand}, {McCully}, {Pastorello},
  {Piro}, {Sand}, {Smartt}, {Terreran}, {Baltay}, {Benetti}, {Brown},
  {Filippenko}, {Fraser}, {Rabinowitz}, {Sullivan}, \& {Yuan}}]{valenti+16}
{Valenti}, S., {Howell}, D.~A., {Stritzinger}, M.~D., {et~al.} 2016, \mnras,
  459, 3939

\bibitem[{Van Der~Walt {et~al.}(2011)Van Der~Walt, Colbert, \&
  Varoquaux}]{numpy2011}
Van Der~Walt, S., Colbert, S.~C., \& Varoquaux, G. 2011, Computing in Science
  \& Engineering, 13, 22

\bibitem[{{Van Dyk} {et~al.}(2012){Van Dyk}, {Davidge}, {Elias-Rosa},
  {Taubenberger}, {Li}, {Levesque}, {Howerton}, {Pignata}, {Morrell}, {Hamuy},
  \& {Filippenko}}]{vandyk+12a}
{Van Dyk}, S.~D., {Davidge}, T.~J., {Elias-Rosa}, N., {et~al.} 2012, \aj, 143,
  19

\bibitem[{{Van Dyk} {et~al.}(2003){Van Dyk}, {Li}, \& {Filippenko}}]{vandyk+03}
{Van Dyk}, S.~D., {Li}, W., \& {Filippenko}, A.~V. 2003, \pasp, 115, 1289

\bibitem[{{Van Dyk} {et~al.}(1999){Van Dyk}, {Peng}, {Barth}, \&
  {Filippenko}}]{vandyk+99}
{Van Dyk}, S.~D., {Peng}, C.~Y., {Barth}, A.~J., \& {Filippenko}, A.~V. 1999,
  \aj, 118, 2331

\bibitem[{{van Loon}(2006)}]{vanloon06}
{van Loon}, J.~T. 2006, in Astronomical Society of the Pacific Conference
  Series, Vol. 353, Stellar Evolution at Low Metallicity: Mass Loss,
  Explosions, Cosmology, ed. H.~J.~G.~L.~M. {Lamers}, N.~{Langer}, T.~{Nugis},
  \& K.~{Annuk}, 211

\bibitem[{{Vink} {et~al.}(2001){Vink}, {de Koter}, \& {Lamers}}]{vink+01}
{Vink}, J.~S., {de Koter}, A., \& {Lamers}, H.~J.~G.~L.~M. 2001, \aap, 369, 574

\bibitem[{{Virtanen} {et~al.}(2020){Virtanen}, {Gommers}, {Oliphant},
  {Haberland}, {Reddy}, {Cournapeau}, {Burovski}, {Peterson}, {Weckesser},
  {Bright}, {van der Walt}, {Brett}, {Wilson}, {Jarrod Millman}, {Mayorov},
  {Nelson}, {Jones}, {Kern}, {Larson}, {Carey}, {Polat}, {Feng}, {Moore}, {Vand
  erPlas}, {Laxalde}, {Perktold}, {Cimrman}, {Henriksen}, {Quintero}, {Harris},
  {Archibald}, {Ribeiro}, {Pedregosa}, {van Mulbregt}, \&
  {Contributors}}]{scipy2020}
{Virtanen}, P., {Gommers}, R., {Oliphant}, T.~E., {et~al.} 2020, Nature
  Methods, 17, 261

\bibitem[{{Wagle} {et~al.}(2020){Wagle}, {Ray}, \& {Raghu}}]{wagle+20}
{Wagle}, G.~A., {Ray}, A., \& {Raghu}, A. 2020, \apj, 894, 118

\bibitem[{{Williams} {et~al.}(2014){Williams}, {Peterson}, {Murphy}, {Gilbert},
  {Dalcanton}, {Dolphin}, \& {Jennings}}]{williams+14}
{Williams}, B.~F., {Peterson}, S., {Murphy}, J., {et~al.} 2014, \apj, 791, 105

\bibitem[{{Wongwathanarat} {et~al.}(2015){Wongwathanarat}, {M{\"u}ller}, \&
  {Janka}}]{wongwathanarat+15}
{Wongwathanarat}, A., {M{\"u}ller}, E., \& {Janka}, H.~T. 2015, \aap, 577, A48

\bibitem[{{Woosley} {et~al.}(2002){Woosley}, {Heger}, \& {Weaver}}]{woosley+02}
{Woosley}, S.~E., {Heger}, A., \& {Weaver}, T.~A. 2002, Reviews of Modern
  Physics, 74, 1015

\bibitem[{{Woosley} \& {Weaver}(1995)}]{woosley+95}
{Woosley}, S.~E. \& {Weaver}, T.~A. 1995, \apjs, 101, 181

\bibitem[{{Yaron} {et~al.}(2017){Yaron}, {Perley}, {Gal-Yam}, {Groh}, {Horesh},
  {Ofek}, {Kulkarni}, {Sollerman}, {Fransson}, {Rubin}, {Szabo}, {Sapir},
  {Taddia}, {Cenko}, {Valenti}, {Arcavi}, {Howell}, {Kasliwal}, {Vreeswijk},
  {Khazov}, {Fox}, {Cao}, {Gnat}, {Kelly}, {Nugent}, {Filippenko}, {Laher},
  {Wozniak}, {Lee}, {Rebbapragada}, {Maguire}, {Sullivan}, \&
  {Soumagnac}}]{yaron+17}
{Yaron}, O., {Perley}, D.~A., {Gal-Yam}, A., {et~al.} 2017, Nature Physics, 13,
  510

\bibitem[{{Yoon} \& {Cantiello}(2010)}]{yoon+10}
{Yoon}, S.-C. \& {Cantiello}, M. 2010, \apjl, 717, L62

\bibitem[{{Zampieri}(2017)}]{zampieri17book}
{Zampieri}, L. 2017, {Light Curves of Type II Supernovae}, ed. A.~W. {Alsabti}
  \& P.~{Murdin}, 737

\bibitem[{{Zampieri} {et~al.}(2003){Zampieri}, {Pastorello}, {Turatto},
  {Cappellaro}, {Benetti}, {Altavilla}, {Mazzali}, \& {Hamuy}}]{zampieri+03}
{Zampieri}, L., {Pastorello}, A., {Turatto}, M., {et~al.} 2003, \mnras, 338,
  711

\bibitem[{{Zapartas} {et~al.}(2019){Zapartas}, {de Mink}, {Justham}, {Smith},
  {de Koter}, {Renzo}, {Arcavi}, {Farmer}, {G{\"o}tberg}, \&
  {Toonen}}]{zapartas+19}
{Zapartas}, E., {de Mink}, S.~E., {Justham}, S., {et~al.} 2019, \aap, 631, A5

\end{thebibliography}

\begin{appendix}
\section{Influence of the gold sample selection on the results}
\label{app:gold_sample}

As described in detail in Sect.~\ref{sec:results}, a gold \snii\ requires both good data coverage and good fits to our models.
The classification of the gold sample was achieved visually by the first three authors of this paper.
In most cases those authors agreed in the classification, but a small number of events were controversial.
In this section, we analyse whether the inclusion or exclusion of some debatable cases affect our results on the IMF parameters.

We started by including three additional \sneii\ to the gold sample reaching a total of 27 events. 
These \sneii\ were not initially included in the gold sample for the following reasons: SN~2005dw presents two \ion{Fe}{ii} line velocity measurements but very close in time; SN~2007hm does not show a clear end of the plateau phase; and SN~2008K shows offsets in the fits to both bolometric LC and photospheric velocities.
The inclusion of these \sneii\ in the gold sample leads to the following IMF parameters: \mmin~=~9.3$^{+0.1}_{-0.1}$~\ms, \mmax~=~24.5$^{+4.3}_{-3.0}$~\ms\ ($M_{\rm high}^{\rm mode}$~=~21.0$^{+3.4}_{-0.6}$~\ms), and $\Gamma$~=~$-$6.48$^{+0.51}_{-0.52}$.
We next excluded three events from the authentic gold sample defined in Sect~\ref{sec:results}. SN~2006be and SN~2007W have observations at the beginning of the transition from the plateau to the tail phase, although it might not be clear.
In addition, we excluded SN~2008bk, which does not fulfil the conditions to be considered a gold event since there is no CSP-I photometry during the transition from the plateau to the radioactive tail phase (published data are used to restrict the end of the plateau). With the exclusion of these three \sneii\ from the gold sample, the results are: \mmin~=~9.3$^{+0.1}_{-0.1}$~\ms, \mmax~=~25.1$^{+4.0}_{-3.5}$~\ms\ ($M_{\rm high}^{\rm mode}$~=~22.7$^{+2.8}_{-1.7}$~\ms), and $\Gamma$~=~$-$6.39$^{+0.65}_{-0.64}$.
In conclusion, this analysis suggests that even if those limiting cases are included/excluded from the gold sample, the parameters derived for the IMF are not statistical different with the results presented in Sect.~\ref{sec:imf}.

\section{Physical parameters of SNe II}
\label{app:physical_pars}

\longtab[1]{
\begin{landscape}
\begin{longtable}{lccccccccccc}
\caption{SN~II physical parameters.}
\label{table:results}
\\ \hline\hline\noalign{\smallskip}
SN & Gold? & $t_{\rm exp}$ & scale & $M_{\rm ZAMS}$ & $M_{\rm ej}$ & $M_{\rm H,env}$ & $M_{\rm H}$ & \emph{R} & $\emph{E}$ & $M_{\rm Ni}$ & $^{56}\rm Ni$ mixing \\
 & & (MJD) & & ($M_{\odot}$) & ($M_{\odot}$) & ($M_{\odot}$) & ($M_{\odot}$) & ($R_{\odot}$) & (foe) & ($M_{\odot}$) & \\
\hline\noalign{\smallskip}
\endfirsthead
\caption{continued.}\\
\hline\hline\noalign{\smallskip}
SN & Gold? & $t_{\rm exp}$ & scale & $M_{\rm ZAMS}$ & $M_{\rm ej}$ & $M_{\rm H,env}$ & $M_{\rm H}$ & \emph{R} & $\emph{E}$ & $M_{\rm Ni}$ & $^{56}\rm Ni$ mixing \\
 & & (MJD) & & ($M_{\odot}$) & ($M_{\odot}$) & ($M_{\odot}$) & ($M_{\odot}$) & ($R_{\odot}$) & (foe) & ($M_{\odot}$) & \\
\hline\noalign{\smallskip}
\endhead
\hline\noalign{\smallskip}
\endfoot
2004ej & YES & 53232.5$^{+0.3}_{-0.5}$ & 0.81$^{+0.01}_{-0.01}$ & 12.01$^{+0.05}_{-0.04}$ & 9.39$^{+0.04}_{-0.03}$ & 8.03$^{+0.02}_{-0.02}$ & 5.42$^{+0.01}_{-0.01}$ & 595$^{+5}_{-2}$ & 0.53$^{+0.01}_{-0.01}$ & 0.018$^{+0.001}_{-0.001}$ & 0.72$^{+0.06}_{-0.10}$ \\
\noalign{\smallskip}
2004er & NO & 53273.7$^{+0.1}_{-0.1}$ & 0.85$^{+0.01}_{-0.01}$ & 15.92$^{+0.08}_{-0.13}$ & 13.01$^{+0.06}_{-0.09}$ & 10.31$^{+0.03}_{-0.06}$ & 6.65$^{+0.02}_{-0.03}$ & 810$^{+3}_{-5}$ & 0.64$^{+0.01}_{-0.01}$ & 0.063$^{+0.001}_{-0.001}$ & 0.80$^{+0.01}_{-0.01}$ \\
\noalign{\smallskip}
2004fb & NO & 53260.3$^{+4.0}_{-5.1}$ & 1.08$^{+0.13}_{-0.12}$ & 13.58$^{+2.36}_{-3.10}$ & 10.99$^{+2.04}_{-2.57}$ & 8.99$^{+1.33}_{-1.69}$ & 5.94$^{+0.71}_{-1.00}$ & 720$^{+91}_{-215}$ & 0.54$^{+0.07}_{-0.08}$ & --- & 0.39$^{+0.20}_{-0.11}$ \\
\noalign{\smallskip}
2004fc & YES & 53293.3$^{+0.7}_{-0.6}$ & 0.87$^{+0.05}_{-0.04}$ & 11.08$^{+0.22}_{-0.15}$ & 8.73$^{+0.15}_{-0.09}$ & 7.58$^{+0.11}_{-0.07}$ & 5.13$^{+0.07}_{-0.05}$ & 554$^{+9}_{-10}$ & 0.29$^{+0.01}_{-0.01}$ & --- & 0.51$^{+0.02}_{-0.02}$ \\
\noalign{\smallskip}
2004fx & YES & 53300.0$^{+0.8}_{-0.4}$ & 1.33$^{+0.05}_{-0.04}$ & 9.98$^{+0.14}_{-0.15}$ & 8.17$^{+0.06}_{-0.05}$ & 7.07$^{+0.05}_{-0.01}$ & 4.78$^{+0.04}_{-0.05}$ & 462$^{+11}_{-2}$ & 0.38$^{+0.01}_{-0.01}$ & 0.012$^{+0.001}_{-0.001}$ & 0.26$^{+0.05}_{-0.04}$ \\
\noalign{\smallskip}
2005J & YES & 53386.2$^{+0.5}_{-1.1}$ & 1.03$^{+0.13}_{-0.12}$ & 15.44$^{+0.93}_{-1.35}$ & 12.66$^{+0.39}_{-1.04}$ & 10.09$^{+0.16}_{-0.69}$ & 6.52$^{+0.06}_{-0.38}$ & 790$^{+35}_{-45}$ & 0.75$^{+0.07}_{-0.09}$ & --- & 0.71$^{+0.05}_{-0.07}$ \\
\noalign{\smallskip}
2005Z & NO & 53400.1$^{+2.0}_{-2.0}$ & 1.22$^{+0.04}_{-0.10}$ & 12.49$^{+2.35}_{-1.90}$ & 9.79$^{+2.41}_{-1.32}$ & 8.25$^{+1.56}_{-0.91}$ & 5.55$^{+0.81}_{-0.57}$ & 641$^{+127}_{-126}$ & 1.01$^{+0.06}_{-0.07}$ & --- & 0.28$^{+0.11}_{-0.05}$ \\
\noalign{\smallskip}
2005af & NO & 53308.9$^{+7.7}_{-3.6}$ & 1.00$^{+0.02}_{-0.01}$ & --- & --- & --- & --- & --- & --- & 0.044$^{+0.003}_{-0.002}$ & --- \\
\noalign{\smallskip}
2005an & NO & 53430.3$^{+2.4}_{-2.3}$ & 1.34$^{+0.02}_{-0.04}$ & 10.06$^{+0.14}_{-0.07}$ & 8.21$^{+0.07}_{-0.03}$ & 7.09$^{+0.07}_{-0.03}$ & 4.81$^{+0.05}_{-0.02}$ & 468$^{+13}_{-6}$ & 0.77$^{+0.04}_{-0.05}$ & --- & 0.22$^{+0.04}_{-0.01}$ \\
\noalign{\smallskip}
2005dk & YES & 53604.9$^{+0.6}_{-0.9}$ & 0.91$^{+0.05}_{-0.03}$ & 12.02$^{+0.09}_{-0.06}$ & 9.40$^{+0.07}_{-0.04}$ & 8.04$^{+0.04}_{-0.03}$ & 5.43$^{+0.02}_{-0.02}$ & 597$^{+8}_{-4}$ & 0.85$^{+0.04}_{-0.03}$ & --- & 0.52$^{+0.05}_{-0.02}$ \\
\noalign{\smallskip}
2005dn & YES & 53605.1$^{+2.6}_{-5.4}$ & 1.13$^{+0.16}_{-0.15}$ & 10.69$^{+1.67}_{-0.60}$ & 8.52$^{+1.16}_{-0.29}$ & 7.39$^{+0.80}_{-0.29}$ & 5.01$^{+0.51}_{-0.19}$ & 523$^{+105}_{-54}$ & 0.94$^{+0.10}_{-0.12}$ & 0.041$^{+0.009}_{-0.010}$ & 0.40$^{+0.08}_{-0.10}$ \\
\noalign{\smallskip}
2005dt & NO & 53602.2$^{+9.1}_{-4.6}$ & 0.99$^{+0.14}_{-0.04}$ & 18.04$^{+2.25}_{-4.96}$ & 13.30$^{+1.22}_{-2.96}$ & 9.88$^{+0.38}_{-1.33}$ & 6.08$^{+0.04}_{-0.36}$ & 981$^{+87}_{-287}$ & 0.58$^{+0.12}_{-0.15}$ & --- & 0.76$^{+0.03}_{-0.06}$ \\
\noalign{\smallskip}
2005dw & NO & 53597.4$^{+6.7}_{-2.2}$ & 0.96$^{+0.13}_{-0.06}$ & 13.29$^{+0.56}_{-2.46}$ & 10.60$^{+0.74}_{-2.01}$ & 8.73$^{+0.49}_{-1.27}$ & 5.81$^{+0.24}_{-0.76}$ & 704$^{+30}_{-169}$ & 0.49$^{+0.07}_{-0.04}$ & --- & 0.34$^{+0.28}_{-0.06}$ \\
\noalign{\smallskip}
2005dx & NO & 53609.9$^{+3.7}_{-3.3}$ & 1.01$^{+0.17}_{-0.06}$ & 10.70$^{+0.49}_{-0.56}$ & 8.53$^{+0.28}_{-0.28}$ & 7.40$^{+0.24}_{-0.27}$ & 5.02$^{+0.15}_{-0.18}$ & 525$^{+35}_{-50}$ & 0.30$^{+0.03}_{-0.02}$ & --- & 0.33$^{+0.30}_{-0.10}$ \\
\noalign{\smallskip}
2005dz & YES & 53622.9$^{+0.5}_{-1.0}$ & 1.10$^{+0.08}_{-0.13}$ & 10.07$^{+0.19}_{-0.22}$ & 8.21$^{+0.09}_{-0.08}$ & 7.09$^{+0.09}_{-0.01}$ & 4.81$^{+0.06}_{-0.07}$ & 468$^{+17}_{-9}$ & 0.50$^{+0.03}_{-0.05}$ & --- & 0.55$^{+0.09}_{-0.07}$ \\
\noalign{\smallskip}
2005lw & NO & 53707.6$^{+0.8}_{-0.5}$ & 0.94$^{+0.01}_{-0.01}$ & 16.96$^{+3.12}_{-0.22}$ & 13.00$^{+1.40}_{-0.01}$ & 10.13$^{+0.08}_{-0.01}$ & 6.44$^{+0.05}_{-0.34}$ & 843$^{+220}_{-7}$ & 0.96$^{+0.06}_{-0.02}$ & --- & 0.23$^{+0.01}_{-0.02}$ \\
\noalign{\smallskip}
2005me & NO & 53711.5$^{+4.7}_{-2.7}$ & 1.07$^{+0.15}_{-0.11}$ & 11.93$^{+2.44}_{-1.55}$ & 9.33$^{+2.50}_{-0.96}$ & 8.00$^{+1.56}_{-0.75}$ & 5.40$^{+0.83}_{-0.49}$ & 592$^{+162}_{-96}$ & 0.55$^{+0.10}_{-0.07}$ & --- & 0.34$^{+0.26}_{-0.11}$ \\
\noalign{\smallskip}
2006ai & YES & 53776.8$^{+0.4}_{-0.2}$ & 1.04$^{+0.06}_{-0.06}$ & 10.06$^{+0.12}_{-0.08}$ & 8.21$^{+0.06}_{-0.03}$ & 7.09$^{+0.06}_{-0.02}$ & 4.81$^{+0.04}_{-0.03}$ & 467$^{+11}_{-6}$ & 1.25$^{+0.08}_{-0.08}$ & 0.047$^{+0.005}_{-0.005}$ & 0.34$^{+0.03}_{-0.03}$ \\
\noalign{\smallskip}
2006be & YES & 53809.3$^{+1.6}_{-2.8}$ & 1.15$^{+0.16}_{-0.16}$ & 9.51$^{+0.43}_{-0.23}$ & 8.03$^{+0.13}_{-0.07}$ & 7.20$^{+0.06}_{-0.12}$ & 4.62$^{+0.15}_{-0.08}$ & 454$^{+7}_{-4}$ & 0.60$^{+0.08}_{-0.07}$ & --- & 0.54$^{+0.19}_{-0.21}$ \\
\noalign{\smallskip}
2006bl & NO & 53818.3$^{+3.0}_{-3.1}$ & 0.97$^{+0.02}_{-0.02}$ & 15.47$^{+1.70}_{-2.49}$ & 12.68$^{+0.37}_{-2.47}$ & 10.11$^{+0.01}_{-1.63}$ & 6.53$^{+0.01}_{-0.85}$ & 792$^{+76}_{-104}$ & 1.40$^{+0.07}_{-0.08}$ & --- & 0.41$^{+0.24}_{-0.15}$ \\
\noalign{\smallskip}
2006ee & YES & 53964.5$^{+1.0}_{-1.6}$ & 0.89$^{+0.03}_{-0.02}$ & 9.27$^{+0.17}_{-0.16}$ & 7.96$^{+0.05}_{-0.05}$ & 7.26$^{+0.04}_{-0.05}$ & 4.54$^{+0.06}_{-0.05}$ & 450$^{+3}_{-3}$ & 0.43$^{+0.02}_{-0.02}$ & --- & 0.69$^{+0.07}_{-0.13}$ \\
\noalign{\smallskip}
2006iw & NO & 54010.9$^{+0.5}_{-0.7}$ & 0.97$^{+0.03}_{-0.02}$ & 12.54$^{+2.69}_{-0.50}$ & 9.83$^{+2.67}_{-0.42}$ & 8.27$^{+1.73}_{-0.23}$ & 5.57$^{+0.90}_{-0.14}$ & 646$^{+136}_{-47}$ & 0.56$^{+0.11}_{-0.05}$ & --- & 0.51$^{+0.09}_{-0.07}$ \\
\noalign{\smallskip}
2006qr & YES & 54069.0$^{+0.6}_{-1.2}$ & 1.28$^{+0.03}_{-0.05}$ & 10.21$^{+0.42}_{-0.22}$ & 8.29$^{+0.21}_{-0.11}$ & 7.16$^{+0.20}_{-0.10}$ & 4.86$^{+0.14}_{-0.07}$ & 481$^{+38}_{-19}$ & 0.36$^{+0.02}_{-0.02}$ & --- & 0.26$^{+0.11}_{-0.05}$ \\
\noalign{\smallskip}
2007P & NO & 54122.7$^{+0.7}_{-1.4}$ & 0.99$^{+0.12}_{-0.02}$ & 11.73$^{+1.23}_{-0.61}$ & 9.19$^{+1.00}_{-0.43}$ & 7.90$^{+0.56}_{-0.30}$ & 5.34$^{+0.34}_{-0.19}$ & 583$^{+102}_{-27}$ & 1.03$^{+0.11}_{-0.06}$ & --- & 0.50$^{+0.10}_{-0.09}$ \\
\noalign{\smallskip}
2007U & NO & 54134.2$^{+1.3}_{-1.7}$ & 1.08$^{+0.08}_{-0.12}$ & 12.00$^{+0.71}_{-1.36}$ & 9.38$^{+0.59}_{-0.89}$ & 8.03$^{+0.32}_{-0.66}$ & 5.42$^{+0.19}_{-0.43}$ & 595$^{+66}_{-76}$ & 1.38$^{+0.08}_{-0.11}$ & --- & 0.34$^{+0.10}_{-0.05}$ \\
\noalign{\smallskip}
2007W & YES & 54136.5$^{+0.9}_{-1.7}$ & 0.88$^{+0.08}_{-0.05}$ & 11.15$^{+0.19}_{-0.39}$ & 8.77$^{+0.13}_{-0.22}$ & 7.61$^{+0.09}_{-0.19}$ & 5.16$^{+0.06}_{-0.12}$ & 557$^{+8}_{-28}$ & 0.22$^{+0.02}_{-0.01}$ & --- & 0.68$^{+0.06}_{-0.07}$ \\
\noalign{\smallskip}
2007X & YES & 54148.0$^{+0.4}_{-0.6}$ & 0.85$^{+0.04}_{-0.03}$ & 14.99$^{+0.23}_{-0.22}$ & 12.32$^{+0.17}_{-0.18}$ & 9.90$^{+0.10}_{-0.12}$ & 6.41$^{+0.06}_{-0.06}$ & 772$^{+9}_{-7}$ & 1.13$^{+0.04}_{-0.04}$ & 0.069$^{+0.004}_{-0.003}$ & 0.54$^{+0.07}_{-0.05}$ \\
\noalign{\smallskip}
2007aa & YES & 54122.3$^{+2.0}_{-1.9}$ & 0.90$^{+0.07}_{-0.07}$ & 9.35$^{+0.27}_{-0.23}$ & 7.98$^{+0.08}_{-0.07}$ & 7.24$^{+0.06}_{-0.08}$ & 4.57$^{+0.09}_{-0.08}$ & 451$^{+4}_{-4}$ & 0.46$^{+0.02}_{-0.02}$ & 0.030$^{+0.004}_{-0.004}$ & 0.62$^{+0.12}_{-0.19}$ \\
\noalign{\smallskip}
2007ab & YES & 54116.6$^{+1.6}_{-1.4}$ & 0.98$^{+0.09}_{-0.04}$ & 11.20$^{+0.26}_{-0.24}$ & 8.81$^{+0.18}_{-0.16}$ & 7.64$^{+0.13}_{-0.12}$ & 5.17$^{+0.08}_{-0.07}$ & 560$^{+11}_{-12}$ & 1.15$^{+0.09}_{-0.04}$ & 0.033$^{+0.003}_{-0.002}$ & 0.22$^{+0.03}_{-0.02}$ \\
\noalign{\smallskip}
2007av & NO & 54177.3$^{+1.1}_{-2.3}$ & 1.31$^{+0.06}_{-0.11}$ & 11.03$^{+3.15}_{-1.08}$ & 8.69$^{+2.99}_{-0.53}$ & 7.55$^{+1.90}_{-0.48}$ & 5.12$^{+1.05}_{-0.35}$ & 552$^{+196}_{-91}$ & 0.57$^{+0.06}_{-0.05}$ & --- & 0.34$^{+0.20}_{-0.10}$ \\
\noalign{\smallskip}
2007hm & NO & 54330.8$^{+0.3}_{-0.2}$ & 1.26$^{+0.01}_{-0.02}$ & 10.05$^{+0.09}_{-0.04}$ & 8.21$^{+0.04}_{-0.02}$ & 7.09$^{+0.04}_{-0.02}$ & 4.81$^{+0.03}_{-0.01}$ & 467$^{+8}_{-4}$ & 0.54$^{+0.01}_{-0.02}$ & --- & 0.20$^{+0.01}_{-0.01}$ \\
\noalign{\smallskip}
2007il & NO & 54353.7$^{+0.1}_{-0.1}$ & 0.90$^{+0.01}_{-0.01}$ & 15.68$^{+0.26}_{-0.29}$ & 12.83$^{+0.19}_{-0.21}$ & 10.20$^{+0.11}_{-0.13}$ & 6.59$^{+0.07}_{-0.08}$ & 800$^{+11}_{-12}$ & 0.80$^{+0.01}_{-0.01}$ & --- & 0.50$^{+0.01}_{-0.01}$ \\
\noalign{\smallskip}
2007it & NO & 54347.9$^{+0.4}_{-0.3}$ & 1.00$^{+0.02}_{-0.01}$ & --- & --- & --- & --- & --- & --- & 0.060$^{+0.004}_{-0.003}$ & --- \\
\noalign{\smallskip}
2007ld & NO & 54375.1$^{+5.2}_{-4.4}$ & 1.01$^{+0.22}_{-0.06}$ & 14.78$^{+3.38}_{-3.85}$ & 12.16$^{+1.21}_{-3.52}$ & 9.78$^{+0.13}_{-2.27}$ & 6.35$^{+0.01}_{-1.26}$ & 766$^{+221}_{-220}$ & 0.71$^{+0.19}_{-0.09}$ & --- & 0.45$^{+0.20}_{-0.18}$ \\
\noalign{\smallskip}
2007oc & YES & 54386.8$^{+0.7}_{-0.8}$ & 1.29$^{+0.02}_{-0.03}$ & 10.03$^{+0.15}_{-0.11}$ & 8.19$^{+0.07}_{-0.04}$ & 7.07$^{+0.07}_{-0.01}$ & 4.80$^{+0.05}_{-0.04}$ & 465$^{+13}_{-4}$ & 0.66$^{+0.02}_{-0.02}$ & --- & 0.41$^{+0.25}_{-0.15}$ \\
\noalign{\smallskip}
2007od & NO & 54399.0$^{+4.5}_{-2.5}$ & 0.55$^{+0.08}_{-0.07}$ & 15.98$^{+1.46}_{-3.45}$ & 13.06$^{+0.06}_{-3.23}$ & 10.33$^{+0.01}_{-2.06}$ & 6.67$^{+0.01}_{-1.10}$ & 813$^{+91}_{-168}$ & 0.66$^{+0.12}_{-0.08}$ & --- & 0.41$^{+0.22}_{-0.15}$ \\
\noalign{\smallskip}
2007sq & NO & 54425.5$^{+2.7}_{-3.7}$ & 1.07$^{+0.11}_{-0.12}$ & 10.90$^{+0.68}_{-0.49}$ & 8.62$^{+0.46}_{-0.24}$ & 7.49$^{+0.33}_{-0.24}$ & 5.08$^{+0.21}_{-0.16}$ & 542$^{+35}_{-44}$ & 0.46$^{+0.04}_{-0.04}$ & --- & 0.56$^{+0.15}_{-0.22}$ \\
\noalign{\smallskip}
2008K & NO & 54481.4$^{+0.1}_{-0.1}$ & 1.19$^{+0.03}_{-0.03}$ & 11.01$^{+0.07}_{-0.05}$ & 8.68$^{+0.05}_{-0.03}$ & 7.55$^{+0.04}_{-0.02}$ & 5.11$^{+0.02}_{-0.02}$ & 552$^{+3}_{-4}$ & 0.81$^{+0.02}_{-0.02}$ & 0.023$^{+0.001}_{-0.001}$ & 0.23$^{+0.04}_{-0.02}$ \\
\noalign{\smallskip}
2008M & YES & 54463.9$^{+1.6}_{-0.9}$ & 1.24$^{+0.09}_{-0.11}$ & 10.37$^{+0.08}_{-0.13}$ & 8.36$^{+0.04}_{-0.06}$ & 7.24$^{+0.04}_{-0.06}$ & 4.91$^{+0.03}_{-0.04}$ & 495$^{+7}_{-12}$ & 0.62$^{+0.05}_{-0.05}$ & 0.021$^{+0.002}_{-0.002}$ & 0.55$^{+0.09}_{-0.06}$ \\
\noalign{\smallskip}
2008W & YES & 54489.4$^{+1.4}_{-1.5}$ & 1.15$^{+0.07}_{-0.09}$ & 10.37$^{+0.17}_{-0.32}$ & 8.36$^{+0.08}_{-0.16}$ & 7.24$^{+0.08}_{-0.15}$ & 4.91$^{+0.06}_{-0.10}$ & 495$^{+15}_{-29}$ & 0.72$^{+0.04}_{-0.07}$ & 0.036$^{+0.006}_{-0.008}$ & 0.48$^{+0.07}_{-0.08}$ \\
\noalign{\smallskip}
2008ag & YES & 54483.1$^{+1.5}_{-0.5}$ & 0.87$^{+0.01}_{-0.01}$ & 20.92$^{+0.08}_{-1.76}$ & 14.85$^{+0.04}_{-0.89}$ & 10.42$^{+0.02}_{-0.31}$ & 6.16$^{+0.01}_{-0.03}$ & 1077$^{+1}_{-44}$ & 0.90$^{+0.01}_{-0.05}$ & 0.065$^{+0.002}_{-0.001}$ & 0.80$^{+0.01}_{-0.01}$ \\
\noalign{\smallskip}
2008aw & YES & 54511.6$^{+4.7}_{-2.4}$ & 0.99$^{+0.14}_{-0.09}$ & 12.42$^{+3.07}_{-0.89}$ & 9.74$^{+2.96}_{-0.69}$ & 8.22$^{+1.90}_{-0.42}$ & 5.53$^{+1.00}_{-0.26}$ & 635$^{+158}_{-60}$ & 1.02$^{+0.17}_{-0.13}$ & 0.036$^{+0.006}_{-0.005}$ & 0.29$^{+0.09}_{-0.06}$ \\
\noalign{\smallskip}
2008bk & YES & 54546.6$^{+1.6}_{-1.2}$ & 0.99$^{+0.07}_{-0.02}$ & 12.05$^{+0.20}_{-0.27}$ & 9.42$^{+0.17}_{-0.20}$ & 8.05$^{+0.09}_{-0.13}$ & 5.43$^{+0.06}_{-0.08}$ & 600$^{+19}_{-14}$ & 0.15$^{+0.01}_{-0.01}$ & 0.006$^{+0.001}_{-0.001}$ & 0.72$^{+0.06}_{-0.11}$ \\
\noalign{\smallskip}
2008br & NO & 54550.8$^{+3.8}_{-2.7}$ & 0.89$^{+0.08}_{-0.05}$ & 11.96$^{+1.46}_{-1.38}$ & 9.35$^{+1.42}_{-0.89}$ & 8.01$^{+0.83}_{-0.67}$ & 5.41$^{+0.46}_{-0.43}$ & 593$^{+118}_{-79}$ & 0.18$^{+0.05}_{-0.01}$ & --- & 0.60$^{+0.14}_{-0.20}$ \\
\noalign{\smallskip}
2008ga & NO & 54709.7$^{+3.6}_{-3.5}$ & 0.97$^{+0.12}_{-0.06}$ & 10.88$^{+0.32}_{-0.40}$ & 8.61$^{+0.20}_{-0.20}$ & 7.48$^{+0.16}_{-0.19}$ & 5.07$^{+0.10}_{-0.13}$ & 540$^{+20}_{-35}$ & 0.63$^{+0.07}_{-0.04}$ & --- & 0.53$^{+0.20}_{-0.21}$ \\
\noalign{\smallskip}
2008gi & NO & 54735.7$^{+3.2}_{-1.5}$ & 1.20$^{+0.05}_{-0.11}$ & 16.82$^{+0.37}_{-1.52}$ & 13.01$^{+0.04}_{-0.46}$ & 10.16$^{+0.01}_{-0.12}$ & 6.47$^{+0.02}_{-0.11}$ & 839$^{+32}_{-54}$ & 1.02$^{+0.06}_{-0.10}$ & --- & 0.41$^{+0.07}_{-0.07}$ \\
\noalign{\smallskip}
2008gr & NO & 54767.8$^{+1.6}_{-2.4}$ & 1.10$^{+0.15}_{-0.13}$ & 14.42$^{+2.29}_{-2.60}$ & 11.87$^{+1.15}_{-2.62}$ & 9.58$^{+0.60}_{-1.64}$ & 6.24$^{+0.26}_{-0.88}$ & 755$^{+81}_{-168}$ & 1.22$^{+0.16}_{-0.16}$ & --- & 0.37$^{+0.21}_{-0.12}$ \\
\noalign{\smallskip}
2008if & YES & 54806.6$^{+1.3}_{-1.2}$ & 0.98$^{+0.07}_{-0.06}$ & 10.10$^{+0.17}_{-0.09}$ & 8.23$^{+0.09}_{-0.04}$ & 7.11$^{+0.08}_{-0.04}$ & 4.82$^{+0.06}_{-0.03}$ & 471$^{+15}_{-8}$ & 1.00$^{+0.06}_{-0.06}$ & 0.042$^{+0.003}_{-0.002}$ & 0.21$^{+0.02}_{-0.01}$ \\
\noalign{\smallskip}
2008in & YES & 54826.9$^{+0.3}_{-0.1}$ & 1.63$^{+0.01}_{-0.01}$ & 9.96$^{+0.04}_{-0.08}$ & 8.17$^{+0.01}_{-0.02}$ & 7.07$^{+0.02}_{-0.01}$ & 4.78$^{+0.01}_{-0.03}$ & 461$^{+1}_{-1}$ & 0.29$^{+0.01}_{-0.01}$ & --- & 0.51$^{+0.02}_{-0.02}$ \\
\noalign{\smallskip}
2009N & YES & 54845.8$^{+4.3}_{-2.9}$ & 1.23$^{+0.08}_{-0.18}$ & 9.23$^{+0.20}_{-0.15}$ & 7.95$^{+0.06}_{-0.05}$ & 7.28$^{+0.04}_{-0.05}$ & 4.53$^{+0.07}_{-0.05}$ & 450$^{+3}_{-3}$ & 0.31$^{+0.02}_{-0.04}$ & --- & 0.67$^{+0.10}_{-0.17}$ \\
\noalign{\smallskip}
2009ao & NO & 54892.9$^{+1.3}_{-2.4}$ & 1.22$^{+0.07}_{-0.13}$ & 12.89$^{+1.54}_{-2.82}$ & 10.13$^{+1.76}_{-1.92}$ & 8.43$^{+1.16}_{-1.34}$ & 5.66$^{+0.59}_{-0.85}$ & 679$^{+77}_{-211}$ & 0.56$^{+0.05}_{-0.05}$ & --- & 0.42$^{+0.21}_{-0.14}$ \\
\noalign{\smallskip}
2009bu & NO & 54909.2$^{+0.5}_{-1.2}$ & 1.28$^{+0.04}_{-0.08}$ & 9.55$^{+0.43}_{-0.34}$ & 8.04$^{+0.13}_{-0.10}$ & 7.19$^{+0.10}_{-0.12}$ & 4.64$^{+0.15}_{-0.12}$ & 455$^{+7}_{-6}$ & 0.58$^{+0.04}_{-0.05}$ & --- & 0.49$^{+0.13}_{-0.14}$ \\
\noalign{\smallskip}
2009bz & NO & 54919.0$^{+0.6}_{-1.4}$ & 1.34$^{+0.05}_{-0.09}$ & 13.36$^{+1.16}_{-3.35}$ & 10.70$^{+1.25}_{-2.52}$ & 8.80$^{+0.84}_{-1.73}$ & 5.85$^{+0.42}_{-1.05}$ & 708$^{+49}_{-245}$ & 0.66$^{+0.05}_{-0.04}$ & --- & 0.52$^{+0.18}_{-0.17}$ \\
\noalign{\smallskip}
\noalign{\smallskip}
\end{longtable}
\tablefoot{Results are characterised by the median of the marginal distributions, adopting the 16th and 84th percentiles as the lower and upper uncertainties. Columns: (1) SN name; (2) indicates whether the SN is a gold event or not; (3) explosion epoch; (4) scale factor; (5) progenitor initial mass; (6) ejecta mass; (7) hydrogen-rich envelope mass; (8) total mass of hydrogen; (9) progenitor radius; (10) explosion energy; (11) $^{56}$Ni mass; (12) $^{56}$Ni mixing as a fraction of the pre-SN mass.}
\end{landscape}
}

\begin{figure}
\centering
\includegraphics[width=0.47\textwidth]{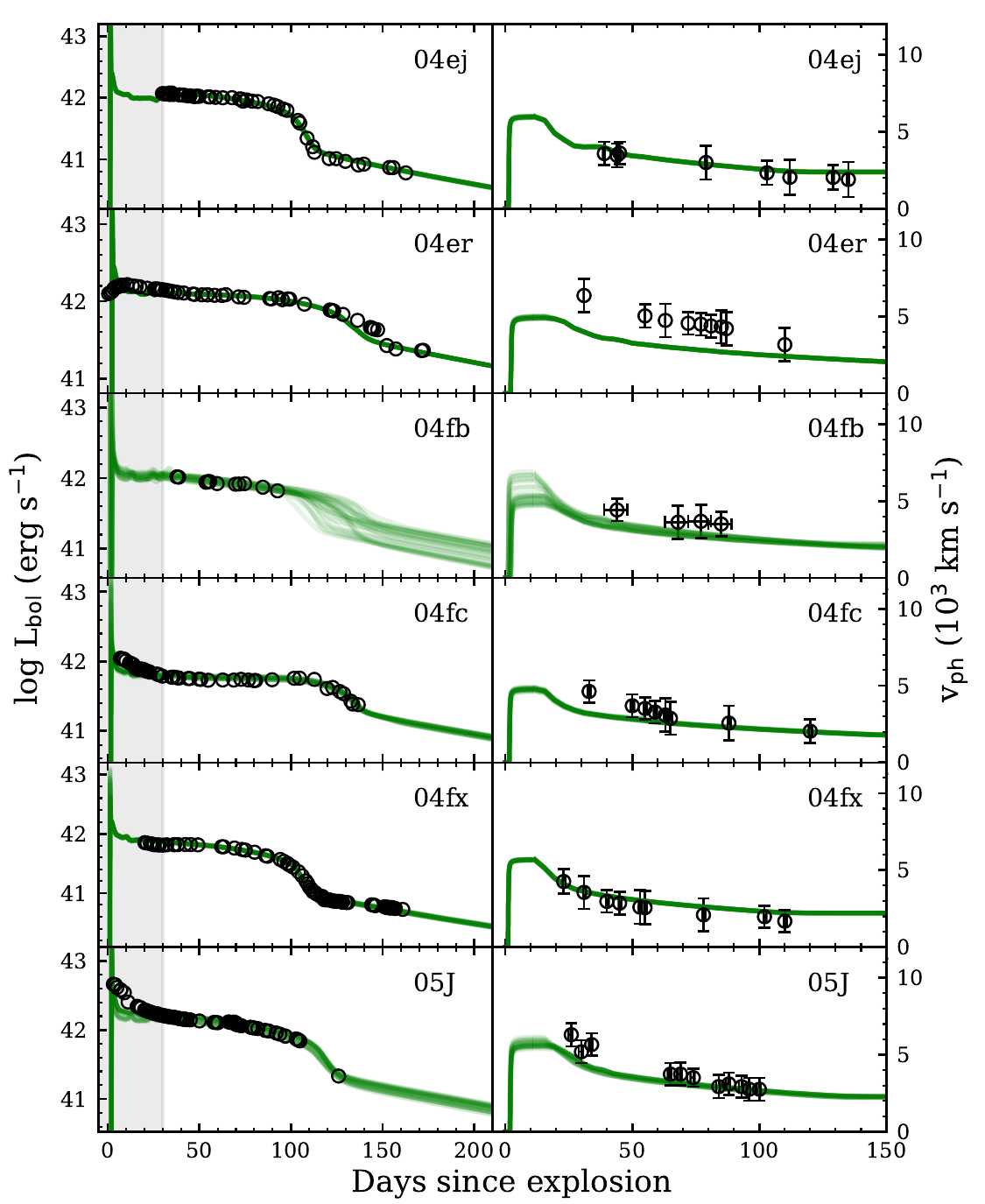} \includegraphics[width=0.47\textwidth]{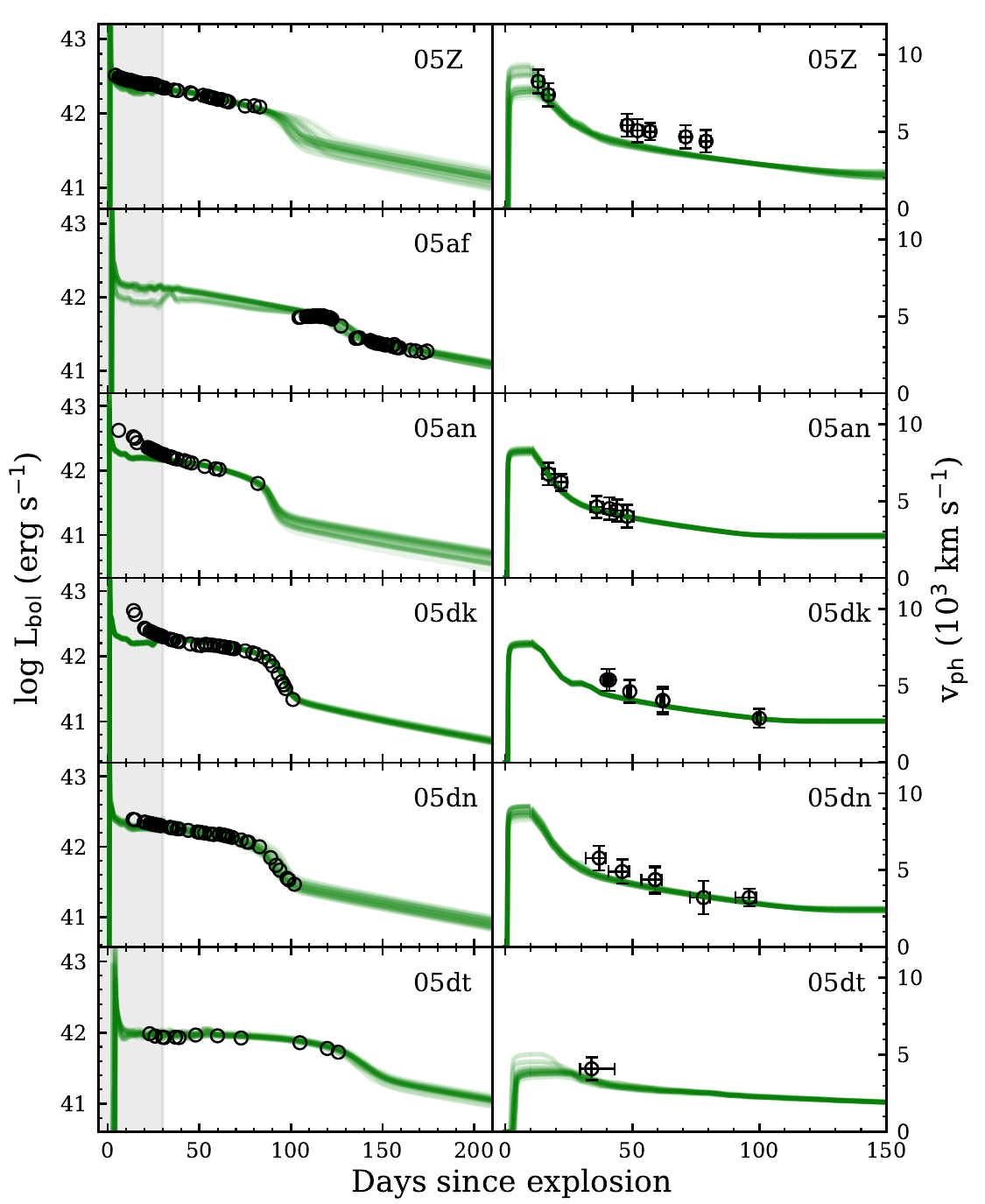}
\caption{Comparison between models and observations. The open circles show the observed bolometric LCs (left) and \ion{Fe}{ii} velocities (right). Solid lines represent 30 models randomly chosen from the probability distribution. The panels present SNe in order of their discovery dates, from SN~2004ej to SN~2005dt. Grey shaded regions show the early data we removed from the fitting. The errors in the observed bolometric LCs are not plotted, for better visualisation.}
\label{fig:fits1}
\end{figure}

\begin{figure*}
\centering
\includegraphics[width=0.47\textwidth]{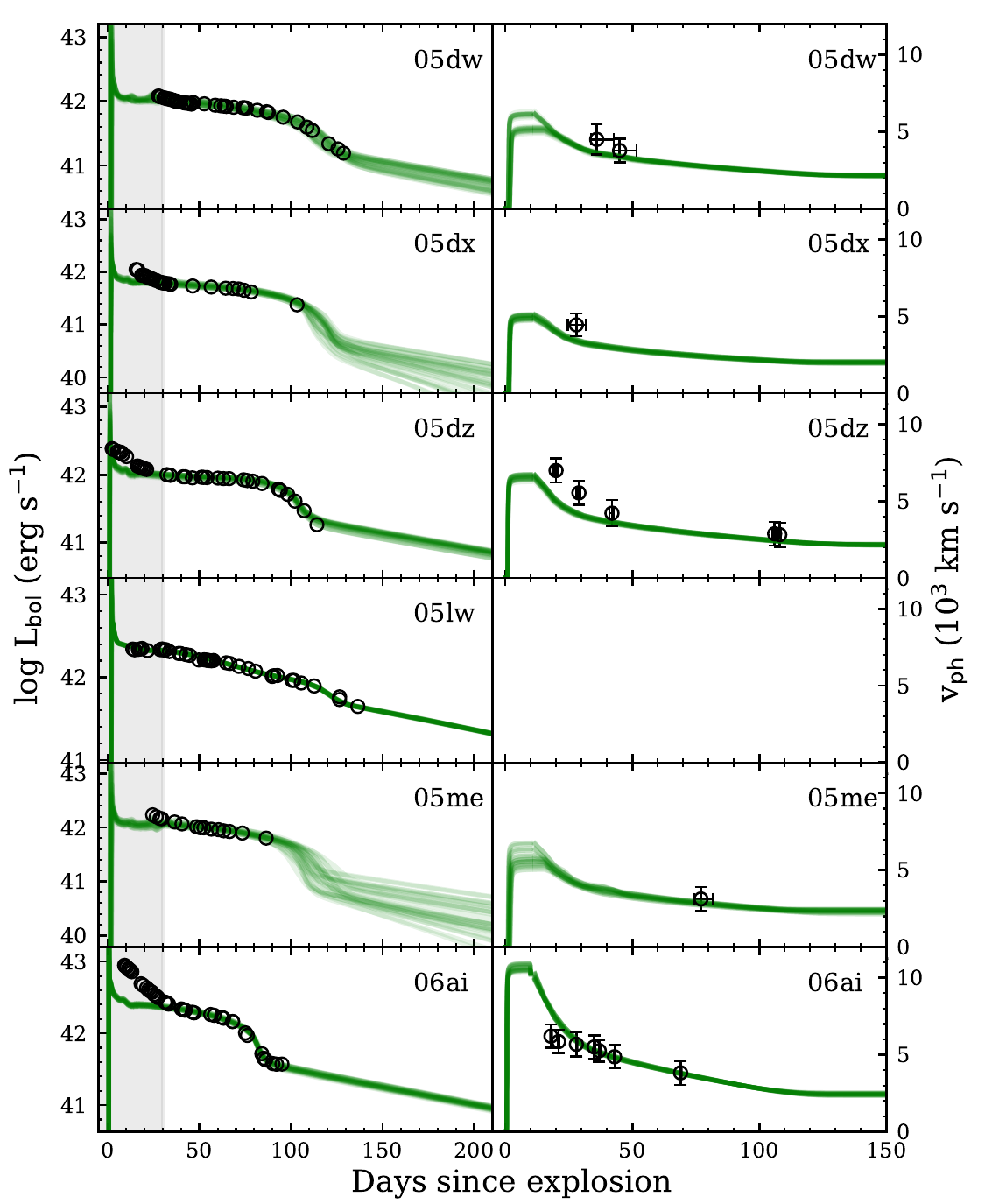}
\includegraphics[width=0.47\textwidth]{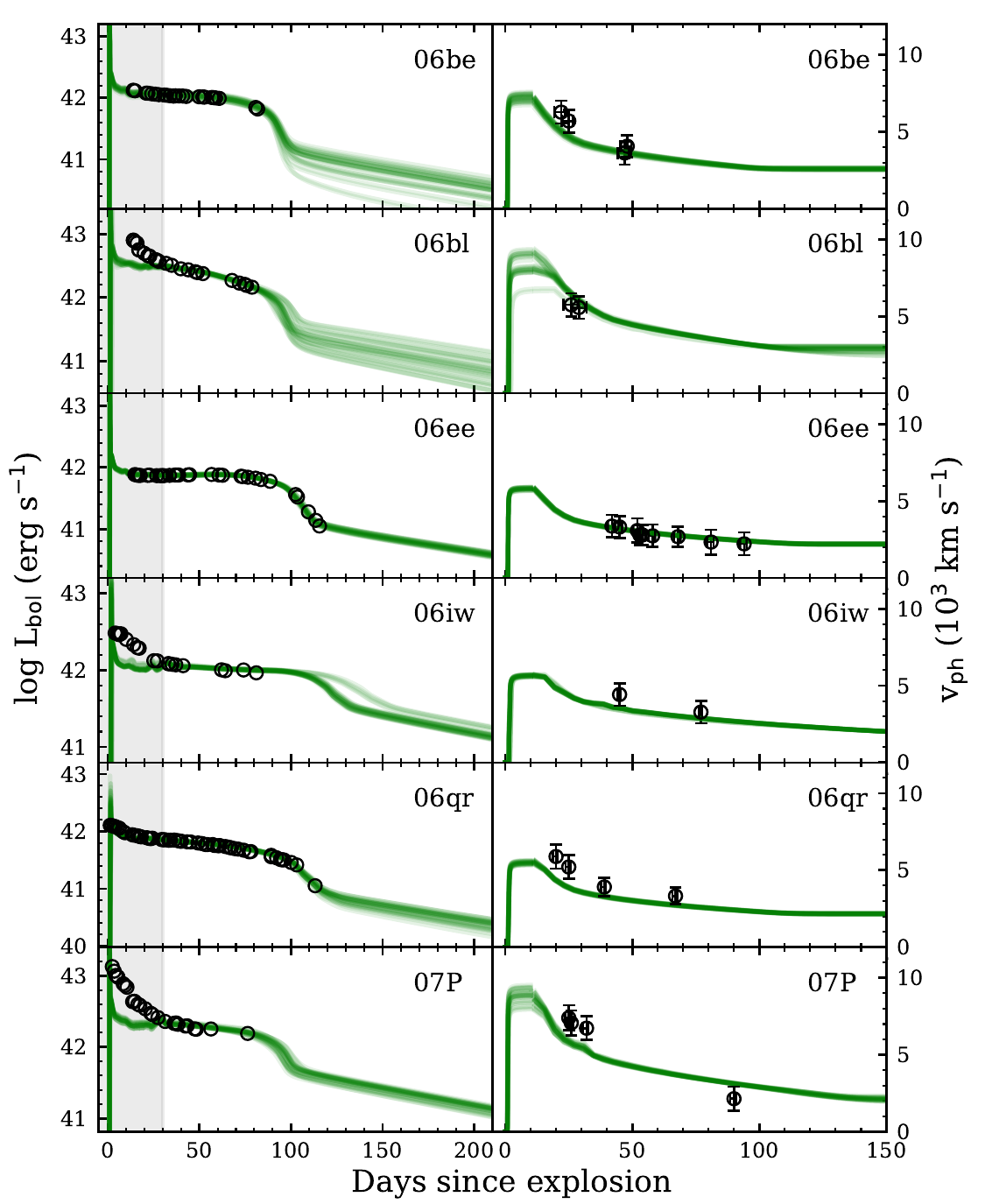}
\includegraphics[width=0.47\textwidth]{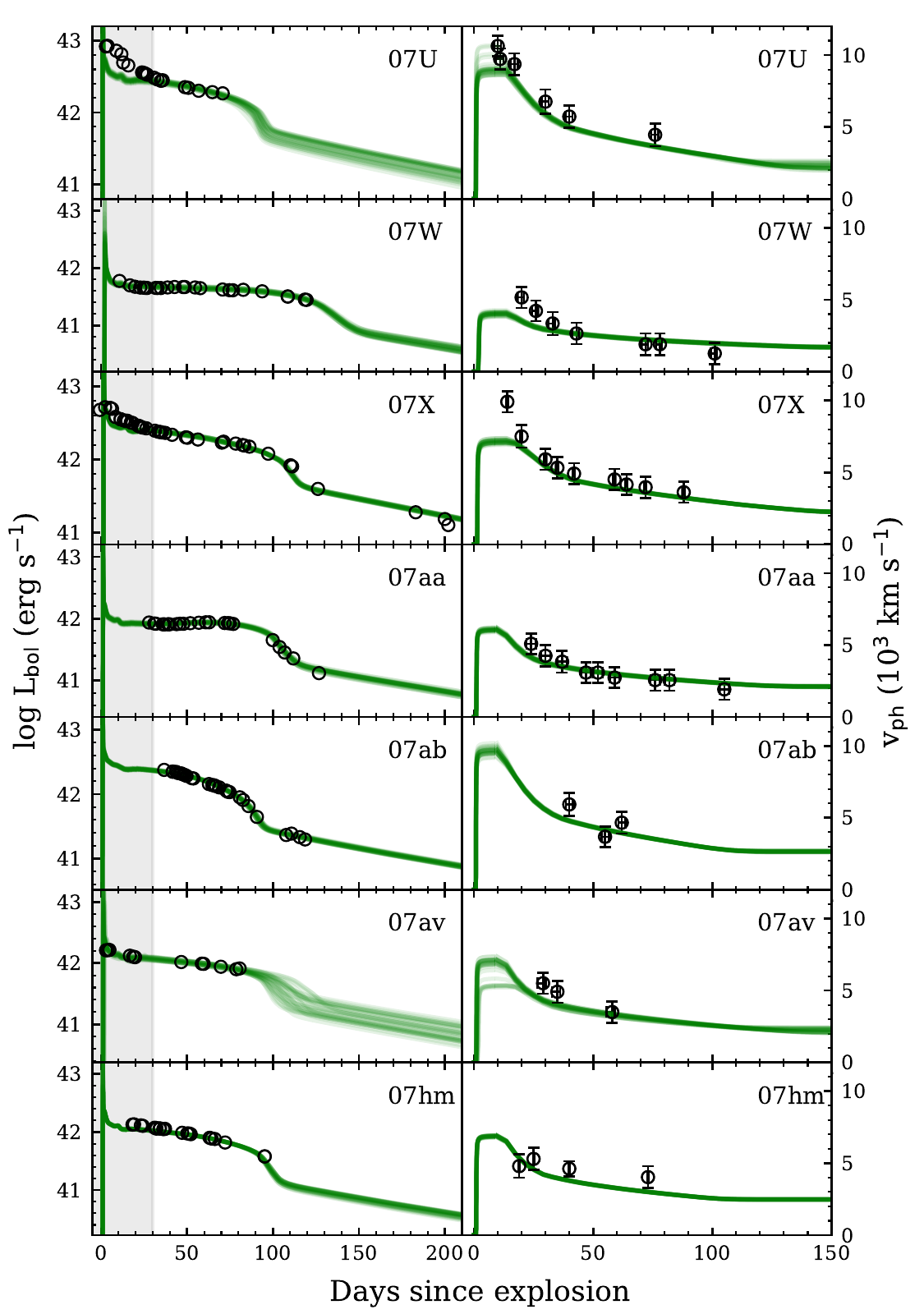} \includegraphics[width=0.47\textwidth]{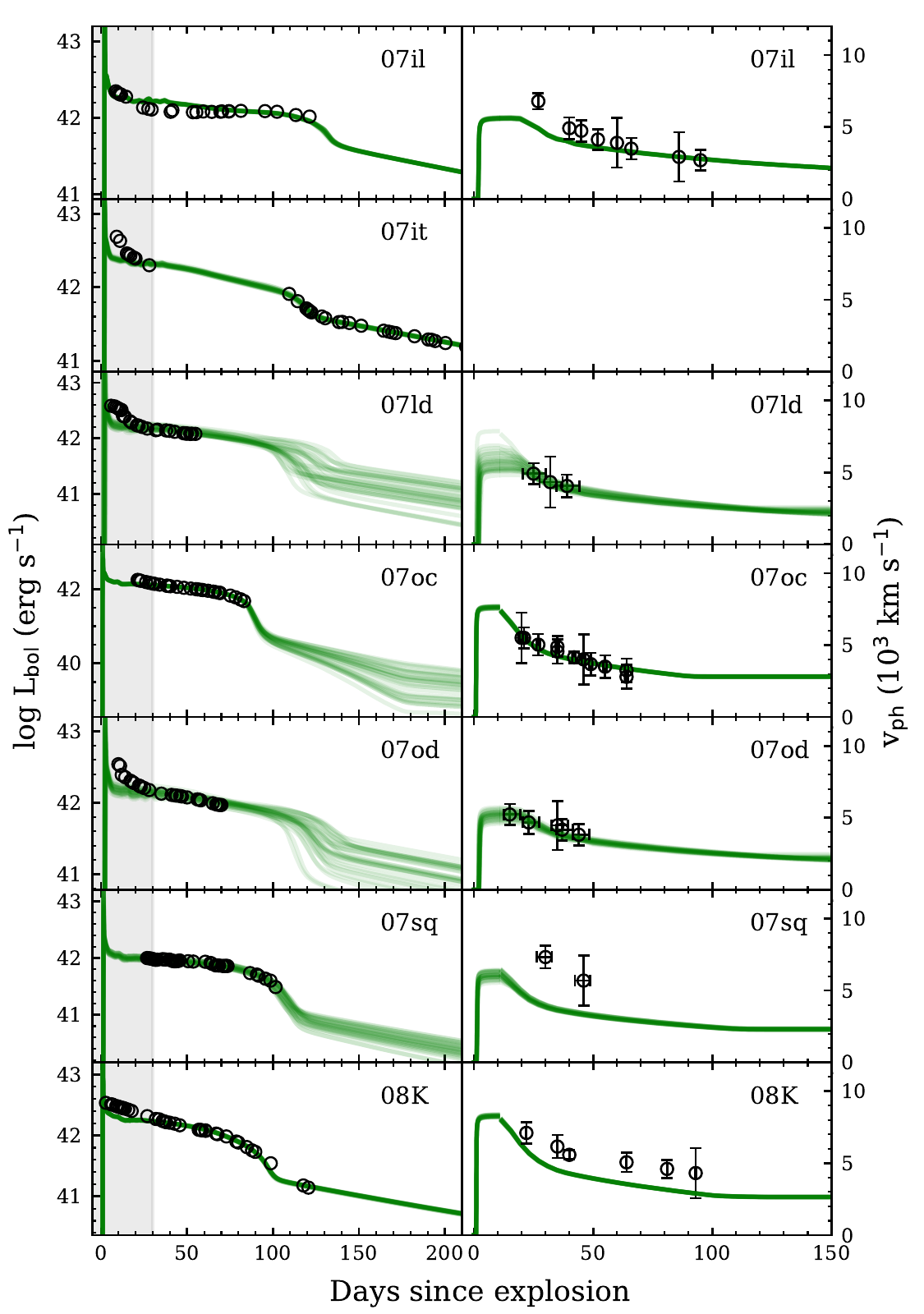}
\caption{Same as in Fig.~\ref{fig:fits1}, but from SN~2005dw to SN~2008K.}
\label{fig:fits2}
\end{figure*}

\begin{figure}
\centering
\includegraphics[width=0.47\textwidth]{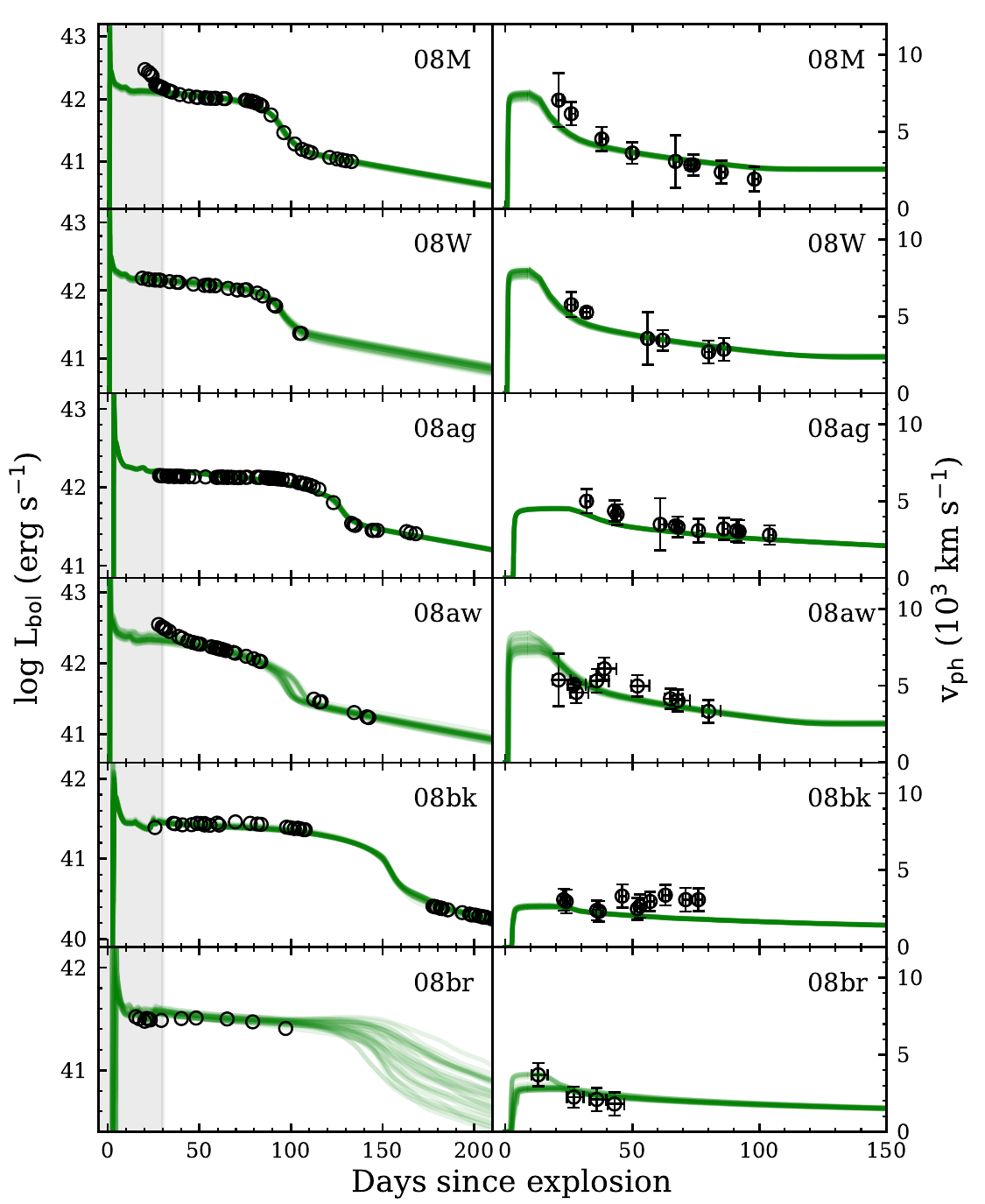}
\includegraphics[width=0.47\textwidth]{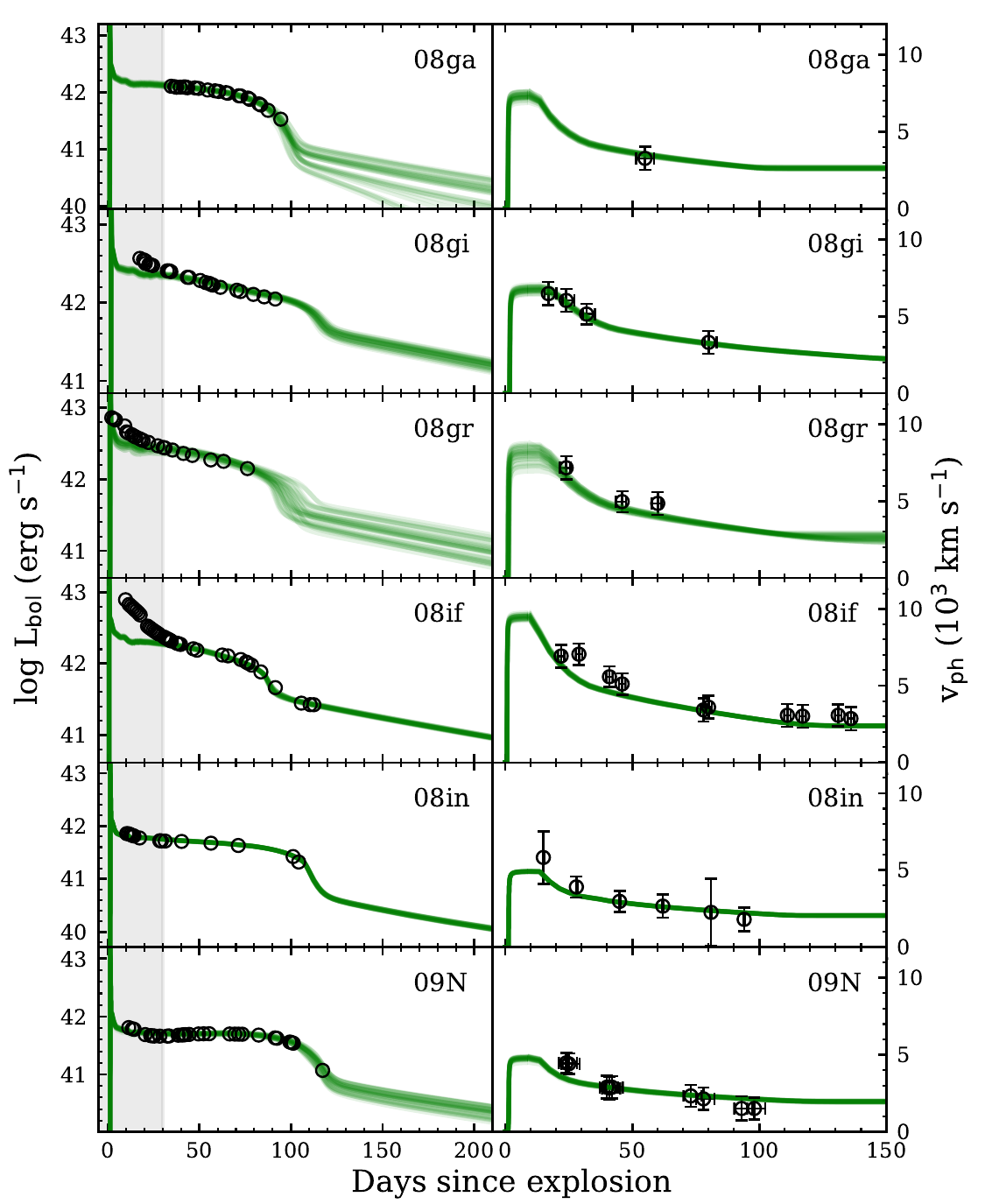}
\caption{Same as in Fig.~\ref{fig:fits1}, but from SN~2008M to SN~2009N.}
\label{fig:fits3}
\end{figure}

\begin{figure}
\centering
\includegraphics[width=0.47\textwidth]{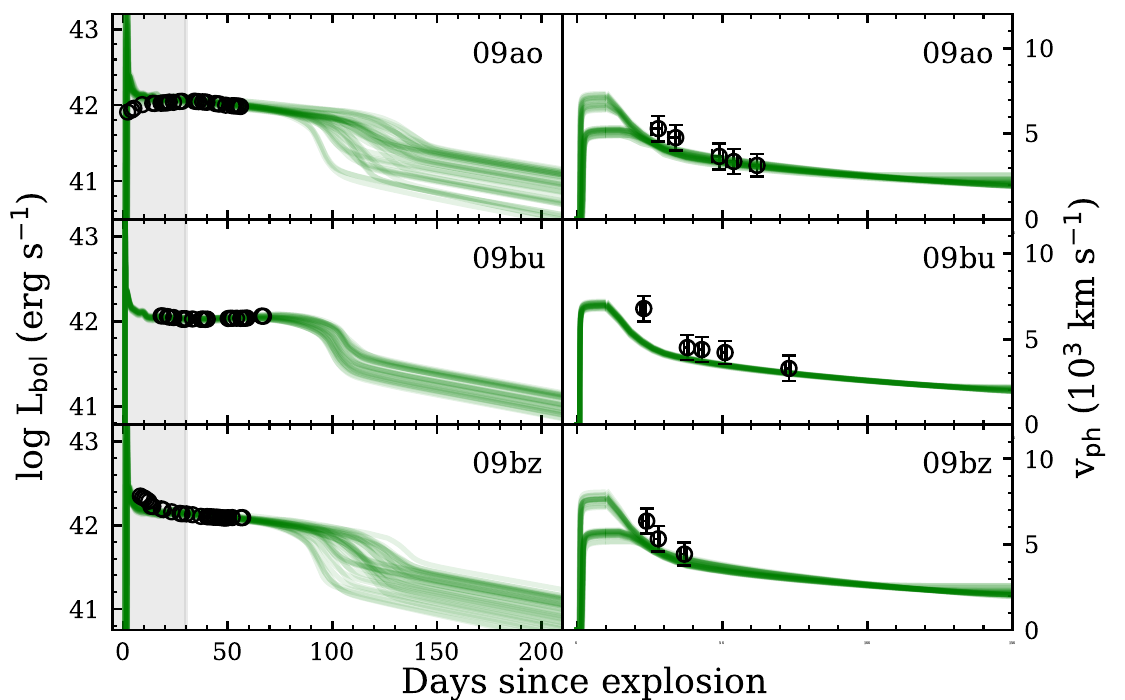}
\caption{Same as in Fig.~\ref{fig:fits1}, but from SN~2009ao to SN~2009bz.}
\label{fig:fits4}
\end{figure}

\begin{figure}
\centering
\includegraphics[width=0.4\textwidth]{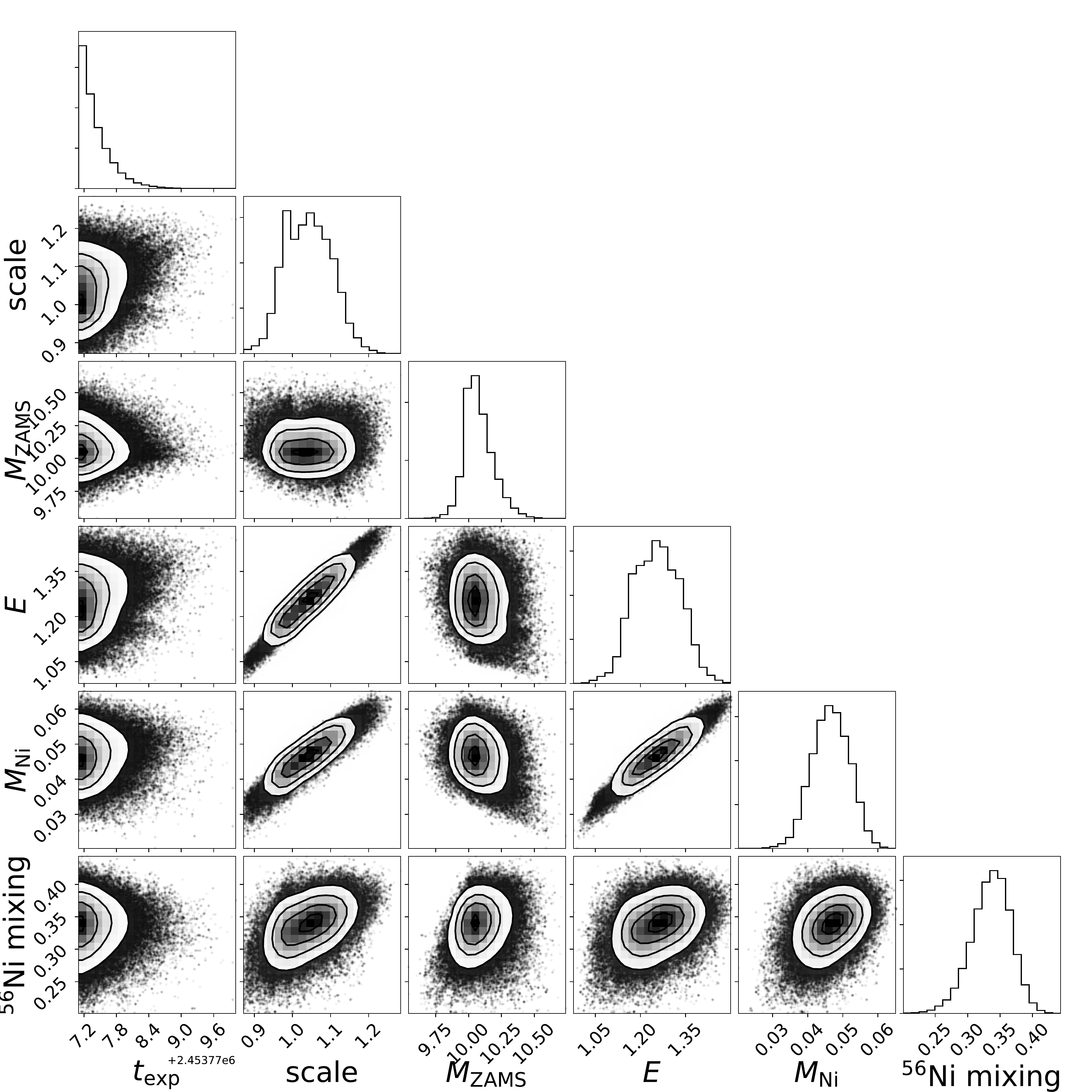}
\caption{Corner plot of the joint posterior probability distribution of the parameters for SN~2006ai.}
\label{fig:corner}
\end{figure}

Figures~\ref{fig:fits1}, \ref{fig:fits2}, \ref{fig:fits3}, and \ref{fig:fits4} present bolometric LCs and \ion{Fe}{ii} velocities for the 53~\sneii\ included in this study in comparison with the models drawn from the posterior distribution of the parameters. 
\ion{Fe}{ii} velocities are not available for SNe~2005af, 2005lw, and 2007it.
Table~\ref{table:results} presents the physical parameters derived from the hydrodynamical modelling of \snii\ bolometric LCs and expansion velocities.
In Table~\ref{table:results}, we also include the estimates of \mej, \mhenv, \mh, and \ra. We note that these progenitor properties are not model parameters and, therefore, they are not fitted. These values are linearly interpolated from the \mzams\ derived from the fitting.
Additionally, an example corner plot with the posterior probability distribution of the parameters is presented in Fig.~\ref{fig:corner}.

\FloatBarrier

\section{High-mass progenitor fits}
\label{app:high_mass}

\begin{figure}[h]
\centering
\includegraphics[width=0.49\textwidth]{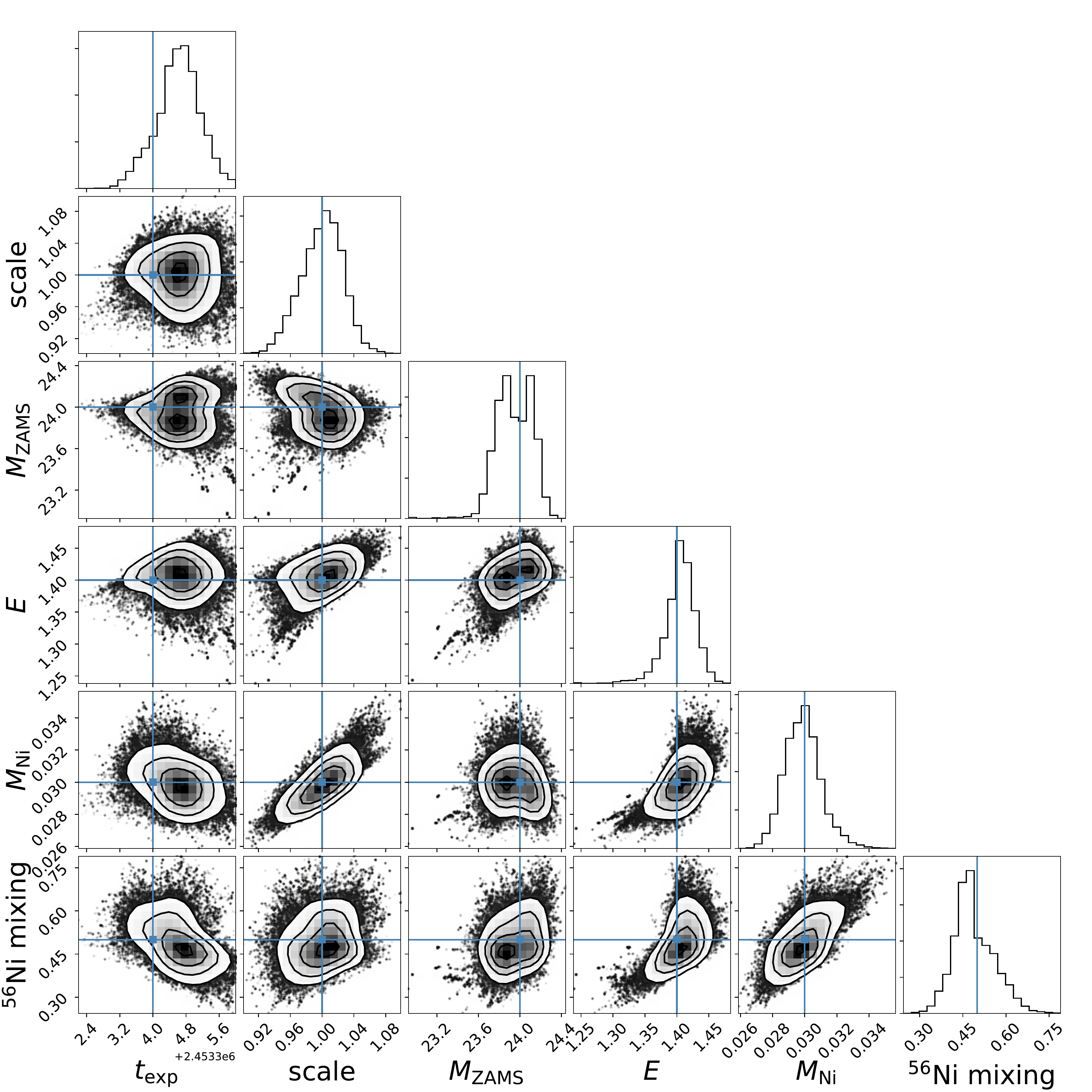}
\caption{Corner plot of the joint posterior probability distribution of the parameters when a high-mass progenitor model is considered as observations for our fitting technique. Blue squares represent the input physical parameters of the model.}
\label{fig:high_mass_corner}
\end{figure}

We mapped LCs and expansion velocities from high-mass models as observations to test our fitting technique.
Figure~\ref{fig:high_mass_corner} shows a corner plot of the joint posterior probability distribution of the parameters for a model of \mzams~=~24~\ms, \e~=~1.4~foe, \mni~=~0.03~\ms, and \mix~=~0.5. The blue squares represent the input physical parameters of the model.

\FloatBarrier

\section{High-mass progenitor models evolved with enhanced mass loss}
\label{app:nonstd}

\begin{figure}[h]
\centering
\includegraphics[width=0.47\textwidth]{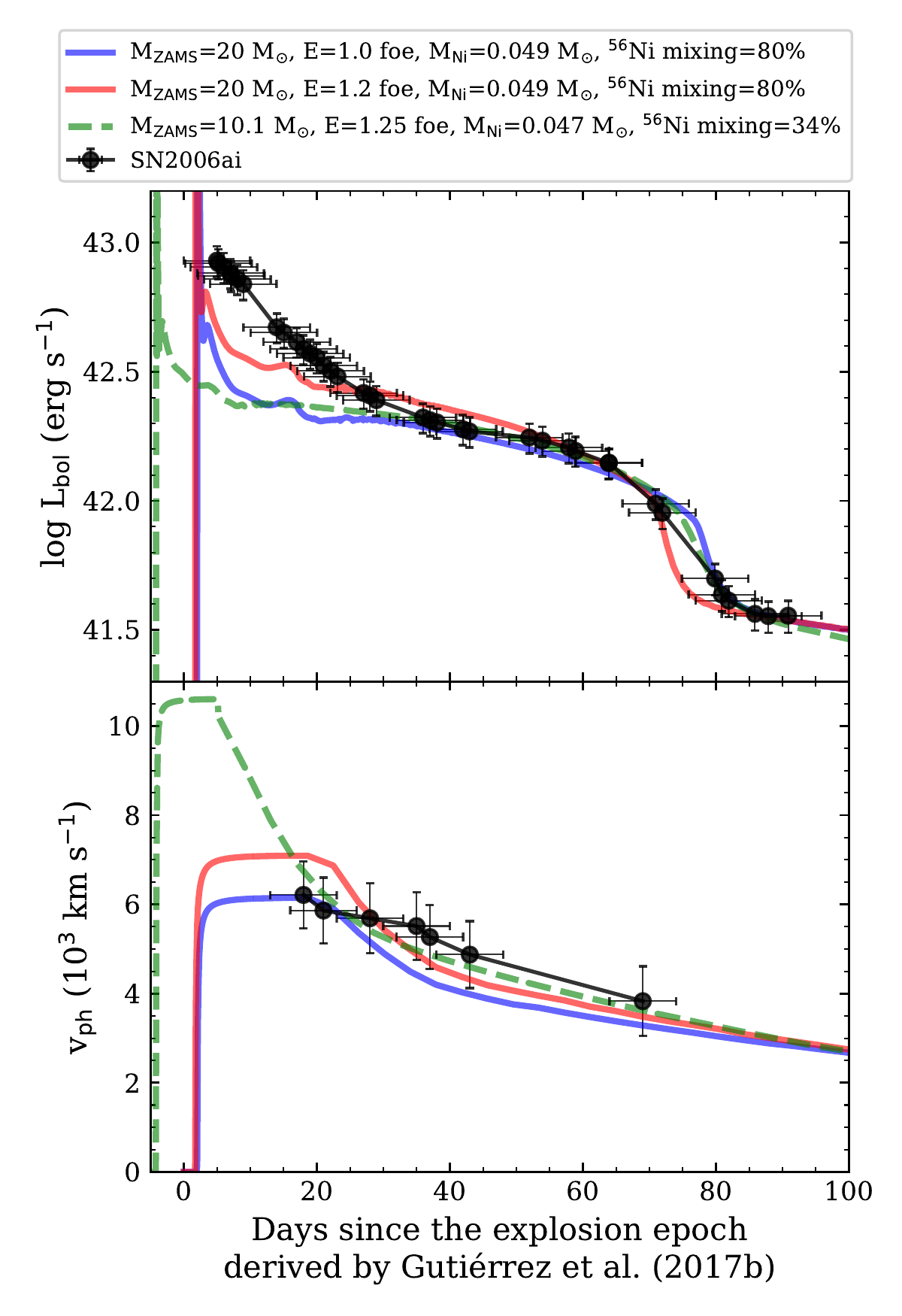}
\caption{Observed bolometric LC (top panel) and \ion{Fe}{ii} $\lambda$5169 line velocities of SN~2006ai (black markers). Solid lines represent the high-\mzams\ models with increased mass loss that best reproduce the observations. Dashed lines represent the maximum a posteriori model using our standard grid of models. This model is corrected by the scale factor and the explosion epoch found in Sect.~\ref{sec:results}. Since our estimation suggests an earlier explosion epoch, the model shown as dashed lines starts before zero on the $x$ axis (where zero corresponds to the explosion epoch derived by \citealt{gutierrez+17I}).}
\label{fig:06ai_high_mass}
\end{figure}

In Sect.~\ref{sec:imf_incompatibility} we proposed that one possible solution for the IMF incompatibility could be to consider a higher rate of mass loss, in addition to a larger mixing-length parameter. A detailed investigation is beyond the scope of this work; however, we give an example to show that our suggestion allows high-\mzams\ progenitors to fit some \sneii\ in the CSP-I sample.

We used \texttt{MESA} \citep{paxton+11,paxton+13,paxton+15,paxton+18,paxton+19} to obtain pre-SN models employing the same configurations detailed in Sect.~\ref{sec:methods}, with two exceptions. We adopted a wind scaling factor of $\eta$\,=\,1.5 (i.e. 50\% higher than standard) and a mixing-length parameter of three.
Then, we used the code presented in \citet{bersten+11} to obtain bolometric LC and photospheric velocity models.
We generated a small grid of models for \mzams\,=\,20 and 22~\ms, \e\,=\,0.8, 1.0, and 1.2~foe, \mni\,=\,0.03~\ms, and \mix\,=\,0.8 (given in fraction of the pre-SN mass).
Thus, we obtained a total of six hydrodynamical models.

We used these six models to compare with all CSP-I \sneii\ with bolometric luminosities during the photospheric phase and -- at least -- the transition to the radioactive tail and velocity measurements.
The observations of a small number of \sneii\ in the CSP-I sample can be reasonably reproduced by some of these models. We discuss the case of SN~2006ai as an example.
The 20~\ms\ models reproduce the plateau phase and the transition to the radioactive phase. The models display lower luminosities during the radioactive tail phase, indicative of a higher \Ni\ mass in the ejecta of SN~2006ai than that assumed in the models. For this reason, we carried out new models with higher \Ni\ masses until the observed radioactive tail phase is reproduced by our models.
Figure~\ref{fig:06ai_high_mass} shows the models that reproduce the observations of SN~2006ai\footnote{We note that the first 30~days of evolution are not considered.}.
Both models assume \mzams\,=\,20~\ms, \mni\,=\,0.049~\ms, and \mix\,=\,0.8. Model~1 and Model~2 were exploded with \e\,=\,1.0 and 1.2~foe, respectively.
We stress that this test seeks reasonable agreement between models and observations and not detailed fits to observations. A fine tuning of the model parameters is required to find the best-fitted model.

This test shows that a high-\mzams\ progenitor evolved with enhanced mass loss and larger mixing-length parameter can reproduce the observations of a \snii\ (i.e. in the direction to solve the IMF incompatibility). However, a deeper analysis is necessary for stronger conclusions.

\end{appendix}
\end{document}